\documentclass[11pt,a4paper]{article}
\usepackage[intlimits]{amsmath}
\usepackage{calc}

\usepackage{jheppub}
\makeatletter
\renewcommand{\@fpheader}{Published in
  \href{http://dx.doi.org/10.1007/JHEP12(2011)076}{JHEP 12 (2011) 076}}
\makeatother

\graphicspath{{axodraw/}}
\DeclareGraphicsExtensions{.eps,{}}
\newcommand{\gscale}{0.7}
\newcommand{\gscaler}{0.65}

\allowdisplaybreaks

\newcommand{\Oc}{\mathcal{O}}
\DeclareMathOperator{\Rep}{Re}
\DeclareMathOperator{\Res}{Res}
\newcommand{\rd}{\mathrm{d}}
\newcommand{\rD}{\mathrm{D}}
\newcommand{\Rb}{\mathbb{R}}
\newcommand{\Zb}{\mathbb{Z}}
\newcommand{\eps}{\epsilon}
\newcommand{\gammaE}{\gamma_{\text{E}}}
\newcommand{\hyperF}[2]{{}_{#1}F_{#2}}
\newcommand{\Li}[1]{\mathrm{Li}_{#1}}
\newcommand{\UV}{{\text{UV}}}
\newcommand{\IR}{{\text{IR}}}
\newcommand{\Rc}{R_{\text{c}}}
\newcommand{\Rnc}{R_{\text{nc}}}
\newcommand{\Nc}{N_{\text{c}}}

\begin{document}

\begin{flushright}
  \begin{tabular}{@{}l@{}}
  TTK-11-53 \\
  SFB/CPP-11-61
  \end{tabular}
\end{flushright}

\title{Foundation and generalization of the expansion by regions}
\author{Bernd Jantzen}
\affiliation{
  Institut f\"ur Theoretische Teilchenphysik und Kosmologie,
  RWTH Aachen University, \\
  52056 Aachen, Germany}
\emailAdd{jantzen@physik.rwth-aachen.de}
\abstract{
  The ``expansion by regions'' is a method of asymptotic expansion
  developed by Beneke and Smirnov in 1997. It expands the integrand
  according to the scaling prescriptions of a set of regions and integrates
  all expanded terms over the whole integration domain. This method has
  been applied successfully to many complicated loop integrals, but a
  general proof for its correctness has still been missing.

  This paper shows how the expansion by regions manages to reproduce the
  exact result correctly in an expanded form and clarifies the conditions
  on the choice and completeness of the considered regions. A generalized
  expression for the full result is presented that involves additional
  overlap contributions. These extra pieces normally yield scaleless
  integrals which are consistently set to zero, but they may be needed
  depending on the choice of the regularization scheme.

  While the main proofs and formulae are presented in a general and concise
  form, a large portion of the paper is filled with simple, pedagogical
  one-loop examples which illustrate the peculiarities of the expansion by
  regions, explain its application and show how to evaluate contributions
  within this method.
}
\keywords{NLO Computations, Standard Model}
\arxivnumber{1111.2589}
\maketitle

\section{Introduction}
\label{sec:intro}

When loop integrals involve many different scales from masses and
kinematical parameters, it can be hard or even impossible to evaluate them
exactly. The integrand may be simplified before integration by exploiting
hierarchies of parameters and expanding in powers of small parameter
ratios. When these expansions are done naively, neglecting their breakdown
in certain parts of the integration domain, new singularities may be
generated and important contributions to the full result can be missed. A
proper treatment requires sophisticated methods of \emph{asymptotic
  expansions}. One of them is the so-called ``strategy of regions'' or
``expansion by regions'' developed by M.~Beneke and
V.A.~Smirnov~\cite{Beneke:1997zp}. The recipe for applying the expansion by
regions to a loop integral reads as
follows~\cite{Beneke:1997zp,Smirnov:1998vk,Smirnov:1999bza,Smirnov:2002pj}:
\begin{enumerate}
\item Divide the space of the loop momenta into various regions and, in
  every region, expand the integrand into a Taylor series with respect to
  the parameters that are considered small there.
\item Integrate the integrand, expanded in the appropriate way in every
  region, over the \emph{whole integration domain} of the loop momenta.
\item Set to zero any scaleless integral.
\end{enumerate}
The sum of these contributions yields the full result of the loop integral
in an expanded form. This recipe will be illustrated in the examples of the
following sections.

However, despite the successful application of the expansion by regions in
many loop calculations, each of the steps in the recipe above raises
questions:
\begin{itemize}
\item In step~2 all expanded terms are integrated over the whole
  integration domain, neglecting the domains of convergence of the
  expansions. This generally leads to new singularities which obviously
  must be cancelled between the contributions of the various regions. How
  is this cancellation ensured?
\item In the original integral each point of the integration domain
  contributes exactly once. According to step~2 each point of the
  integration domain contributes once per region. How can this double- or
  multiple-counting of contributions be correct?
\item How do we have to choose the regions in step~1? And how do we know
  that the chosen set of regions is complete?
\item Although it seems natural to eliminate scaleless integrals when using
  dimensional or analytic regularization, we can ask: What is the role of
  the scaleless integrals in step~3?
\end{itemize}
These questions will be addressed in the current paper.

The developers of the expansion by regions have introduced their method
using examples and ``heuristic motivations'' (e.g.\ related to analogies
between the regions and degrees of freedom in effective theories). They
have shown the validity of the method for some types of expansions (in
particular Euclidean-type limits like off-shell large-momentum expansion or
large-mass expansion) through the agreement with existing and proven
expansions by subgraphs~\cite{Smirnov:2002pj}. They admit, however, that
they cannot give a general mathematical proof of their
prescriptions~\cite{Beneke:1997zp} and that ``it is not guaranteed that
expansion by regions works in all situations''~\cite{Smirnov:2002pj}.

A practical problem in the application of the expansion by regions is the
correct choice of the regions. In the original paper~\cite{Beneke:1997zp}
the authors determine the relevant regions from the structure of the poles
of the propagators in the loops. They close the contour of the integration
over the zero-component of the loop momentum and study its scaling at the
residues depending on the size of the spatial components. In general,
relevant regions can often be found by looking at the structure of the
integrand and at singularities which arise in the given parameter limit. It
does not matter to consider more regions than necessary: The irrelevant
regions will simply produce scaleless contributions which are set to
zero. The tricky point here is to avoid double-counting of regions which
look different but yield equivalent expansions.

Alternatively, the expansion by regions has been applied to the
alpha-parameter representation of loop
integrals~\cite{Smirnov:1999bza,Smirnov:2002pj}. The double-counting of
equivalent regions should not occur here, but there exist at least some
regions (in particular the ``potential region'' in threshold expansions)
which cannot easily be identified in the language of the alpha
parameters. Based on this approach, an algorithm for finding the regions in
alpha-parameter representations has recently been
developed~\cite{Pak:2010pt}.
In the present paper I stick exclusively to the version of the expansion by
regions which is applied directly to loop integrals because I find this
original variant of the method more natural. However, the formalism which I
present is general and can be applied to the asymptotic expansion of many
types of integrals, including alpha-parameter representations.

An important check for the completeness of the regions is the cancellation
of singularities. The sum of all contributions must not be more singular
than the original integral, and if additional singularities persist, then
probably the contribution from another region is missing. This check is,
however, not sufficient to guarantee the completeness of regions, because
some regions or a subset of them can also yield non-singular contributions
such that their absence could remain undetected.

Another possibility of finding all relevant regions for an integral in a
given parameter limit is the one which I have successfully used in many
calculations
(e.g.~\cite{Jantzen:2006jv,Jantzen:2006ys,Denner:2006jr,Denner:2008yn}):
Evaluate the full integral with the propagators raised to generic powers in
terms of as many Mellin--Barnes representations~\cite{Usyukina:1975yg} as
needed. Extract the leading-order asymptotic expansion by closing those
Mellin--Barnes integrals which involve the expansion parameter and taking
the residue of the first pole next to the contour. This yields a sum of
terms reproducing the exact result up to higher powers of the expansion
parameter than those already present. The terms often still contain
Mellin--Barnes integrals and may be too complicated to evaluate them
further. But each term is characterized by a homogeneous dependence on the
expansion parameter raised to some power which is a function of the
space--time dimension~$d=4-2\eps$ and of the propagator powers. An
examination of each term's dependence on the expansion parameter is usually
able to tell the corresponding region needed for producing this
contribution. If the asymptotic expansion by Mellin--Barnes representations
is performed correctly, the prescription described here yields all the
relevant regions for a subsequent application of the expansion by regions.

The correspondence between the contributions of the expansion by regions
and the asymptotic expansion via Mellin--Barnes representations is, of
course, as heuristic as the derivations and justifications known so far for
the expansion by regions.
In the present paper I follow a different approach: I start from a general
integral for which an expansion is required and transform the expression
step by step, in a mathematically well-defined way, into a sum of
expansions which can be identified with the ones originating from the
expansion by regions. The resulting expression contains additional
\emph{overlap contributions} which are absent in the usual prescription for
the expansion by regions. While these extra pieces normally yield scaleless
integrals which are consistently set to zero according to step~3 above,
there are cases --- depending on the choice of the regularization
scheme --- where these overlap contributions are present and required for the
correct result.
Such overlap contributions have also been introduced in effective-theory
treatments as ``zero-bin subtractions'' (see
e.g.~\cite{Manohar:2006nz,Chiu:2009yx}).

The basic idea of the formalism which I develop in this paper goes back to
a one-dimensional toy example~\cite{Beneke:1997note} (see also section~3.2
of~\cite{Smirnov:2002pj}). I do not treat this one-dimensional toy
integral, but start directly with a simple \mbox{$d$-dimensional} loop
integral in section~\ref{sec:2point} before presenting a first version of
the formalism for general integrals in section~\ref{sec:formalism}. A
generalized version of the formalism is elaborated in
section~\ref{sec:formalismnc}, where one restriction of the first version
is relaxed. Examples which illustrate the application of the formalism and
explain the evaluation of loop integrals with the expansion by regions are
shown in sections \ref{sec:threshold}, \ref{sec:Sudakov}
and~\ref{sec:forward}. The last of these examples demonstrates the
relevance of overlap contributions under certain circumstances.
The main statements and formulae are summarized and discussed in the
conclusions of section~\ref{sec:conclusions}.
Details from the evaluation of the expanded loop integrals have been
shifted to appendix~\ref{app:details}. Finally
appendix~\ref{app:finiteboundary} treats an example with a finite
integration boundary which illustrates a subtlety raised in
section~\ref{sec:formalism}.

The order of the general parts (sections \ref{sec:formalism},
\ref{sec:formalismnc} and~\ref{sec:conclusions}) and illustrating examples
(sections \ref{sec:2point}, \ref{sec:threshold}, \ref{sec:Sudakov}
and~\ref{sec:forward}) has been chosen from a pedagogic viewpoint in order
to facilitate the understanding especially for readers who are not very
familiar with the expansion by regions. More experienced readers may skip
the examples and concentrate on the general sections for a quick study of
the main statements. The examples, however, also show how the general
formalism can actually be applied to loop integrals and how its conditions
on the regions are checked. Later examples use notations and conventions
introduced in earlier examples.

\section{Example: off-shell large-momentum expansion}
\label{sec:2point}

The two-point one-loop integral of the first example is defined by the
expression
\begin{equation}
  \label{eq:2point}
  F = \int\!\rD k \, I
\end{equation}
with the integrand $I = I_1 I_2$, the propagators
\begin{equation}
  \label{eq:2pointprops}
  I_1 = \frac{1}{\bigl((k+p)^2\bigr)^{n_1}}
      = \frac{1}{(k^2 + 2k\cdot p + p^2)^{n_1}}
  \quad \text{and} \quad
  I_2 = \frac{1}{(k^2 - m^2)^{n_2}}
\end{equation}
and the integration measure
\begin{equation}
  \label{eq:Dk}
  \int\!\rD k \equiv \mu^{2\eps} \, e^{\eps\gammaE}
    \int\!\frac{\rd^d k}{i\pi^{d/2}}
  \,,
\end{equation}
where $d=4-2\eps$ is the space--time dimension, $\mu$ is the scale of
dimensional regularization, and $\gammaE \approx 0.577216$ is Euler's
constant. The usual infinitesimal imaginary part in the Feynman propagators
is understood and has been dropped in the notation for brevity.

\begin{figure}[t]
  \centering
  \includegraphics[scale=\gscale]{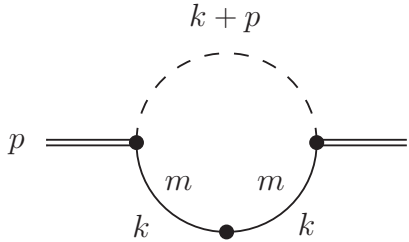}
  \caption{Two-point loop integral for off-shell large-momentum expansion.}
  \label{fig:2point}
\end{figure}

The integral~$F$ depends on the external momentum~$p$ and the mass~$m$. We
are interested in the off-shell large-momentum limit
\[
  |p^2| \gg m^2
\]
and look for an asymptotic expansion in powers of $m^2/p^2$.  The integral
also depends on the propagator powers $n_1$ and $n_2$, and we focus
particularly on the case $n_1=1$, $n_2=2$ which is depicted in
figure~\ref{fig:2point}
and for which the integral is finite for both small and large~$k$.
Exact results for this integral can easily be obtained and expanded in
$m^2/p^2$ for comparison with the asymptotic expansion.

In the following two subsections we first treat this loop integral
according to the recipe formulated in section~\ref{sec:intro} for the
expansion by regions before proving the correctness of this approach by
independent mathematical transformations of the integral.

\subsection{Expansion by regions in the large-momentum limit}
\label{sec:2pointEbR}

The first propagator in~(\ref{eq:2pointprops}) is characterized by the
large momentum~$p$, whereas the second propagator is characterized by the
small mass~$m$. It is therefore natural to assume that the two regions of
relevance to this problem are
\begin{itemize}
\item the \textbf{hard region \boldmath$(h)$}, where $k \sim p$,
\item and the \textbf{soft region \boldmath$(s)$}, where $k \sim m$.
\end{itemize}
By $k \sim p$ we mean that all components of the loop momentum~$k$ are of
the order of (``scale like'') the (overall size of) the momentum~$p$, and
similarly for $k \sim m$. Instead of dividing the integration domain into
subdomains with explicit boundaries, the regions simply define scaling
prescriptions for the loop momentum on the basis of which we are able to
perform expansions of the integrand.

For the hard region we consider $|k^2| \gg m^2$ and expand the second
propagator in~(\ref{eq:2pointprops}), while the first propagator remains
unchanged:
\begin{equation}
  \label{eq:2pointhexp}
  I_1 \to T^{(h)} I_1 = I_1 \,,\qquad
  I_2 \to T^{(h)} I_2 \equiv \sum_j T^{(h)}_j I_2
    = \sum_{j=0}^\infty \frac{(n_2)_j}{j!} \, \frac{(m^2)^j}{(k^2)^{n_2+j}}
    \,,
\end{equation}
where $T^{(h)}$ is the expansion operator of the hard region, and
$T^{(h)}_j$ generates its \mbox{$j$-th} order expansion term.
For the soft region we consider $|k^2| \ll |p^2|$ and $|2k\cdot p| \ll
|p^2|$, permitting the expansion of the first propagator, while the second
remains unchanged:
\begin{align}
  \label{eq:2pointsexp}
  I_1 &\to T^{(s)} I_1 \equiv \sum_j T^{(s)}_j I_1
    \equiv \sum_{j_1,j_2} T^{(s)}_{j_1,j_2} I_1
    = \sum_{j_1,j_2=0}^\infty \frac{(n_1)_{j_{12}}}{j_1! \, j_2!} \,
      \frac{(-k^2)^{j_1} \, (-2k\cdot p)^{j_2}}{(p^2)^{n_1+j_{12}}}
      \,,
  \nonumber \\*
  I_2 &\to T^{(s)} I_2 = I_2 \,,
\end{align}
where
\begin{equation}
  \label{eq:shorthandsum}
  j_{\alpha\beta\cdots} \equiv j_\alpha + j_\beta + \ldots
\end{equation}
is introduced as a shorthand notation, which is also used for other
symbols.
Here $T^{(s)}_j$ generates the \mbox{$j$-th} order expansion term of the
soft region, and it is \emph{a priori} not clear how $j$ relates to $j_1$
and $j_2$ because $k^2$ and $2k\cdot p$ involve different powers of the
soft momentum~$k$. So we postpone this question until after the loop
integration when we will rearrange the summation over $j_1$ and $j_2$.
Both the hard and the soft expansions have employed the generic Taylor
expansion
\begin{equation}
  \label{eq:Taylor}
  \frac{1}{(X+y_1+\ldots+y_m)^n} = \sum_{j_1,\ldots,j_m=0}^\infty
    \frac{(n)_{j_1+\ldots+j_m}}{j_1! \cdots j_m!} \,
    \frac{(-y_1)^{j_1} \cdots (-y_m)^{j_m}}{X^{n+j_1+\ldots+j_m}}
    \quad \text{if } |y_j| \ll |X| \, \forall j
    \,,
\end{equation}
with the Pochhammer symbol $(\alpha)_j \equiv
\Gamma(\alpha+j)/\Gamma(\alpha)$.

According to step~2 of the recipe in section~\ref{sec:intro}, each expanded
term has to be integrated over the \emph{whole integration domain}. The
\mbox{$j$-th} order contribution of the hard region reads
\begin{equation}
  \label{eq:2pointhint}
  F^{(h)}_j = \int\!\rD k \, T^{(h)}_j I
    = \frac{(n_2)_j}{j!} \, (m^2)^j \int\! \frac{\rD k}{
        \bigl((k+p)^2\bigr)^{n_1} \, (k^2)^{n_2+j}}
    \,,
\end{equation}
and the soft-region contribution with indices $j_1,j_2$ is given by
\begin{equation}
  \label{eq:2pointsint}
  F^{(s)}_{j_1,j_2} = \int\!\rD k \, T^{(s)}_{j_1,j_2} I
    = \frac{(n_1)_{j_{12}}}{j_1! \, j_2!} \,
      \frac{(-1)^{j_{12}}}{(p^2)^{n_1+j_{12}}}
      \int\!\rD k \, \frac{(k^2)^{j_1} \, (2k\cdot p)^{j_2}}{(k^2 - m^2)^{n_2}}
    \,.
\end{equation}
These new integrals are simpler than the original
integral~(\ref{eq:2point}): The hard contributions~(\ref{eq:2pointhint})
are massless integrals and functions only of~$p^2$. And the soft
contributions~(\ref{eq:2pointsint}), once each scalar product in the
numerator has been separated into $k\cdot p = p_\mu k^\mu$, are massive
tadpole tensor integrals of rank~$j_2$ and functions only of $m^2$. These
are all one-scale integrals, and it is already clear at this step from a
dimensional analysis that the hard and soft contributions are
\emph{homogeneous functions} of $m^2/p^2$ with
\begin{equation}
  F^{(h)}_j \propto (m^2)^j \, |p^2|^{\frac{d}{2}-n_{12}-j}
  \,, \qquad
  F^{(s)}_{j_1,j_2} \propto (m^2)^{\frac{d}{2}-n_2+j_1+j_2/2} \,
    |p^2|^{-n_1-j_1-j_2/2}
  \,.
\end{equation}
We also see that the expansions generate new singularities: The hard
contributions~(\ref{eq:2pointhint}) will become infrared-singular (for $k
\to 0$) due to the increasing number of $k^2$-terms in the denominator, and
the soft contributions~(\ref{eq:2pointsint}) will become
ultraviolet-singular (for $k \to \infty$) with more and more powers of the
loop momentum in the numerator.
In the particular case $n_1=1$, $n_2=2$, the original integral is finite,
but all terms from the hard region are infrared-singular and all terms from
the soft region are ultraviolet-singular, even the leading-order
contributions with $j=0$ and $j_1=j_2=0$.

The contributions (\ref{eq:2pointhint}) and (\ref{eq:2pointsint}) can
easily be evaluated. The hard-region integrals are straightforward by
standard methods. They yield
\begin{align}
  \label{eq:2pointhres}
  F^{(h)}_j &= \mu^{2\eps} \, e^{\eps\gammaE} \, e^{-i\pi n_{12}} \,
    (-p^2-i0)^{2-n_{12}-\eps} \left(\frac{m^2}{p^2}\right)^j \,
    \frac{\Gamma(2-n_1-\eps)}{\Gamma(n_1) \, \Gamma(n_2)}
\nonumber \\* &\qquad{} \times
    \frac{\Gamma(n_{12}-2+\eps+j) \, \Gamma(2-n_2-\eps-j)}{
      j! \, \Gamma(4-n_{12}-2\eps-j)}
\end{align}
and can be summed up to the all-order hard contribution
\begin{align}
  \label{eq:2pointhressum}
  F^{(h)} = \sum_{j=0}^\infty F^{(h)}_j
  &= \mu^{2\eps} \, e^{\eps\gammaE} \, e^{-i\pi n_{12}} \,
    (-p^2-i0)^{2-n_{12}-\eps}
\nonumber \\* &\qquad{} \times
    \frac{\Gamma(n_{12}-2+\eps) \, \Gamma(2-n_1-\eps) \, \Gamma(2-n_2-\eps)}{
      \Gamma(n_1) \, \Gamma(n_2) \, \Gamma(4-n_{12}-2\eps)}
\nonumber \\* &\qquad{} \times
    \hyperF21\!\left(n_{12}-2+\eps, n_{12}-3+2\eps; n_2-1+\eps;
      \frac{m^2}{p^2}\right)
  ,
\end{align}
where $\hyperF21$ is the hypergeometric function and ``$-i0$'' indicates
the side of the branch cut for the analytic continuation in the case $p^2 >
0$.

The soft-region contributions can e.g.\ be solved by tensor reduction (for
$j_2 > 0$). The integrals are only non-zero for even~$j_2$. We can identify
$j = j_1 + \frac{j_2}{2}$ as the number of additional powers of $m^2/p^2$
compared to the leading order and rewrite
\begin{equation}
  \label{eq:2pointsj}
  \sum_{j_1,j_2=0}^\infty F^{(s)}_{j_1,j_2}
  \equiv \sum_{j=0}^\infty F^{(s)}_j
  \,.
\end{equation}
An evaluation of the soft contributions~(\ref{eq:2pointsint}) in closed
form without the need for tensor reduction is shown in
appendix~\ref{app:2pointscalc}. The result reads
\begin{align}
  \label{eq:2pointsres}
  F^{(s)}_j &= \mu^{2\eps} \, e^{\eps\gammaE} \, e^{-i\pi n_{12}} \,
    (m^2)^{2-n_2-\eps} \, (-p^2-i0)^{-n_1} \left(\frac{m^2}{p^2}\right)^j \,
    \frac{\Gamma(2-n_1-\eps)}{\Gamma(n_1) \, \Gamma(n_2)}
\nonumber \\* &\qquad{} \times
    \frac{\Gamma(n_1+j) \, \Gamma(n_2-2+\eps-j)}{
      j! \, \Gamma(2-n_1-\eps-j)}
\end{align}
and, summed up to all orders,
\begin{align}
  \label{eq:2pointsressum}
  F^{(s)} = \sum_{j=0}^\infty F^{(s)}_j
  &= \mu^{2\eps} \, e^{\eps\gammaE} \, e^{-i\pi n_{12}} \,
    (m^2)^{2-n_2-\eps} \, (-p^2-i0)^{-n_1} \,
    \frac{\Gamma(n_2-2+\eps)}{\Gamma(n_2)}
\nonumber \\* &\qquad{} \times
    \hyperF21\!\left(n_1, n_1-1+\eps; 3-n_2-\eps; \frac{m^2}{p^2}\right)
  .
\end{align}

The asymptotic expansion of the original integral~(\ref{eq:2point}) in
powers of $m^2/p^2$ is obtained by adding the contributions of the hard and
the soft region. Which of the terms $F^{(h)}_j$ and $F^{(s)}_j$ are of the
same order in $m^2/p^2$ depends on $n_2$ (and $\eps$). For the special
choice $n_1=1$, $n_2=2$, we have
\begin{align}
  F^{(h)}_j &= \frac{1}{p^2} \left(\frac{m^2}{p^2}\right)^j
    \left(\frac{\mu^2}{-p^2-i0}\right)^\eps \,
    \frac{e^{\eps\gammaE} \, \Gamma(1+\eps) \, \Gamma(1-\eps) \,
        \Gamma(-\eps)}{\Gamma(1-2\eps)} \,
    \frac{(2\eps)_j}{j!}
    \,,
\nonumber \\
  F^{(s)}_j &= \frac{1}{p^2} \left(\frac{m^2}{p^2}\right)^j
    \left(\frac{\mu^2}{m^2}\right)^\eps \,
    e^{\eps\gammaE} \, \Gamma(\eps) \,
    \frac{(\eps)_j}{(1-\eps)_j}
    \,.
\end{align}
All terms $F^{(h)}_j$ are infrared-singular and all terms $F^{(s)}_j$ are
ultraviolet-singular, but only the leading-order terms $F^{(h)}_0$ and
$F^{(s)}_0$ exhibit an explicit $1/\eps$ singularity:
\begin{align}
  \label{eq:2point12hsreseps}
  F^{(h)}_0 &= \frac{1}{p^2} \left[
    -\frac{1}{\eps} + \ln\!\left(\frac{-p^2-i0}{\mu^2}\right)
    \right] + \Oc(\eps)
    \,, &
  F^{(h)}_j &= -\frac{2}{p^2} \left(\frac{m^2}{p^2}\right)^j \, \frac{1}{j}
    + \Oc(\eps)
    \,,
\nonumber \\
  F^{(s)}_0 &= \frac{1}{p^2} \left[
    \frac{1}{\eps} + \ln\!\left(\frac{\mu^2}{m^2}\right)
    \right] + \Oc(\eps)
    \,, &
  F^{(s)}_j &= \frac{1}{p^2} \left(\frac{m^2}{p^2}\right)^j \, \frac{1}{j}
    + \Oc(\eps)
    \,,
\end{align}
where the column to the right is valid for $j \ge 1$. The $1/\eps$ poles
cancel each other such that the complete expansion terms
\begin{align}
  F_j = F^{(h)}_j + F^{(s)}_j
\end{align}
are all as finite in the limit $\eps \to 0$ as the original integral:
\begin{align}
  \label{eq:2point12reseps}
  F_0 &= \frac{1}{p^2} \, \ln\!\left(\frac{-p^2-i0}{m^2}\right) + \Oc(\eps)
    \,, &
  F_j &= -\frac{1}{p^2} \left(\frac{m^2}{p^2}\right)^j \, \frac{1}{j}
    + \Oc(\eps)
    \,, \; j \ge 1 \,.
\end{align}
But remember that in~$F_0$ an infrared $1/\eps$ pole has been cancelled by
an ultraviolet pole, which looks unnatural. We will come back to this point
in the next section.

The known exact result is reproduced by summing up all orders of the
asymptotic expansion, either from~(\ref{eq:2point12reseps}) or by expanding
the summed contributions (\ref{eq:2pointhressum}) and
(\ref{eq:2pointsressum}) from the regions about $\eps=0$:
\begin{align}
  F = \sum_{j=0}^\infty F_j
    = \frac{1}{p^2} \left[ \ln\!\left(\frac{-p^2-i0}{m^2}\right)
      + \ln\!\left(1-\frac{m^2}{p^2}\right) \right]
      + \Oc(\eps)
  \,.
\end{align}
For general $n_1$ and $n_2$, the original integral can also be evaluated
directly by standard methods, yielding
\begin{multline}
  \label{eq:2pointdirect}
  F = \mu^{2\eps} \, e^{\eps\gammaE} \, e^{-i\pi n_{12}} \,
    (m^2)^{2-n_{12}-\eps} \,
    \frac{\Gamma(2-n_1-\eps) \, \Gamma(n_{12}-2+\eps)}{
      \Gamma(n_2) \, \Gamma(2-\eps)} \,
\\* {} \times
    \hyperF21\!\left(n_1, n_{12}-2+\eps; 2-\eps; \frac{p^2+i0}{m^2}\right)
    \,.
\end{multline}
By applying the transformation formula for the hypergeometric function
which inverses its argument (see e.g.~\cite{Gradshteyn:2007}), the sum of
$F^{(h)}$~(\ref{eq:2pointhressum}) and $F^{(s)}$~(\ref{eq:2pointsressum})
is exactly recovered from~(\ref{eq:2pointdirect}).
Another check is obtained by evaluating the full integral with the help of
a Mellin--Barnes representation and extracting the complete series of
residues with rising powers of $m^2/p^2$. By doing so, the expansion terms
(\ref{eq:2pointhres}) and (\ref{eq:2pointsres}) are reproduced.

As a final remark for this section I want to emphasize that --- although
the evaluation of the integrals originating from the expansion by regions
\emph{to all orders} in the expansion parameter can be quite tedious --- it
is usually rather easy to extract just the leading-order contributions from
all regions. This method is therefore particularly well suited for
obtaining the leading order or the first few terms of an asymptotic
expansion.

\subsection{Proof of the large-momentum expansion}
\label{sec:2pointproof}

While the previous section has introduced the use of the expansion by
region, we restart in this section from the original integral and transform
it in a mathematically well-defined way until we arrive at the expanded
expressions which have already been employed and evaluated in the previous
section.

We want to use the expansions of the hard~(\ref{eq:2pointhexp}) and
soft~(\ref{eq:2pointsexp}) region which, obviously, do not converge towards
the original integrand throughout the complete integration domain. But we
can divide the integration domain into a hard domain~$D_h$ and a soft
domain~$D_s$,
\begin{align}
  \label{eq:2pointDdef}
  D_h &= \bigl\{ k \in \Rb^d : \, |k^2| \ge \Lambda^2 \bigr\} \,, &
  D_s &= \bigl\{ k \in \Rb^d : \, |k^2| < \Lambda^2 \bigr\} \,,
\end{align}
with some intermediate scale~$\Lambda^2$ chosen such that $m^2 \ll
\Lambda^2 \ll |p^2|$. These two domains are non-intersecting and cover the
complete integration domain:
\begin{align}
  D_h \cap D_s &= \emptyset \,, &
  D_h \cup D_s &= \Rb^d \,.
\end{align}
When the loop momentum~$k$ is restricted to one of these domains, the
corresponding expansion \emph{converges absolutely} and the integrand is
identical to its series expansion as long as the latter is summed to all
orders:
\begin{align}
  \label{eq:2pointconv}
  I = T^{(h)} I \equiv \sum_j T^{(h)}_j I
    \quad \text{for } k \in D_h \,,
\nonumber \\
  I = T^{(s)} I \equiv \sum_j T^{(s)}_j I
    \quad \text{for } k \in D_s \,.
\end{align}
For the soft expansion, the summation index~$j$ symbolically represents a
proper combination of the indices $j_1$ and $j_2$
in~(\ref{eq:2pointsexp}).
The hard expansion~$T^{(h)}$ only requires $|k^2| \gg m^2$ which is
certainly fulfilled for $k \in D_h$. But the statement
in~(\ref{eq:2pointconv}) about the soft expansion~$T^{(s)}$ is less
trivial: While one of the conditions, $|k^2| \ll |p^2|$, surely holds for
$k \in D_s$, the other condition, $|2k\cdot p| \ll |p^2|$, can still be
violated. We have to remember, though, that the expansions are always
performed under the loop integral~$\int\!\rD k$. By tensor reduction, each
power of $(k\cdot p)^2$ in the numerator of the soft-region
integrals~(\ref{eq:2pointsint}) gives a contribution proportional to $k^2
p^2$, while odd powers of $k \cdot p$ vanish under the integration. This is
still true if we restrict the integration by the Lorentz-invariant
condition $|k^2| < \Lambda^2$. So we can safely count $|2k\cdot p| \sim
|k^2 p^2|^{1/2}$, and the condition $|2k\cdot p| \ll |p^2|$ holds under the
integral over the soft domain~$D_s$.

Alternatively, for ensuring the convergence of~$T^{(s)}$ within~$D_s$, one
may choose a reference frame in which either the zero-component or the
spatial components of the vector~$p$ vanish, depending on the sign of
$p^2$, and define the boundaries of $D_h$ and $D_s$ appropriately in this
reference frame. Or, for $p^2 < 0$, one may perform a Wick rotation and
define the boundaries of the domains as relations between positive-definite
norms of Euclidean vectors.

A consequence of the absolute convergence of the expansions is that the
summation~$\sum_j$ commutes with the integration and can safely be pulled
out of the integral if the latter is restricted to the corresponding
domain:
\begin{align}
  \label{eq:2pointexpcomm}
  \int_{k \in D_h}\!\rD k \, I
    = \sum_j \int_{k \in D_h}\!\rD k \, T^{(h)}_j I
    \,,
  \qquad
  \int_{k \in D_s}\!\rD k \, I
    = \sum_j \int_{k \in D_s}\!\rD k \, T^{(s)}_j I
    \,.
\end{align}
After these preliminaries we can start transforming the original integral
by splitting the integration into the two domains and performing the
appropriate expansions in each of them:
\begin{align}
  \label{eq:2pointsplitD}
  F = \int\!\rD k \, I
    = \int_{k \in D_h}\!\rD k \, I + \int_{k \in D_s}\!\rD k \, I
    = \sum_j \int_{k \in D_h}\!\rD k \, T^{(h)}_j I
      + \sum_j \int_{k \in D_s}\!\rD k \, T^{(s)}_j I
  \,.
\end{align}
With dimensional regularization at hand, we can also perform the
integration of each expanded term over the complete integration domain, but
we have to compensate for this by subtracting the surplus from the added
domain:
\begin{align}
  \label{eq:2pointextend}
  \int_{k \in D_h}\!\rD k \, T^{(h)}_j I
    &= \int\!\rD k \, T^{(h)}_j I
      - \int_{k \in D_s}\!\rD k \, T^{(h)}_j I
    \,,
\nonumber \\
  \int_{k \in D_s}\!\rD k \, T^{(s)}_j I
    &= \int\!\rD k \, T^{(s)}_j I
      - \int_{k \in D_h}\!\rD k \, T^{(s)}_j I
    \,.
\end{align}
Without the indication of a restriction, the integrations are understood as
being performed over the complete integration
domain~$\Rb^d$. Relation~(\ref{eq:2pointexpcomm}) also holds if the
integrand~$I$ is not the original one, but a term from a previous
expansion, i.e.\ each expansion can be applied to any integrand if the
integral is restricted to the corresponding domain:
\begin{align}
  \label{eq:2pointexpcommdouble}
  \int_{k \in D_h}\!\rD k \, T^{(s)}_j I
    = \sum_i \int_{k \in D_h}\!\rD k \, T^{(h)}_i T^{(s)}_j I
    \,,
  \qquad
  \int_{k \in D_s}\!\rD k \, T^{(h)}_j I
    = \sum_i \int_{k \in D_s}\!\rD k \, T^{(s)}_i T^{(h)}_j I
    \,.
\end{align}
Let us have a look at the newly generated double expansions. The order in
which the hard and soft expansions are applied is irrelevant because the
doubly expanded integrand is the same in both cases (written here with two
separate indices for the soft expansion to be specific):
\begin{align}
  \label{eq:2pointdoubleexp}
  T^{(h)}_i T^{(s)}_{j_1,j_2} I
  = T^{(s)}_{j_1,j_2} T^{(h)}_i I
  = \frac{(n_2)_i}{i!} \, \frac{(n_1)_{j_{12}}}{j_1! \, j_2!} \,
    \frac{(m^2)^i \, (-1)^{j_{12}}}{(p^2)^{n_1+j_{12}}} \,
    \frac{(2k\cdot p)^{j_2}}{(k^2)^{n_2+i-j_1}}
  \,.
\end{align}
In such cases of \emph{commuting expansions}, we label multiple expansions
by a comma-separated list in the round brackets:
\begin{align}
  \label{eq:2pointdoubleexpdef}
  T^{(h)}_i T^{(s)}_j = T^{(s)}_j T^{(h)}_i \equiv T^{(h,s)}_{i,j}
  \,.
\end{align}
After an appropriate relabelling of the summation indices, the two
contributions with double expansions can be added together,
\begin{align}
  \label{eq:2pointdoubleexpcomb}
  \sum_i \sum_j \int_{k \in D_s}\!\rD k \, T^{(h,s)}_{i,j} I
  + \sum_j \sum_i \int_{k \in D_h}\!\rD k \, T^{(h,s)}_{i,j} I
  = \sum_{i,j} \int\!\rD k \, T^{(h,s)}_{i,j} I
  \,,
\end{align}
arriving at an integral over the complete integration domain. The
non-trivial point here is that we have to exchange the order of the two
summations in one of the contributions. While e.g.\ in the first term with
$k$ restricted to the soft domain~$D_s$, the summation $\sum_j$ of the soft
expansion is absolutely convergent, we cannot easily claim convergence for
the summation $\sum_i$ of the hard expansion when $k \in D_s$. However, we
are not summing expanded integrands here, but integrals. And the only scale
involved in the integrals over the doubly-expanded integrand
$T^{(h,s)}_{i,j} I$~(\ref{eq:2pointdoubleexp}) originates from the boundary
of the integration domain, as all occurrences of the momentum~$p$ in the
scalar products $k \cdot p$ in the numerator can be pulled out of the
integral. In fact, for dimensional reasons, we know that
\begin{align}
  \label{eq:2pointdoublescale}
  \int_{k \in D_s}\!\rD k \, T^{(h,s)}_{i,j_1,j_2} I
  \propto
    |p^2|^{-n_1} \, (\Lambda^2)^{\frac{d}{2}-n_2}
    \left(\frac{m^2}{\Lambda^2}\right)^i
    \left(\frac{\Lambda^2}{|p^2|}\right)^{j_1+j_2/2}
  \,,
\end{align}
because the only dimensionful parameter in the definition of the
domain~$D_s$ is $\Lambda^2$. By the same reasoning,
(\ref{eq:2pointdoublescale}) holds if the integral is restricted to $k \in
D_h$ instead. As the boundary has been chosen to obey $m^2 \ll \Lambda^2
\ll |p^2|$, both the summations over~$i$ and over $j_1,j_2$
in~(\ref{eq:2pointdoublescale}) converge absolutely, and their order can be
exchanged at will.

We are now able to collect all pieces contributing to the
integral~$F$. Writing
\begin{align}
  F^{(h)}_j &= \int\!\rD k \, T^{(h)}_j I \,,
& 
  F^{(s)}_j &= \int\!\rD k \, T^{(s)}_j I \,,
& 
  F^{(h,s)}_{i,j} &= \int\!\rD k \, T^{(h,s)}_{i,j} I
\end{align}
for the individual contributions and
\begin{align}
  F^{(h)} &= \sum_j F^{(h)}_j \,, &
  F^{(s)} &= \sum_j F^{(s)}_j \,, &
  F^{(h,s)} &= \sum_{i,j} F^{(h,s)}_{i,j}
\end{align}
for the summed-up series, we obtain
\begin{align}
  \label{eq:2pointidentity}
  F = F^{(h)} + F^{(s)} - F^{(h,s)}
\end{align}
for the original integral after the above transformations.
Note that all integrations involved in~(\ref{eq:2pointidentity}) are
performed over the whole integration domain~$\Rb^d$. So all restrictions to
the two individual domains $D_h$ and $D_s$ drop out and the final terms
in~(\ref{eq:2pointidentity}) are individually independent of the separating
scale~$\Lambda^2$. Thus the exact position of the boundary between the
domains is irrelevant, and we could have defined the domains e.g.\ in the
following way:
\begin{align}
  \label{eq:2pointDsloppy}
  D_h &= \bigl\{ k \in \Rb^d : \, |k^2| \gg m^2 \bigr\} \,, &
  D_s &= \bigl\{ k \in \Rb^d : \, |k^2| \lesssim m^2 \bigr\} \,,
\end{align}
where ``$\lesssim$'' is understood as the negation of ``$\gg$'', such that
$D_h \cap D_s = \emptyset$ and $D_h \cup D_s = \Rb^d$ hold. In later
examples we will not introduce specific boundaries between convergence
domains, but use rather ``sloppy'' specifications of the domains as
above. It is understood, however, that exact positions of the boundaries
exist and could be specified if needed.

The first two terms in the final identity~(\ref{eq:2pointidentity})
correspond exactly to the contributions from the hard~(\ref{eq:2pointhint})
and soft~(\ref{eq:2pointsint}) regions prescribed by the expansion by
regions and evaluated in the previous section. But now we have obtained a
third term, subtracted from the first two. This additional \emph{overlap
  contribution}~$F^{(h,s)}$ is absent in the recipe formulated in
section~\ref{sec:intro}. Let us have a look at its terms:
\begin{align}
  \label{eq:2pointoverlap}
  F^{(h,s)}_{i,j_1,j_2} = \int\!\rD k \, T^{(h,s)}_{i,j_1,j_2} I
  = \frac{(n_2)_i}{i!} \, \frac{(n_1)_{j_{12}}}{j_1! \, j_2!} \,
    \frac{(m^2)^i \, (-1)^{j_{12}}}{(p^2)^{n_1+j_{12}}}
    \int\!\rD k \, \frac{(2k\cdot p)^{j_2}}{(k^2)^{n_2+i-j_1}}
  = 0
  \,.
\end{align}
These are \emph{scaleless integrals}, which must consistently be set to
zero when using dimensional regularization. In fact, each of the integrals
in~(\ref{eq:2pointoverlap}) can be transformed by tensor reduction into an
integral $\int\!\rD k \, (k^2)^{-n}$ with some power~$n$, and these
massless tadpole integrals exhibit both ultraviolet and infrared
singularities in such a way that they cancel each other. (Individual parts
of the integration have different convergence domains in the complex plane
of the space--time dimension~$d$, but analytic continuation permits to
combine the pieces.) The integral
\begin{align}
  \label{eq:tadpoleUVIR}
  \int\!\frac{\rD k}{(k^2)^2}
  = \frac{1}{\eps_\UV} - \frac{1}{\eps_\IR} = 0
\end{align}
is the only case in this class of integrals where ultraviolet poles
$1/\eps_\UV$ or infrared poles $1/\eps_\IR$ appear.
In the special case $n_1 = 1$, $n_2 = 2$, where the original integral~$F$
is finite, the integrals of the overlap
contribution~(\ref{eq:2pointoverlap}) exhibit exactly those ultraviolet
singularities to cancel the ones in the soft contribution~$F^{(s)}$ and
those infrared singularities to cancel the ones in the hard
contribution~$F^{(h)}$. Although the overlap contribution~$F^{(h,s)}$ is
scaleless and vanishes, it is this term which makes the complete
result~(\ref{eq:2pointidentity}) separately ultraviolet-finite and
infrared-finite. One can check explicitly (see
appendix~\ref{app:2pointhscalc}) that the summed-up overlap
contributions~(\ref{eq:2pointoverlap}) with ultraviolet and infrared
$1/\eps$ poles separated cancel the corresponding poles
in~(\ref{eq:2point12hsreseps}).

Note that the overlap contribution terms~(\ref{eq:2pointoverlap}) are the
same as the doubly expanded terms arising in an expansion by
subgraphs~\cite{Smirnov:2002pj}.

Having checked that the overlap contribution $F^{(h,s)}$ is scaleless and
vanishes, the original integral is reproduced by
\begin{align}
  \label{eq:2pointidentity2}
  F = F^{(h)} + F^{(s)}
  \,.
\end{align}
This is exactly the sum of contributions which has been evaluated in
section~\ref{sec:2pointEbR}, where we have assumed that we need these two
regions (hard and soft) and evaluated them according to the recipe of the
expansion by regions. Now we have obtained the same answer by
mathematically transforming the original integral. Let us recapitulate what
we had to check on our way:
\begin{itemize}
\item For the two regions we had to find domains ($D_h$ and $D_s$) where
  their expansions converge absolutely. These domains have to be
  non-intersecting ($D_h \cap D_s = \emptyset$) and cover the complete
  integration domain ($D_h \cup D_s = \Rb^d$).
\item In the double expansion
  the order of the two expansions has to be irrelevant (``commuting
  expansions'').
\item The overlap contribution from the double expansion involves only
  scaleless integrals.
\end{itemize}
These three points had to be proven explicitly for the example integral.
The rest of the transformations used in this section is general and applies
to any other integral with a set of regions and domains obeying analogous
conditions. We will work this out in the next section.

Note that we did not have to evaluate any of the integrals in $F^{(h)}$,
$F^{(s)}$ or $F^{(h,s)}$ in order to prove the
identity~(\ref{eq:2pointidentity}). It is sufficient to study the
expansions at the integrand level. And even for the
form~(\ref{eq:2pointidentity2}), where the scaleless overlap contribution
has been dropped, a look at the expanded integrand has been enough
(although in other cases it can be more involved to show that the overlap
contributions are scaleless).

Remember how important it is within the framework of the expansion by
regions that scaleless integrals can be set to zero. In our example it is
dimensional regularization which regularizes scaleless integrals in such a
way that they vanish. In some cases (see in particular the examples in
sections \ref{sec:Sudakov} and~\ref{sec:forward}) this is not sufficient,
and we have to use analytic regularization as well. In the absence of such
nicely behaved regularization schemes, however, interesting patterns appear
(see section~\ref{sec:forward1111} and appendix~\ref{app:finiteboundary})
where overlap contributions play an important role.

\section{General formalism with commuting expansions}
\label{sec:formalism}

In this section the proof of section~\ref{sec:2pointproof} is generalized
to arbitrary integrals, with some restrictions. Consider the following
situation:
\begin{itemize}
\item We want to expand the \emph{integral}
  \begin{equation}
  \label{eq:intorig}
    F = \int\!\rD k \, I
    \,,
  \end{equation}
  where the integrand~$I$ is integrated over the integration
  domain~$D$. This can be a one-loop integral (with $D = \Rb^d$), a
  multi-loop integral ($D = \Rb^{n\cdot d}$) or any arbitrary integral.
\item We have identified a set
  \begin{equation}
    R = \{ x_1, \ldots, x_N \}
  \end{equation}
  of $N$~\emph{regions}~$x_i$. Each region~$x$ is characterized by an
  \emph{expansion}
  \begin{equation}
    T^{(x)} \equiv \sum_j T^{(x)}_j
  \end{equation}
  which, when applied to the integrand, replaces the latter by a series of
  expanded terms. The summation index~$j$ can also represent a set of
  indices $j_1,j_2,\ldots$, but we only write one index per expansion.
  These expansion operators also have to be defined when they are applied
  to terms resulting from previous expansions (when multiple expansions are
  generated). In such cases it may happen that a certain expansion is an
  identity transformation and the set of summation indices represented
  by~$j$ is empty.
\item For each region~$x$ there is a \emph{domain} $D_x \subset D$ such
  that the expansion~$T^{(x)}$ \emph{converges absolutely} when the
  integration variable is restricted to $k \in D_x$.
\end{itemize}
Let us assume that the regions, expansions and domains fulfill the
following \emph{conditions}:
\newcounter{EbRnoconditions}
\newcounter{EbRcondcomm}
\begin{enumerate}
\item\label{item:EbRcondD}%
  The domains are non-intersecting and cover the complete integration
  domain:
  \begin{equation}
    D_x \cap D_{x'} = \emptyset \; \forall x \ne x'
    \,, \qquad
    \bigcup_{x\in R} D_x = D
    \,.
  \end{equation}
\item\label{item:EbRcondcomm}\setcounter{EbRcondcomm}{\value{enumi}}%
  All expansions \emph{commute} with each other:
  \begin{equation}
    T^{(x)} T^{(x')} = T^{(x')} T^{(x)} \; \forall x,x' \in R
    \,.
  \end{equation}
\item\label{item:EbRcondreg}%
  The original integral itself and all integrals over expanded terms ---
  whether restricted to some convergence domain~$D_x$ or not --- are
  regularized.
\item\label{item:EbRcondsums}%
  The series expansions~$T^{(x)}$ converge absolutely (or are properly
  regularized) even when the expanded terms are integrated over the whole
  integration domain~$D$ instead of just their convergence domain~$D_x$.
\setcounter{EbRnoconditions}{\value{enumi}}
\end{enumerate}
Condition~\ref{item:EbRcondD} ensures that any integral over the complete
integration domain can be split into integrals restricted to the
domains~$D_x$:
\begin{equation}
  \label{eq:intsplit}
  \int\!\rD k \equiv \int_{k \in D}\!\rD k
  = \sum_{x \in R} \; \int_{k \in D_x}\!\rD k
  \,.
\end{equation}
The consequence of this condition is that we might have to invent
``auxiliary'' regions (which do not contribute to the final result)
in order to cover the complete integration domain~$D$ with the convergence
domains~$D_x$.

Condition~\ref{item:EbRcondcomm} is to be understood at the operator level
of the expansions: Whatever integrand the double expansion $T^{(x)}
T^{(x')}$ is applied to, whether the original integrand~$I$ or a term
resulting from previous expansions, the order in which the expansions are
performed must be irrelevant, i.e.\ in all cases the same (multiple) series
of doubly expanded integrand terms is established.
This condition, however, cannot be fulfilled in all cases where the
expansion by regions has been applied successfully. In
section~\ref{sec:formalismnc} this restriction will be relaxed and the
treatment of non-commuting expansions will be presented.

Condition~\ref{item:EbRcondreg} implies that we use a regularization
prescription which provides a mathematically well-defined meaning to all
integrals occurring in the formalism described in this section. Usually
this is dimensional regularization, eventually combined with analytic
regularization, but other schemes are possible.

Finally condition~\ref{item:EbRcondsums} requires a certain mechanism that
makes series expansions converge even outside their convergence
domains. For loop integrals with dimensional regularization this is usually
the case. More generally, this mostly works for integrals where the
boundaries either lie at zero or at infinity such that they do not
introduce new scales. Then the integration over the complete domain~$D$,
although formally divergent, is regularized and determined only by the
scaleful parameters within the convergence domain, keeping the series
expansions as convergent as with integrals restricted to the convergence
domains~$D_x$.
It is possible to apply the expansion by regions to other integrals, e.g.\
with domains~$D$ involving finite boundaries, where
condition~\ref{item:EbRcondsums} is possibly violated. But then the
summation of the series expansions has to be done with care, and it might
be necessary to combine certain terms which are individually divergent.
This is a subtle issue in the expansion by region and in the formalism
presented here. Its consequences will be pointed out at the relevant steps
later in this section, and appendix~\ref{app:finiteboundary} presents an
example illustrating this behaviour.

Let us introduce the notations to be used in this section. They have
partially already been defined for the example in section~\ref{sec:2point}.
Multiple expansions (replacing some integrand first by its expansion terms
according to one region and repeating this step with the resulting terms
for other regions) are denoted by
\begin{equation}
  \label{eq:Tmult}
  T^{(x_1,x_2,\ldots)}_{j_1,j_2,\ldots}
    \equiv T^{(x_1)}_{j_1} T^{(x_2)}_{j_2} \cdots
  \,,\qquad
  T^{(x_1,x_2,\ldots)} \equiv
    \sum_{j_1,j_2,\ldots} T^{(x_1,x_2,\ldots)}_{j_1,j_2,\ldots}
  \,,
\end{equation}
if the expansions are commuting, i.e.\ their order is irrelevant.
The $j$-th order expanded integral according to the region~$x$ is denoted
by
\begin{equation}
  \label{eq:Fexp}
  F^{(x)}_j \equiv \int\!\rD k \, T^{(x)}_j I
  \,,
\end{equation}
and its summation to all orders by
\begin{equation}
  \label{eq:Fexpsum}
  F^{(x)} \equiv \sum_j F^{(x)}_j = \sum_j \int\!\rD k \, T^{(x)}_j I
  \,,
\end{equation}
where the integrals are performed over the complete integration domain~$D$.
Analogous notations are used for multiple expansions:
\begin{equation}
  \label{eq:Fexpmult}
  F^{(x_1,x_2,\ldots)}_{j_1,j_2,\ldots} \equiv
    \int\!\rD k \, T^{(x_1,x_2,\ldots)}_{j_1,j_2,\ldots} I
  \,, \qquad
  F^{(x_1,x_2,\ldots)} \equiv
    \sum_{j_1,j_2,\ldots} F^{(x_1,x_2,\ldots)}_{j_1,j_2,\ldots}
  \,.
\end{equation}
The restriction of an integration to a domain~$D_x$ is indicated by a lower
index in square brackets:
\begin{equation}
  \label{eq:Frestrict}
  F_{[x]} \equiv \int_{k \in D_x}\!\rD k \, I \,,\qquad
  F^{(x',\ldots)}_{j,\ldots\,[x]} \equiv
    \int_{k \in D_x}\!\rD k \, T^{(x',\ldots)}_{j,\ldots} I \,,\qquad
  \text{etc.}
\end{equation}
If the integration is performed over the combination of several domains, we
write
\begin{equation}
  \label{eq:Frestrict2}
  F_{[x_1+\ldots+x_n]} \equiv \sum_{i=1}^n F_{[x_i]}
  \,,\qquad
  F^{(x',\ldots)}_{j,\ldots\,[x_1+\ldots+x_n]}
    \equiv \sum_{i=1}^n F^{(x',\ldots)}_{j,\ldots\,[x_i]}
  \,.
\end{equation}

The absolute convergence of each expansion~$T^{(x)}$ within the
corresponding domain~$D_x$ implies that we can safely replace any
integrand~$I'$ (the original integrand~$I$ or the result of previous
expansions) by its expanded series if the integration variable is
restricted to $k \in D_x$:
\begin{equation}
  I' = T^{(x)} I' = \sum_j T^{(x)}_j I'
    \quad \text{for } k \in D_x
  \,.
\end{equation}
Absolute convergence also implies that we can pull the summation of such an
expansion out of an adequately restricted integral, using the notation of
(\ref{eq:Fexp})--(\ref{eq:Frestrict}):
\begin{equation}
  \label{eq:Frestrictexpand}
  F_{[x]} = F^{(x)}_{[x]}
    = \sum_j F^{(x)}_{j\,[x]}
  \,,\qquad
  F^{(x',\ldots)}_{j',\ldots\,[x]}
    = \sum_j F^{(x,x',\ldots)}_{j,j',\ldots\,[x]}
  \,.
\end{equation}

Now we can start with the original integral~(\ref{eq:intorig}) and split
the integration into the $N$~domains corresponding to the $N$~regions,
according to~(\ref{eq:intsplit}). In each domain we replace the integral by
its series expansion according to~(\ref{eq:Frestrictexpand}):
\begin{equation}
  \label{eq:formalism1}
  F = \sum_{x \in R} F_{[x]}
  = \sum_{x \in R} F^{(x)}_{[x]}
  \,.
\end{equation}
The right-hand side of~(\ref{eq:formalism1}) involves a sum of $N$ series
expansions with each integral restricted to the corresponding convergence
domain. This is a special case (with $n=1$) of the expression
\begin{equation}
  \label{eq:formalismn}
  \sum_{\{x'_1,\ldots,x'_n\} \subset R}
    F^{(x'_1,\ldots,x'_n)}_{[x'_1+\ldots+x'_n]}
  \,,
\end{equation}
where the sum runs over all subsets of $n$~distinct regions out of the
$N$~regions in~$R$ ($1 \le n \le N$). Each integrand is multiply expanded
according to these $n$~regions, and the integrals are performed over the
combination of the $n$~corresponding domains.
Let us postpone for a few lines the question whether the
expression~(\ref{eq:formalismn}) is a convergent series expansion, despite
the fact that (for $n>1$) the integrations are performed over larger
domains than the convergence domain of each of the $n$~individual
expansions.

If $n < N$, i.e.\ if the integrations in~(\ref{eq:formalismn}) are not
performed over the complete integration domain~$D$ yet, the regularization
of the integrals (see condition~\ref{item:EbRcondreg} above) allows to
extend all these integrations to $k \in D$ when compensating for this by
subtracting the integrations over the additional domains:
\begin{align}
  \label{eq:formalismextend}
  F^{(x'_1,\ldots,x'_n)}_{[x'_1+\ldots+x'_n]}
  &= \sum_{j_1,\ldots,j_n}
    F^{(x'_1,\ldots,x'_n)}_{j_1,\ldots,j_n\,[x'_1+\ldots+x'_n]}
\nonumber \\*
  &= \sum_{j_1,\ldots,j_n} \biggl(
    F^{(x'_1,\ldots,x'_n)}_{j_1,\ldots,j_n}
    - \sum_{x'_{n+1} \in R \setminus \{x'_1,\ldots,x'_n\}}
      F^{(x'_1,\ldots,x'_n)}_{j_1,\ldots,j_n\,[x'_{n+1}]}
    \biggr)
  \,.
\end{align}
The subtraction terms are integrals performed over one
domain~$D_{x'_{n+1}}$ each (where $x'_{n+1}$ runs over all regions which
are absent in the subset $\{x'_1,\ldots,x'_n\}$).
These subtraction terms can be replaced by their expansions according
to~(\ref{eq:Frestrictexpand}):
\begin{equation}
  \label{eq:formalismexpn1}
  F^{(x'_1,\ldots,x'_n)}_{j_1,\ldots,j_n\,[x'_{n+1}]}
  = \sum_{j_{n+1}}
    F^{(x'_1,\ldots,x'_n,x'_{n+1})}_{j_1,\ldots,j_n,j_{n+1}\,[x'_{n+1}]}
  \,,
\end{equation}
where we have already used condition~\ref{item:EbRcondcomm} above that the
expansions commute.
Now we sum the individual terms in~(\ref{eq:formalismextend}) separately
and write
\begin{align}
  \label{eq:formalismextsums}
  F^{(x'_1,\ldots,x'_n)}_{[x'_1+\ldots+x'_n]}
  = F^{(x'_1,\ldots,x'_n)}
    - \sum_{x'_{n+1} \in R \setminus \{x'_1,\ldots,x'_n\}}
      F^{(x'_1,\ldots,x'_n,x'_{n+1})}_{[x'_{n+1}]}
  \,.
\end{align}
This is a non-trivial step. Even if the complete
expression~(\ref{eq:formalismextend}) is a finite series expansion, this
does not necessarily mean that all summations
in~(\ref{eq:formalismextsums}) are individually convergent. Depending on
the parameters involved, especially the boundaries of the complete
domain~$D$ and the boundaries between the individual domains~$D_x$, we
might have to regularize the summations in~(\ref{eq:formalismextend}). We
may e.g.\ think of truncating the summations by some upper limit ($j_i \le
j_{\text{max}} \, \forall i$), thus limiting the accuracy of the expanded
expressions, but dealing only with finite sums. This truncation is removed
(by $j_{\text{max}} \to \infty$) only in the end when the summations have
been combined into convergent ones (cf.\ condition~\ref{item:EbRcondsums}
above).

For reproducing~(\ref{eq:formalismn}) we finally we have to
sum~(\ref{eq:formalismextsums}) over all subsets of $n$~regions
$\{x'_1,\ldots,x'_n\}$. The subtraction terms yield summations over subsets
of $(n+1)$~regions, where each term appears $(n+1)$~times with different
integration domains~$D_{x'_{n+1}}$:
\begin{align}
  \label{eq:formalismsumscomb}
  \sum_{\{x'_1,\ldots,x'_n\} \subset R} \;\;
    \sum_{x'_{n+1} \in R \setminus \{x'_1,\ldots,x'_n\}}
      F^{(x'_1,\ldots,x'_n,x'_{n+1})}_{[x'_{n+1}]}
  &= \sum_{\{x'_1,\ldots,x'_{n+1}\} \subset R} \;\;
    \sum_{i=1}^{n+1} F^{(x'_1,\ldots,x'_{n+1})}_{[x'_i]}
\nonumber \\*
  &= \sum_{\{x'_1,\ldots,x'_{n+1}\} \subset R}
    F^{(x'_1,\ldots,x'_{n+1})}_{[x'_1+\ldots+x'_{n+1}]}
  \,.
\end{align}
This requires not only that the expansion commute
(condition~\ref{item:EbRcondcomm}), but also that the order of the series
summations in~(\ref{eq:formalismsumscomb}) can be exchanged, which in turn
requires their absolute convergence. For the example presented in
section~\ref{sec:2pointproof} this has been shown explicitly. I cannot
provide a rigorous proof for this convergence issue in the general case
treated here, but I am convinced that even divergent series in intermediate
steps of this derivation (which then have to be regularized) are not
problematic when condition~\ref{item:EbRcondsums} ensures the convergence
of those sums which remain in the final result.
Note that this problem only arises when the series expansions are summed up
to all orders. If an approximation with limited accuracy is sought and
the series expansions are truncated at some finite order of the expansion
parameter, then convergence problems of individual (intermediary or final)
terms are absent.

Combining all terms from (\ref{eq:formalismextsums}) and
(\ref{eq:formalismsumscomb}), the expression~(\ref{eq:formalismn}) yields
\begin{align}
  \label{eq:formalisminduction}
  \sum_{\{x'_1,\ldots,x'_n\} \subset R}
    F^{(x'_1,\ldots,x'_n)}_{[x'_1+\ldots+x'_n]}
  = \sum_{\{x'_1,\ldots,x'_n\} \subset R} F^{(x'_1,\ldots,x'_n)}
    - \sum_{\{x'_1,\ldots,x'_{n+1}\} \subset R}
      F^{(x'_1,\ldots,x'_{n+1})}_{[x'_1+\ldots+x'_{n+1}]}
  \,.
\end{align}
The first term on the right-hand side consists of integrals performed over
the complete domain~$D$. We want to keep such terms for the final
result. The second term is exactly the same as the one on the left-hand
side, but with $n$ replaced by $(n+1)$ and with opposite sign. Thus
(\ref{eq:formalisminduction}) represents a recursion formula which can be
iterated from $n=1$, as in~(\ref{eq:formalism1}), up to $n=N-1$.
This allows us to write the original integral in the following form:
\begin{align}
  \label{eq:formalismidentity}
    F &=
      \sum_{x \in R} F^{(x)}
      - \sum_{\{x'_1,x'_2\} \subset R} F^{(x'_1,x'_2)}
      + \ldots
      - (-1)^n \sum_{\{x'_1,\ldots,x'_n\} \subset R} F^{(x'_1,\ldots,x'_n)}
    \nonumber \\* &\qquad{}
      + \ldots
      - (-1)^N \, F^{(x_1,\ldots,x_N)}
    \,.
\end{align}
This is the master identity for the expansion by regions in the formalism
presented in this section. It involves only integrations over the complete
domain~$D$, so at least the series expansions in this final result are all
convergent if condition~\ref{item:EbRcondsums} holds.

The master identity~(\ref{eq:formalismidentity}) involves single and
multiple expansions, according to the $N$~regions $x_1,\ldots,x_N$ and all
of their combinations, with alternating signs.
The recipe for the expansion by regions presented in
section~\ref{sec:intro} only knows about the first term on the right-hand
side of~(\ref{eq:formalismidentity}), where a single-expanded integrand
according to each region is integrated over the complete domain. This means
that all other terms with multiple expansions must vanish in ``normal''
situations for the recipe to be valid.

Indeed, if the regularization of a loop integral and the regions are chosen
properly, then each of the terms~$F^{(x)}_j$ (i.e.\ the single-expanded
terms present in the known recipe) is a single-scale integral yielding a
\emph{homogeneous} function of the expansion parameter with a \emph{unique
  scaling}\footnote{%
  This means that each term in the series of expanded integrals depends on
  the expansion parameter by a simple power which is different for each
  region. If two regions share the same dependence on the expansion
  parameter, then their overlap contribution does not have to be
  scaleless. Note that single-scale integrals as understood above may
  exhibit a non-trivial dependence on additional parameters or ratios of
  $\Oc(1)$ which are considered neither small nor large in the expansion.}
(cf.\ section~\ref{sec:2pointEbR}). Every further expansion of such a
single-scale integral according to a different region (which would yield a
different scaling with the expansion parameter) makes the integral
scaleless such that it is set to zero. This is why usually the terms with
multiple expansions do not contribute to the asymptotic expansion of loop
integrals.

But the identity~(\ref{eq:formalismidentity}) is more general: It is
independent of the chosen regularization scheme, as long as all individual
terms are mathematically well-defined (as required by
conditions~\ref{item:EbRcondreg} and~\ref{item:EbRcondsums}). The
identity~(\ref{eq:formalismidentity}) can also be applied to other types of
integrals or to loop integrals which use other regularizations than the
standard dimensional and analytic ones. In such cases the \emph{overlap
  contributions}, i.e.\ the terms in~(\ref{eq:formalismidentity}) with
expansions according to more than one region, may become relevant.

While these overlap contributions arise naturally out of the formalism
presented here, they have already been noted in the context of effective
theories and are called ``zero-bin subtractions'' there (see
e.g.~\cite{Manohar:2006nz,Chiu:2009yx}). The overlap contributions
in~(\ref{eq:formalismidentity}) have exactly the same form of multiple
expansions as what is --- usually only to leading order --- introduced in
the literature under the name zero-bin subtractions. The
identity~(\ref{eq:formalismidentity}) clarifies the whole picture of
subtractions which are needed in the general case.

One example where overlap contributions (or zero-bin subtractions) are
relevant is provided in~~\cite{Chiu:2009yx} where $\Delta$-regulators are
introduced which push all propagator denominators artificially off-shell
by some amount. These $\Delta$-regulators introduce new scales into the
integrals. Therefore the contributions from each region are not homogeneous
functions of the expansion parameter and the overlap contributions are not
scaleless.
Note that the authors of ~\cite{Chiu:2009yx} only consider the overlap of
each collinear region $n,\bar n$ with the soft region~$s$, in the language
of this section $F^{(n,s)}$ and $F^{(\bar n,s)}$. The
identity~(\ref{eq:formalismidentity}) would have told them that the full
result is (assuming that this set of regions is complete)
\begin{align}
  \label{eq:Chiu_identity}
  F = F^{(n)} + F^{(\bar n)} + F^{(s)}
      - F^{(n,s)} - F^{(\bar n,s)} - F^{(n,\bar n)}
      + F^{(n,\bar n,s)}
  \,.
\end{align}
But the soft expansion of their integral is identical to the combination of
the two collinear expansions, such that $F^{(s)} = F^{(n,\bar n)} =
F^{(n,s)} = F^{(\bar n,s)} = F^{(n,\bar n,s)}$. Cancelling the last two
terms in~(\ref{eq:Chiu_identity}), the result can be written as
\begin{align}
  F = \left(F^{(n)} - F^{(n,s)}\right)
      + \left(F^{(\bar n)} - F^{(\bar n,s)}\right) + F^{(s)}
  \,,
\end{align}
as they did, or by omitting the (irrelevant) soft region:
\begin{align}
  F = F^{(n)} + F^{(\bar n)} - F^{(n,\bar n)}
  \,.
\end{align}

The master identity~(\ref{eq:formalismidentity}) is an exact relation for
the original integral~$F$. It involves the summation of the single and
multiple series expansions to all orders according to the definitions
(\ref{eq:Fexpsum}) and~(\ref{eq:Fexpmult}). If instead only the leading
terms of the right-hand side in~(\ref{eq:formalismidentity}) are taken into
account, then an approximation for the integral~$F$ on the left-hand side
is obtained. The expansion parameter by whose powers higher-order terms are
suppressed is related to the parameter hierarchies exploited in the
expansions~$T^{(x)}$. While intermediate expressions in the derivation of
this formula involve the boundaries between the convergence domains, the
final result~(\ref{eq:formalismidentity}) is independent of these
boundaries, and so are the series expansions in the master identity.

A leading-order asymptotic expansion of the integral~$F$ can also be obtained
directly without ever touching infinite series expansions. If, in the steps
(\ref{eq:formalism1}) and~(\ref{eq:formalismexpn1}) above, the integrands are
simply replaced by their leading-order expansion terms according to the
regions $x$ and $x'_{n+1}$, respectively,
\begin{align}
  F \to \sum_{x \in R} F^{(x)}_{0\,[x]}
  \,,\qquad
  F^{(x'_1,\ldots,x'_n)}_{0,\ldots,0\,[x'_{n+1}]}
    \to F^{(x'_1,\ldots,x'_n,x'_{n+1})}_{0,\ldots,0,0\,[x'_{n+1}]}
  \,,
\end{align}
then higher-order terms are neglected which are suppressed either by powers
of the expansion parameter or by other small parameter ratios involving the
boundaries between the convergence domains. Remember that these expansions
are only introduced when they are absolutely convergent because the
integration variable is restricted to the corresponding domain, $D_x$ or
$D_{x'_{n+1}}$, respectively. In the course of the derivation above, with
all summations replaced by their leading terms, the contributions are
combined such that any dependence on the boundaries cancels out. Finally
the leading-order approximation
\begin{align}
  \label{eq:formalismidentity0}
  F_0 &=
    \sum_{x \in R} F^{(x)}_0
    - \sum_{\{x'_1,x'_2\} \subset R} F^{(x'_1,x'_2)}_{0,0}
    + \ldots
    - (-1)^n \sum_{\{x'_1,\ldots,x'_n\} \subset R} F^{(x'_1,\ldots,x'_n)}_{0,\ldots,0}
\nonumber \\* &\qquad{}
    + \ldots
    - (-1)^N \, F^{(x_1,\ldots,x_N)}_{0,\ldots,0}
\end{align}
is obtained, which reproduces the original integral~$F$ up to terms
suppressed by powers of the expansion parameter. The leading contributions
from the different regions may start at different powers of the expansion
parameter such that only some of the terms in~(\ref{eq:formalismidentity0})
are actually leading-order contributions while others are suppressed.

The leading-order expression~(\ref{eq:formalismidentity0}), although it may be
derived from the all-order result~(\ref{eq:formalismidentity}), has a value of
its own because it still holds when the validity of
condition~\ref{item:EbRcondreg} above cannot be verified for higher-order
terms or when condition~\ref{item:EbRcondsums} is violated because the
summations do not converge individually.
Appendix~\ref{app:finiteboundary} illustrates such a behaviour of the
expansion by regions with an example involving a finite integration
boundary and non-converging series expansions.

\section{Example: threshold expansion}
\label{sec:threshold}

Before generalizing the formalism of section~\ref{sec:formalism} to
non-commuting expansions, let us apply it to another example. The threshold
expansion of the one-loop three-point integral presented here is also the
first example treated in the original paper~\cite{Beneke:1997zp}. It is
illustrated in figure~\ref{fig:threshold}. We choose the loop-momentum
parametrization which the authors of~\cite{Beneke:1997zp} only use for the
soft/ultrasoft region because this parametrization is better adapted for
what we want to demonstrate here:
\begin{figure}[t]
  \centering
  \includegraphics[scale=\gscale]{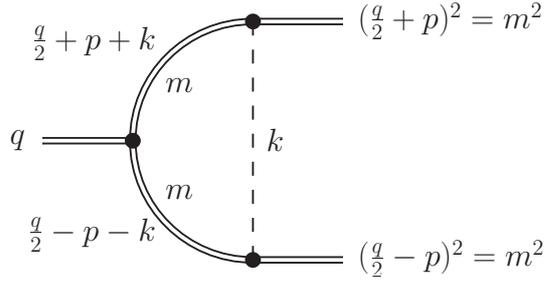}
  \caption{Loop integral for the threshold expansion.}
  \label{fig:threshold}
\end{figure}
\begin{align}
  \label{eq:threshold}
  F &= \int\!\rD k \, I
    \,,\quad \text{with }
    I = I_1 I_2 I_3
    \quad \text{and}
\nonumber \\
  I_1 &= \frac{1}{\bigl( (\frac{q}{2}+p+k)^2 - m^2\bigr)^{n_1}}
      = \frac{1}{(k^2 + q\cdot k + 2p\cdot k)^{n_1}}
    \,,
\nonumber \\
    I_2 &= \frac{1}{\bigl( (\frac{q}{2}-p-k)^2 - m^2\bigr)^{n_2}}
      = \frac{1}{(k^2 - q\cdot k + 2p\cdot k)^{n_2}}
    \,,\qquad
    I_3 = \frac{1}{(k^2)^{n_3}}
  \,.
\end{align}
We will only evaluate this integral for $n_1=n_2=n_3=1$, but we need the
general propagator powers for the analytic regularization of some
contributions.
In the expressions~(\ref{eq:threshold}) for the propagators the on-shell
conditions $(\frac{q}{2} \pm p)^2 = m^2$ have been used. These also imply
$q\cdot p = 0$ and $p^2 = m^2 - q^2/4$.
We are interested in the threshold regime $q^2 \approx (2m)^2$, where
\begin{align}
  q^2 \gg |p^2|
  \,.
\end{align}
Let us choose as a specific reference frame the centre-of-mass system of
the momentum~$q$, where
\begin{align}
  (q^\mu) = (q_0, \vec 0\,)
  \quad \text{and} \quad
  (p^\mu) = (0, \vec p\,)
  \,.
\end{align}
The propagators then read
\begin{align}
  \label{eq:thresholdcms}
  I_1 &= \frac{1}{(k_0^2 - \vec k^2 + q_0 k_0 - 2\vec p\cdot\vec k)^{n_1}}
    \,,\qquad
    I_2 = \frac{1}{(k_0^2 - \vec k^2 - q_0 k_0 - 2\vec p\cdot\vec k)^{n_2}}
    \,,
\nonumber \\*
  I_3 &= \frac{1}{(k_0^2 - \vec k^2)^{n_3}}
  \,,
\end{align}
and we are looking for an expansion in powers of $|\vec p\,|/q_0 \ll 1$.
The loop-momentum components $k_0$ and $\vec k$ are multiplied with
prefactors of different orders of magnitude in the propagators $I_1$ and
$I_2$. Thus it is natural that we also get a region which distinguishes
between $k_0$ and $\vec k$. The three regions needed for this example are
\begin{itemize}
\item the \textbf{hard region \boldmath$(h)$},
  characterized by $k_0 \sim \vec k \sim q_0$, with the expansion
  \begin{align}
    \label{eq:thresholdTh}
    T^{(h)} I_{1,2} = \sum_{j=0}^\infty
      \frac{(n_{1,2})_j}{j!} \,
      \frac{(2\vec p\cdot\vec k)^j}{
        (k_0^2 - \vec k^2 \pm q_0 k_0)^{n_{1,2}+j}}
      \,,\qquad
    T^{(h)} I_3 = I_3
      \,,
  \end{align}
  converging absolutely within $D_h = \bigl\{ k \in D : \,
  |\vec k| \gg |\vec p\,| \,\vee\, |k_0| \gg |\vec p\,| \bigr\}$,
\item the \textbf{soft region \boldmath$(s)$},
  characterized by $k_0 \sim \vec k \sim \vec p$, with the expansion
  \begin{align}
    \label{eq:thresholdTs}
    T^{(s)} I_{1,2} = \sum_{j_1,j_2,j_3=0}^\infty
      \frac{(n_{1,2})_{j_{123}}}{j_1! \, j_2! \, j_3!} \,
      \frac{(-k_0^2)^{j_1} \, (\vec k^2)^{j_2} \, (2\vec p\cdot\vec k)^{j_3}}
        {(\pm q_0 k_0)^{n_{1,2}+j_{123}}}
      \,,\qquad
    T^{(s)} I_3 = I_3
      \,,
  \end{align}
  converging absolutely within $D_s = \bigl\{ k \in D : \,
    |\vec k| \lesssim |k_0| \lesssim |\vec p\,| \bigr\}$,
\item and the \textbf{potential region \boldmath$(p)$},
  characterized by $k_0 \sim \vec p\,^2/q_0$ and $\vec k \sim \vec p$,
  with the expansion
  \begin{align}
    \label{eq:thresholdTp}
    T^{(p)} I_{1,2} &= \sum_{j=0}^\infty
      \frac{(n_{1,2})_j}{j!} \,
      \frac{(-k_0^2)^j}{
        (-\vec k^2 \pm q_0 k_0 - 2\vec p\cdot\vec k)^{n_{1,2}+j}}
      \,,
  \nonumber \\*
    T^{(p)} I_3 &= \sum_{j=0}^\infty
      \frac{(n_3)_j}{j!} \,
      \frac{(-k_0^2)^j}{(-\vec k^2)^{n_3+j}}
      \,,
  \end{align}
  converging absolutely within $D_p = \bigl\{ k \in D : \,
    |k_0| \ll |\vec k| \lesssim |\vec p\,| \bigr\}$,
\end{itemize}
where $D = \Rb^d$ is the complete integration domain. We do not have to
specify the exact positions of the boundaries between the convergence
domains, and the relation ``$\lesssim$'' is understood as the negation of
``$\gg$'' like in~(\ref{eq:2pointDsloppy}). So
condition~\ref{item:EbRcondD} of the formalism in
section~\ref{sec:formalism} holds:
\begin{align}
  D_h \cap D_s = D_h \cap D_p = D_s \cap D_p = \emptyset
  \,,\qquad
  D_h \cup D_s \cup D_p = D
  \,.
\end{align}
All expansions commute with each other (condition~\ref{item:EbRcondcomm}),
and the multiple expansions read:
\begin{align}
  \label{eq:thresholdTmult}
  (h,s): &&
    T^{(h,s)} I_{1,2} &= T^{(s)} I_{1,2}
      \,,&
    T^{(h,s)} I_3 &= I_3
      \,,
\nonumber \\
  (h,p): &&
    T^{(h,p)} I_{1,2} &= \sum_{j_1,j_2=0}^\infty
      \frac{(n_{1,2})_{j_{12}}}{j_1! \, j_2!} \,
      \frac{(-k_0^2)^{j_1} \, (2\vec p\cdot\vec k)^{j_2}}{
        (-\vec k^2 \pm q_0 k_0)^{n_{1,2}+j_{12}}}
      \,,&
    T^{(h,p)} I_3 &= T^{(p)} I_3
      \,,
\nonumber \\
  (s,p): &&
    T^{(s,p)} I_{1,2} &= T^{(s)} I_{1,2}
      \,,&
    T^{(s,p)} I_3 &= T^{(p)} I_3
      \,,
\nonumber \\
  (h,s,p): &&
    T^{(h,s,p)} I_{1,2} &= T^{(s)} I_{1,2}
      \,,&
    T^{(h,s,p)} I_3 &= T^{(p)} I_3
      \,.
\end{align}
Appendix~\ref{app:thresholdconv} demonstrates that the expansions
(\ref{eq:thresholdTh})--(\ref{eq:thresholdTp}) converge absolutely when the
integration variable~$k$ is restricted to the corresponding domain.
The domains $D_h$, $D_s$ and $D_p$ have been chosen as large as possible
within the convergence domains of the expansions. They contain parts which
do not correspond to the scaling of the loop momentum components specified
for each region. We may e.g.\ have $|\vec k| \ll |k_0|$ within~$D_s$, which
contradicts $\vec k \sim k_0 \sim \vec p$. But by choosing such enlarged
domains we avoid the introduction of additional, artificial regions for
covering the complete integration domain.

The expanded integrals read
\begin{align}
  \label{eq:thresholdints}
  F^{(h)} &= \sum_{j_1,j_2=0}^\infty
    \frac{(n_1)_{j_1} \, (n_2)_{j_2}}{j_1! \, j_2!}
  \nonumber \\* &\qquad {}\times
    \int\!
    \frac{\rD k \, (2\vec p\cdot\vec k)^{j_{12}}}{
      (k_0^2 - \vec k^2 + q_0 k_0)^{n_1+j_1} \,
      (k_0^2 - \vec k^2 - q_0 k_0)^{n_2+j_2} \, (k_0^2 - \vec k^2)^{n_3}}
    \,,
\nonumber \\
  F^{(s)} &= \sum_{j_1,\ldots,j_6=0}^\infty
    \frac{(n_1)_{j_{123}} \, (n_2)_{j_{456}}}{j_1! \cdots j_6!} \,
  \nonumber \\* &\qquad {}\times
    \int\!
    \frac{\rD k \, (-k_0^2)^{j_{14}} \, (\vec k^2)^{j_{25}} \,
        (2\vec p\cdot\vec k)^{j_{36}}}{
      (q_0 k_0 + i0)^{n_1+j_{123}} \, (-q_0 k_0 + i0)^{n_2+j_{456}} \,
        (k_0^2 - \vec k^2)^{n_3}}
    \,,
\nonumber \\
  F^{(p)} &= \sum_{j_1,j_2,j_3=0}^\infty
    \frac{(n_1)_{j_1} \, (n_2)_{j_2} \, (n_3)_{j_3}}{j_1! \, j_2! \, j_3!}
  \nonumber \\* &\qquad {}\times
    \int\!
    \frac{\rD k \, (-k_0^2)^{j_{123}}}{
        (-\vec k^2 + q_0 k_0 - 2\vec p\cdot\vec k)^{n_1+j_1} \,
        (-\vec k^2 - q_0 k_0 - 2\vec p\cdot\vec k)^{n_2+j_2} \,
        (-\vec k^2)^{n_3+j_3}}
    \,,
\nonumber \\
  F^{(h,s)} &= F^{(s)}
    \,,
\nonumber \\
  F^{(h,p)} &= \sum_{j_1,\ldots,j_5=0}^\infty
      \frac{(n_1)_{j_{12}} \, (n_2)_{j_{34}} \, (n_3)_{j_5}}{
        j_1! \cdots j_5!} \,
  \nonumber \\* &\qquad {}\times
      \int\!
      \frac{\rD k \, (-k_0^2)^{j_{135}} \, (2\vec p\cdot\vec k)^{j_{24}}}{
        (-\vec k^2 + q_0 k_0)^{n_1+j_{12}} \,
        (-\vec k^2 - q_0 k_0)^{n_2+j_{34}} \,
        (-\vec k^2)^{n_3+j_5}}
    \,,
\nonumber \\
  F^{(s,p)} &= \sum_{j_1,\ldots,j_7=0}^\infty
    \frac{(n_1)_{j_{123}} \, (n_2)_{j_{456}}\, (n_3)_{j_7}}{
      j_1! \cdots j_7!} \,
  \nonumber \\* &\qquad {}\times
    \int\!
    \frac{\rD k \, (-k_0^2)^{j_{147}} \, (\vec k^2)^{j_{25}} \,
        (2\vec p\cdot\vec k)^{j_{36}}}{
      (q_0 k_0 + i0)^{n_1+j_{123}} \, (-q_0 k_0 + i0)^{n_2+j_{456}} \,
        (-\vec k^2)^{n_3+j_7}}
    \,,
\nonumber \\
  F^{(h,s,p)} &= F^{(s,p)}
    \,.
\end{align}
As we will see in a moment when evaluating the expressions
in~(\ref{eq:thresholdints}), the integrals there are well-defined through
dimensional and analytic regularization, and the summations are absolutely
convergent. So also conditions \ref{item:EbRcondreg}
and~\ref{item:EbRcondsums} of section~\ref{sec:formalism} hold, and the
original integral~(\ref{eq:threshold}) can be expressed through the master
identity~(\ref{eq:formalismidentity}):
\begin{align}
  \label{eq:thresholdidentity0}
  F = F^{(h)} + F^{(s)} + F^{(p)}
      - F^{(h,s)} - F^{(h,p)} - F^{(s,p)}
      + F^{(h,s,p)}
  \,.
\end{align}

Now, independent of what the contributions $F^{(s)}$ and $F^{(s,p)}$ are,
we can see that they drop out because the hard expansion~$T^{(h)}$ does not
change the integrand if the latter has been expanded via~$T^{(s)}$ before,
i.e.\ $F^{(h,s)} = F^{(s)}$ and $F^{(h,s,p)} = F^{(s,p)}$, and therefore
\begin{align}
  \bigl( F^{(s)} - F^{(h,s)} \bigr)
  - \bigl( F^{(s,p)} - F^{(h,s,p)} \bigr)
  = 0
  \,.
\end{align}
So all terms including the soft expansion in~(\ref{eq:thresholdidentity0})
do not contribute to the result.

Examining these terms nevertheless, we recognize them as scaleless
contributions: Scaling the loop momentum by $k_0 \to \lambda k_0$ and $\vec
k \to \lambda \vec k$ in the integrals of $F^{(s)}$ and $F^{(s,p)}$
in~(\ref{eq:thresholdints}), we notice that each of these integrals is
identical to itself times $\lambda^{4-n_{12}-2n_3+j_{1245}-2\eps}$. Within
dimensional regularization ($\eps \not\in \Zb$) and analytic regularization
($n_i \not\in \Zb$), this factor is different from~$1$, so the integrals
are scaleless, they either vanish or diverge. Without analytic
regularization, the integrals of $F^{(s)}$ and $F^{(s,p)}$ are regularized
by the $(3-2\eps)$-dimensional $\vec k$-integration, but the
one-dimensional $k_0$-integration can still be divergent when it is
considered separately and evaluated before the $\vec k$-integration. In
particular, the $k_0$-integration is then singular because the integration
contour is pinched between two poles both at $k_0=0$, but on different
sides of the contour.\footnote{%
  In~\cite{Beneke:1997zp} the authors argue that the pinching of poles must
  be ignored in the soft region because these poles have already been taken
  into account through the potential region. In the formalism developed
  here we cannot use this argument because we have to take the integral as
  it arises from the expansions, and all possible double-counting is
  eliminated via the subtractions of the overlap contributions.} %
So, strictly speaking, we need analytic regularization (through non-integer
powers of the first two propagators) or some other additional
regularization here in order to make the integrals well-defined. Then the
contributions $F^{(s)}$, $F^{(h,s)}$, $F^{(s,p)}$ and $F^{(h,s,p)}$ are
scaleless and must be set to zero.

A similar argument shows that the contribution $F^{(h,p)}$
in~(\ref{eq:thresholdints}) is scaleless as well: Scaling the loop momentum
components by $k_0 \to \lambda^2 k_0$ and $\vec k \to \lambda \vec k$, each
of these integrals is found to be identical to itself times
$\lambda^{5-2n_{123}+2j_{135}-j_{24}-2\eps}$. So also $F^{(h,p)}$
vanishes.

Finally we obtain
\begin{align}
  \label{eq:thresholdidentity}
  F = F^{(h)} + F^{(p)}
  \,.
\end{align}
The evaluation of these two remaining contributions is sketched in the
appendices \ref{app:thresholdhcalc} and~\ref{app:thresholdpcalc}. For $n_1
= n_2 = n_3 = 1$ the results read~\cite{Beneke:1997zp}
\begin{align}
  \label{eq:thresholdres}
  F^{(h)} &= -\frac{2}{q^2}
    \left(\frac{4\mu^2}{q^2}\right)^\eps \,
    e^{\eps\gammaE} \, \Gamma(\eps)
    \sum_{j=0}^\infty \left(-\frac{4p^2}{q^2}\right)^j \,
      \frac{(1+\eps)_j}{j! \, (1+2\eps+2j)}
    \,,
\nonumber \\
  F^{(p)} &= \frac{1}{\sqrt{q^2 \, (p^2-i0)}}
    \left(\frac{\mu^2}{p^2-i0}\right)^\eps \,
    \frac{e^{\eps\gammaE} \, \Gamma(\frac{1}{2}+\eps) \, \sqrt{\pi}}{
      2\eps}
    \,.
\end{align}
Note that the potential region only has a leading-order contribution, its
higher-order integrals vanish, cf.\ appendix~\ref{app:thresholdpcalc}. The
series expansion of the hard contribution~$F^{(h)}$ is absolutely
convergent for $|p^2| \ll q^2$ as required by
condition~\ref{item:EbRcondsums} in section~\ref{sec:formalism}.
The hard contribution can be summed up and expressed through a
hypergeometric function:
\begin{align}
  \label{eq:thresholdres2F1}
  F^{(h)} = -\frac{2}{q^2}
    \left(\frac{4\mu^2}{q^2}\right)^\eps \,
    \frac{e^{\eps\gammaE} \, \Gamma(\eps)}{1+2\eps} \,
    \hyperF21\!\left(1+\eps, \frac{1}{2}+\eps; \frac{3}{2}+\eps;
      -\frac{4p^2}{q^2}\right)
  .
\end{align}

The original integral~(\ref{eq:threshold}) can alternatively be solved by
introducing a Mellin--Barnes representation. For $n_1 = n_2 = n_3 = 1$ the
Mellin--Barnes integral reads
\begin{align}
  \label{eq:thresholdMB}
  F = \frac{1}{q^2} \left(\frac{4\mu^2}{q^2}\right)^\eps \,
    \frac{e^{\eps\gammaE}}{\eps}
    \int_{-i\infty}^{i\infty}\! \frac{\rd z}{2i\pi}
    \left(\frac{4p^2-i0}{q^2}\right)^z \,
    \frac{\Gamma(-z) \, \Gamma(1+\eps+z)}{-\frac{1}{2}-\eps-z}
  \,,
\end{align}
where the pole at $z = -\frac{1}{2}-\eps$ lies to the right of the
integration contour. An asymptotic expansion in powers of $p^2/q^2$ is
obtained by closing the $z$-contour to the right. The residues of the poles
of $\Gamma(-z)$ reproduce the hard-region expansion $F^{(h)}$
in~(\ref{eq:thresholdres}), whereas the residue at $z = -\frac{1}{2}-\eps$
yields the potential contribution $F^{(p)}$.

The Mellin--Barnes representation~(\ref{eq:thresholdMB}) also permits to
perform an asymptotic expansion in the opposite case, $|p^2| \gg q^2$, by
closing the contour to the left. Here only one series of residues, from the
poles of $\Gamma(1+\eps+z)$, contributes and is expressed through a single
hypergeometric function:
\begin{align}
  F = \frac{1}{2p^2}
    \left(\frac{\mu^2}{p^2-i0}\right)^\eps \,
    e^{\eps\gammaE} \, \Gamma(\eps) \,
    \hyperF21\!\left(1+\eps, \frac{1}{2}; \frac{3}{2};
      \frac{-q^2}{4p^2-i0}\right)
  .
\end{align}
Through analytic continuation of this result to $q^2 \gg |p^2|$ by
inversing the argument of the hypergeometric
function (see e.g.~\cite{Gradshteyn:2007}), the sum of
$F^{(h)}$~(\ref{eq:thresholdres2F1}) and $F^{(p)}$~(\ref{eq:thresholdres})
is recovered.

\section{Formalism for non-commuting expansions}
\label{sec:formalismnc}

Let us return to the general case. When introducing the general formalism
in section~\ref{sec:formalism}, we required
(condition~\ref{item:EbRcondcomm}) that all expansions commute with each
other. As we will see in examples, this condition cannot always be
fulfilled by a proper choice of the regions. In the following paragraphs a
generalized formalism is developed which relaxes this condition to some
extent.

We start with the same situation as described at the beginning of
section~\ref{sec:formalism}: The integral $F = \int\!\rD k \, I$ with
integration domain~$D$ shall be expanded. We have a set~$R$ of $N$~regions,
$R=\{x_1,\ldots,x_N\}$. Each region~$x$ is characterized by an
expansion~$T^{(x)} \equiv \sum_j T^{(x)}_j$ which converges absolutely
within the domain~$D_x$.
Conditions \ref{item:EbRcondD}, \ref{item:EbRcondreg} and
\ref{item:EbRcondsums} hold, i.e.\ the domains are non-intersecting and
cover the complete integration domain~$D$, all integrals over expanded
terms are regularized, and the series expansions~$F^{(x)}$ (with integrals
over the complete domain~$D$) converge absolutely.

Condition~\ref{item:EbRcondcomm} is relaxed to the new
condition~\ref{item:EbRcondcommnc}:
\begin{enumerate}
\setcounter{enumi}{\value{EbRcondcomm}-1}
\renewcommand{\theenumi}{\arabic{enumi}a}
\item\label{item:EbRcondcommnc}%
  All expansions corresponding to regions within some subset $\Rc \subset
  R$ commute with each other and with expansions of any other region
  in~$R$:
  \begin{equation}
    T^{(x)} T^{(x')} = T^{(x')} T^{(x)} \; \forall x \in \Rc \,,\ x' \in R
    \,.
  \end{equation}
\end{enumerate}
In other words, two expansions can be interchanged if at least one of them
belongs to a region from the subset~$\Rc$. Without loss of generality let
the subset~$\Rc$ contain the first $\Nc$~regions from~$R$:
\begin{equation}
  \Rc = \{ x_1, \ldots, x_{\Nc} \} \subset R
  \,,\qquad
  0 \le \Nc \le N
  \,.
\end{equation}
Let
\begin{equation}
  \Rnc = R \setminus \Rc = \{ x_{\Nc+1}, \ldots, x_N \}
\end{equation}
be the subset of regions with non-commuting expansions. Then two expansions
are non-commuting only if their corresponding regions both belong
to~$\Rnc$. We do not specify how small or large the set $\Rnc$ can be
within~$R$, but the formalism developed in the following will only provide
useful statements if there are still regions left within~$\Rc$. Obviously
the case $\Nc = N-1$ is equivalent to $\Nc=N$ because a single region
within~$\Rnc$ would still provide an expansion commuting with all others,
as specified in condition~\ref{item:EbRcondcommnc} above.

The notations introduced in section~\ref{sec:formalism} are used here as
well, with one addition: According to~(\ref{eq:Tmult}), a multiple
expansion $T^{(x_1,x_2,\ldots)} \equiv \sum_{j_1,j_2,\ldots}
T^{(x_1,x_2,\ldots)}_{j_1,j_2,\ldots}$ implies that all these
expansions~$T^{(x_i)}_{j_i}$ commute with each other, i.e.\ at most one of
them is a non-commuting expansion of a region from~$\Rnc$. Whenever two
non-commuting expansions are applied successively, the order of their
application has to be specified. We define
\begin{equation}
  T^{(x'_2 \leftarrow x'_1)}_{j_2,j_1}
    \equiv T^{(x'_2)}_{j_2} T^{(x'_1)}_{j_1}
  \,,\qquad
  T^{(x'_2 \leftarrow x'_1)} \equiv
    \sum_{j_1,j_2} T^{(x'_2 \leftarrow x'_1)}_{j_2,j_1}
  \,,\qquad
  x'_1,x'_2 \in \Rnc
  \,,
\end{equation}
as the operator which first expands according to the region~$x'_1$, then
according to the region~$x'_2$, indicated by the arrow in the superscript.
Such a pair of non-commuting expansions may be combined with further
commuting ones corresponding to regions from~$\Rc$. They are specified in a
comma-separated list because the order of their application with respect to
each other and to the pair of non-commuting expansions is irrelevant:
\begin{equation}
  T^{(x'_2 \leftarrow x'_1,x'_3,\ldots)}_{j_2,j_1,j_3,\ldots}
    \equiv T^{(x'_2)}_{j_2} T^{(x'_1)}_{j_1} T^{(x'_3)}_{j_3} \cdots
    = T^{(x'_3)}_{j_3} \cdots T^{(x'_2)}_{j_2} T^{(x'_1)}_{j_1}
  \,,\qquad
  x'_3,\ldots \in \Rc
  \,.
\end{equation}

Exactly as in section~\ref{sec:formalism}, we start by splitting the
integration into the $N$~domains and expanding each restricted integral
accordingly, see~(\ref{eq:formalism1}):
\begin{equation}
  \label{eq:formalismnc1}
  F = \sum_{x \in R} F^{(x)}_{[x]}
  \,.
\end{equation}
The right-hand side of~(\ref{eq:formalismnc1}) is a special case (for
$n=1$) of the expression
\begin{equation}
  \label{eq:formalismncn}
  \sum^{\langle\Rc+1\rangle}_{\{x'_1,\ldots,x'_n\} \subset R}
    F^{(x'_1,\ldots,x'_n)}_{[x'_1+\ldots+x'_n]}
  \,,
\end{equation}
where each integrand is multiply expanded according to $n$~regions and
integrated over the combination of the $n$~corresponding domains. The sum
runs over subsets of $n$~distinct regions, as in
section~\ref{sec:formalism}. But the superscript $\langle\Rc+1\rangle$
indicates that the sum is restricted to such subsets which contain at most
one region from~$\Rnc$ with a non-commuting expansion:
\begin{equation}
  \label{eq:sumRcpone}
  \sum^{\langle\Rc+1\rangle}_{\{x'_1,\ldots,x'_n\} \subset R}
  \equiv
  \sum_{\substack{
      \{x'_1,\ldots,x'_n\} \subset R, \\
      \{x'_1,\ldots,x'_n\} \cap \Rnc = \emptyset \text{ or } \{x\}
    }}
  \equiv
  \sum_{\substack{
      \{x'_1,\ldots,x'_{n-1}\} \subset \Rc\,, \\
      x'_n \in R \setminus  \{x'_1,\ldots,x'_{n-1}\}
    }}
  \,.
\end{equation}
This restriction ensures that all expansions $T^{(x'_1)}, \ldots,
T^{(x'_n)}$ in~(\ref{eq:formalismncn}) commute with each other and that the
multiple expansion can be written in the
form~(\ref{eq:formalismncn}). Obviously such a restricted sum over subsets
of regions excludes any pair of regions from~$\Rnc$ with non-commuting
expansions.

Following the steps of section~\ref{sec:formalism} in
(\ref{eq:formalismextend})--(\ref{eq:formalismextsums}), we extend the
integrations in~(\ref{eq:formalismncn}) to the complete integration domain,
compensating for this by subtraction terms:
\begin{align}
  \label{eq:formalismncextsums}
  F^{(x'_1,\ldots,x'_n)}_{[x'_1+\ldots+x'_n]}
  = F^{(x'_1,\ldots,x'_n)}
    - \sum_{x'_{n+1} \in R \setminus \{x'_1,\ldots,x'_n\}}
      F^{(x'_1,\ldots,x'_n)}_{[x'_{n+1}]}
  \,.
\end{align}
Note that, in contrast to~(\ref{eq:formalismextsums}), the subtraction
terms in~(\ref{eq:formalismncextsums}) have not yet been expanded according
to the additional region~$x'_{n+1}$. We can perform this
expansion~$T^{(x'_{n+1})}$, but we have to distinguish two cases here:
If $\{x'_1,\ldots,x'_n\} \subset \Rc$ or $x'_{n+1} \in \Rc$, then all
expansions $T^{(x'_1)},\ldots,T^{(x'_{n+1})}$ commute with each other and
we proceed as in section~\ref{sec:formalism}:
\begin{equation}
  F^{(x'_1,\ldots,x'_n)}_{[x'_{n+1}]}
  = F^{(x'_1,\ldots,x'_n,x'_{n+1})}_{[x'_{n+1}]}
  \,.
\end{equation}
But if $x'_{n+1} \in \Rnc$ and also $\{x'_1,\ldots,x'_n\}$ already contains
one region from~$\Rnc$ (let us assume without loss of generality that $x'_1
\in \Rnc$ and $\{x'_2,\ldots,x'_n\} \subset \Rc$), then
\begin{equation}
  F^{(x'_1,\ldots,x'_n)}_{[x'_{n+1}]}
  = F^{(x'_{n+1} \leftarrow x'_1,x'_2\ldots,x'_n)}_{[x'_{n+1}]}
  \,,
\end{equation}
because the expansion~$T^{(x'_1)}$ has been performed
before~$T^{(x'_{n+1})}$.
Summing these terms according to (\ref{eq:formalismncn})
and~(\ref{eq:formalismncextsums}), we get two contributions, one in analogy
to~(\ref{eq:formalismsumscomb}) in section~\ref{sec:formalism}, the other
involving pairs of non-commuting expansions:
\begin{multline}
  \label{eq:formalismncsumscomb}
  \sum^{\langle\Rc+1\rangle}_{\{x'_1,\ldots,x'_n\} \subset R} \;\;
    \sum_{x'_{n+1} \in R \setminus \{x'_1,\ldots,x'_n\}}
    F^{(x'_1,\ldots,x'_n)}_{[x'_{n+1}]}
\\*
  = \sum^{\langle\Rc+1\rangle}_{\{x'_1,\ldots,x'_{n+1}\} \subset R} \;\;
    \underbrace{\sum_{i=1}^{n+1} F^{(x'_1,\ldots,x'_{n+1})}_{[x'_i]}}_{
      \textstyle
      F^{(x'_1,\ldots,x'_{n+1})}_{[x'_1+\ldots+x'_{n+1}]}}
  + \sum_{x'_1 \in \Rnc} \;\;
    \sum_{\substack{x'_2 \in \Rnc, \\ x'_2 \ne x'_1}} \;\;
    \sum_{\{x'_3,\ldots,x'_{n+1}\} \subset \Rc}
    F^{(x'_2 \leftarrow x'_1,x'_3\ldots,x'_{n+1})}_{[x'_2]}
  \,.
\end{multline}
We arrive at the following recursion formula for the
expression~(\ref{eq:formalismncn}):
\begin{multline}
  \label{eq:formalismncinduction}
  \sum^{\langle\Rc+1\rangle}_{\{x'_1,\ldots,x'_n\} \subset R}
    F^{(x'_1,\ldots,x'_n)}_{[x'_1+\ldots+x'_n]}
  = \sum^{\langle\Rc+1\rangle}_{\{x'_1,\ldots,x'_n\} \subset R}
    F^{(x'_1,\ldots,x'_n)}
  - \sum^{\langle\Rc+1\rangle}_{\{x'_1,\ldots,x'_{n+1}\} \subset R}
    F^{(x'_1,\ldots,x'_{n+1})}_{[x'_1+\ldots+x'_{n+1}]}
\\*
  - \sum_{\substack{x'_1,x'_2 \in \Rnc, \\ x'_1 \ne x'_2}} \;\;
    \sum_{\{x'_3,\ldots,x'_{n+1}\} \subset \Rc}
    F^{(x'_2 \leftarrow x'_1,x'_3\ldots,x'_{n+1})}_{[x'_2]}
  \,.
\end{multline}
The first two terms on the right-hand side
of~(\ref{eq:formalismncinduction}) look similar to the ones
in~(\ref{eq:formalisminduction}), but their sums are restricted to
combinations of expansions which commute with each other. The first term
involves integrals performed over the complete domain~$D$ and will be kept
for the final result. The second term is reinserted on the left-hand side
when the recursion formula~(\ref{eq:formalismncinduction}) is iterated from
$n=1$ as in~(\ref{eq:formalismnc1}) to $n=\min\{\Nc+1,N-1\}$. The recursion
ends when $n=\Nc+1$ because this is the maximal number of regions which can
be combined into a subset with commuting expansions, i.e.\ subsets
containing all regions in~$\Rc$ plus one region out of~$\Rnc$ each. In the
last step of the recursion, when $n=\Nc+1$, the second term on the
right-hand side of~(\ref{eq:formalismncinduction}) is absent.

Finally the third term on the right-hand side
of~(\ref{eq:formalismncinduction}) sums over pairs of regions with
non-commuting expansions plus an arbitrary number of commuting
expansions. The integrals in this term are performed over the convergence
domain~$D_{x'_2}$ of the non-commuting expansion applied last. Also these
terms are kept unchanged for the final result, so no combinations of more
than two non-commuting expansions appear in this formalism.

The recursion results in the following expression for the original
integral~$F$:
\begin{align}
  \label{eq:formalismncidentity}
    F &=
      \sum_{x \in R} F^{(x)}
      - \sum^{\langle\Rc+1\rangle}_{\{x'_1,x'_2\} \subset R} F^{(x'_1,x'_2)}
      + \ldots
      - (-1)^n \sum^{\langle\Rc+1\rangle}_{\{x'_1,\ldots,x'_n\} \subset R}
        F^{(x'_1,\ldots,x'_n)}
    \nonumber \\ &\qquad{}
      + \ldots
      + (-1)^{\Nc} \sum_{x' \in \Rnc} F^{(x',x_1,\ldots,x_{\Nc})}
    \nonumber \\ &\qquad{}
      + \sum_{\substack{x'_1,x'_2 \in \Rnc, \\ x'_1 \ne x'_2}}
        \Biggl(
          -F^{(x'_2 \leftarrow x'_1)}_{[x'_2]}
          + \ldots
          - (-1)^n \sum_{\{x''_1,\ldots,x''_n\} \subset \Rc}
            F^{(x'_2 \leftarrow x'_1,x''_1,\ldots,x''_n)}_{[x'_2]}
    \nonumber \\* &\qquad\qquad\qquad\qquad{}
          + \ldots
          - (-1)^{\Nc} \,
            F^{(x'_2 \leftarrow x'_1,x_1,\ldots,x_{\Nc})}_{[x'_2]}
        \Biggr)
    \,.
\end{align}
This is the generalized master identity for the expansion by regions with
non-commuting expansions. The terms in the first two lines
of~(\ref{eq:formalismncidentity}) look familiar when compared to the ones
from section~\ref{sec:formalism} in~(\ref{eq:formalismidentity}). The first
term is exactly the same: single expansions according to all regions,
integrated over the complete integration domain, corresponding to the known
recipe of the expansion by regions.

Also the following terms in the first two lines
of~(\ref{eq:formalismncidentity}) are similar to the identity from
section~\ref{sec:formalism}: Multiple expansions integrated again over the
complete domain. Here only those combinations of expansions appear which
commute with each other, but otherwise these terms are the same as
in~(\ref{eq:formalismidentity}). Whether they are scaleless or relevant
contributions depends on the same properties of the regularization and
choice of regions as described in section~\ref{sec:formalism}.

The terms in the last two lines of~(\ref{eq:formalismncidentity}), however,
are disturbing. Each of them is an integral performed over the convergence
domain~$D_{x'_2}$ of a non-commuting expansion. We would not expect such
terms to appear from the expansion by regions. And we cannot be sure that
the series expansions in these terms converge separately, because such
restricted integrals are not covered by condition~\ref{item:EbRcondsums}
from section~\ref{sec:formalism}. So we have to clarify why and under which
circumstances these extra terms vanish.

There is a general argument why these extra terms should cancel among
themselves if we only have a small number of non-commuting regions such
that the boundaries of their convergence domains can be varied
independently. The complete integral~$F$ must be independent of the
boundaries between the convergence domains~$D_x$. But all integrals in the
large round brackets of~(\ref{eq:formalismncidentity}) are only performed
over the subdomain~$D_{x'_2}$. If the boundary of $D_{x'_2}$ is varied
infinitesimally without changing the boundaries of the other regions
from~$\Rnc$, then the expression in the brackets and therefore the complete
result changes --- unless the sum of all the integrands within the brackets
vanishes close to the boundary. And because the position of the boundary is
quite arbitrary within some range, we may expect that the sum of the
integrands vanishes all over~$D_{x'_2}$ in order to make the complete
result independent of the boundary.

In the following we use a different, more formal way to show that the extra
terms cancel. We impose an additional condition on the expansions:
\begin{samepage}
  \begin{enumerate}
    \setcounter{enumi}{\value{EbRnoconditions}}
  \item\label{item:EbRcondnc}%
    For every combination of two non-commuting expansions, there is a
    region from~$\Rc$ whose expansion does not further change the doubly
    expanded integrand:
    \begin{equation}
      \label{eq:EbRcondnc}
      \forall x'_1,x'_2 \in \Rnc, \, x'_1 \ne x'_2, \:
      \exists \, x \in \Rc : \,
      T^{(x'_2 \leftarrow x'_1,x)} = T^{(x'_2 \leftarrow x'_1)}
      \,.
    \end{equation}
  \end{enumerate}
\end{samepage}
In every example with non-commuting regions which I have studied so far,
this is indeed the case. A few such examples are presented in sections
\ref{sec:Sudakov} and~\ref{sec:forward} and in
appendix~\ref{app:finiteboundarytheta} of this paper.

If condition~\ref{item:EbRcondnc} holds, then, for every pair $x'_1,x'_2
\in \Rnc$, the extra terms in the round brackets
of~(\ref{eq:formalismncidentity}) can be grouped in pairs of two, with and
without this particular region $x \in \Rc$ involved:
\begin{align}
  &\Bigl( -F^{(x'_2 \leftarrow x'_1)}_{[x'_2]}
    + F^{(x'_2 \leftarrow x'_1,x)}_{[x'_2]} \Bigr)
  + \ldots
\nonumber \\* &{}
  - (-1)^n \sum_{\{x''_1,\ldots,x''_n\} \subset \Rc \setminus \{x\}}
    \Bigl( F^{(x'_2 \leftarrow x'_1,x''_1,\ldots,x''_n)}_{[x'_2]}
      - F^{(x'_2 \leftarrow x'_1,x,x''_1,\ldots,x''_n)}_{[x'_2]} \Bigr)
  + \ldots
\end{align}
With the equality of the expansions imposed in~(\ref{eq:EbRcondnc}), these
combinations of integrals vanish at the integrand level, as
\begin{equation}
  F^{(x'_2 \leftarrow x'_1,x)}_{[x'_2]}
    = F^{(x'_2 \leftarrow x'_1)}_{[x'_2]}
    \,,\qquad
  F^{(x'_2 \leftarrow x'_1,x,x''_1,\ldots,x''_n)}_{[x'_2]}
    = F^{(x'_2 \leftarrow x'_1,x''_1,\ldots,x''_n)}_{[x'_2]}
  \,,
\end{equation}
and therefore all extra terms in the round brackets
of~(\ref{eq:formalismncidentity}) cancel with each other and vanish.

We conclude: If condition~\ref{item:EbRcondnc} holds, the master identity
for the expansion by regions with non-commuting expansions reduces to
\begin{align}
  \label{eq:formalismncidred}
    F &=
      \sum_{x \in R} F^{(x)}
      - \sum^{\langle\Rc+1\rangle}_{\{x'_1,x'_2\} \subset R} F^{(x'_1,x'_2)}
      + \ldots
      - (-1)^n \sum^{\langle\Rc+1\rangle}_{\{x'_1,\ldots,x'_n\} \subset R}
        F^{(x'_1,\ldots,x'_n)}
    \nonumber \\* &\qquad{}
      + \ldots
      + (-1)^{\Nc} \sum_{x' \in \Rnc} F^{(x',x_1,\ldots,x_{\Nc})}
    \,,
\end{align}
which reproduces the identity~(\ref{eq:formalismidentity}) in
section~\ref{sec:formalism} with the difference that the sums over sets of
expansions are now restricted to such combinations which commute with each
other.

Note that, as in section~\ref{sec:formalism}, it is possible to restrict
the derivation of this identity to the leading-order term in each
expansion, yielding the leading-order
approximation $F_0$~(\ref{eq:formalismidentity0}) for the integral~$F$ with
the same restrictions on the sums over regions as
in~(\ref{eq:formalismncidred}).

\section{Example: Sudakov form factor}
\label{sec:Sudakov}

\begin{figure}[t]
  \centering
  \includegraphics[scale=\gscale]{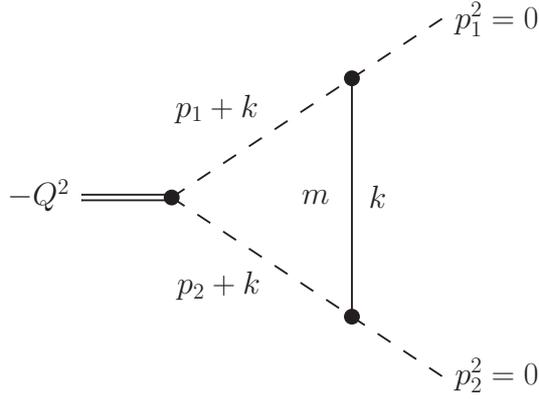}
  \caption{Vertex correction to the Sudakov form factor.}
  \label{fig:Sudakov}
\end{figure}

A toy example for non-commuting expansions can be found in
appendix~\ref{app:finiteboundarytheta}, where it is discussed in the
context of divergence problems with the series expansions.
Let us study here as a ``real'' loop integral with non-commuting expansions
the one-loop vertex correction to the Sudakov form factor: a three-point
function with one massive exchanged particle where one leg is off shell by
a large amount, whereas the other two are on shell (and massless, for
simplicity):
\begin{align}
  \label{eq:Sudakov}
  F &= \int\!\rD k \, I
    \,,\quad \text{with }
    I = I_1 I_2 I_3
    \quad \text{and}
\nonumber \\
  I_1 &= \frac{1}{\bigl( (k + p_1)^2 \bigr)^{n_1}}
      = \frac{1}{(k^2 + 2p_1\cdot k)^{n_1}}
    \,,
\nonumber \\
  I_2 &= \frac{1}{\bigl( (k + p_2)^2 \bigr)^{n_2}}
      = \frac{1}{(k^2 + 2p_2\cdot k)^{n_2}}
    \,,\qquad
  I_3 = \frac{1}{(k^2 - m^2)^{n_3}}
  \,,
\end{align}
with $p_1^2 = p_2^2 = 0$ and $(p_1-p_2)^2 = -2p_1\cdot p_2 = -Q^2 <
0$. This integral is illustrated in figure~\ref{fig:Sudakov}.
We consider the integral in the Sudakov limit
\begin{align}
  Q^2 \gg m^2
  \,,
\end{align}
so we want to expand it in powers of $m^2/Q^2$. We are particularly
interested in the case $n_1=n_2=n_3=1$, but we will use the propagator
powers, especially $n_1$ and $n_2$, as analytic regulators.

In the Sudakov limit the integral~(\ref{eq:Sudakov}) is characterized by
the two light-like directions $p_1$ and $p_2$. It is convenient to
parametrize any momentum~$k$ through the light-cone coordinates
\begin{align}
  \label{eq:lightconecoords}
  k^\pm = \frac{2}{Q} \, p_{1,2}\cdot k
\end{align}
and the perpendicular components~$k_\perp$ such that
\begin{align}
  \label{eq:lightconedecomp}
  k = k^- \, \frac{p_1}{Q} + k^+ \, \frac{p_2}{Q} + k_\perp
  \,,\qquad
  \text{with } p_{1,2}\cdot k_\perp = 0
  \,,\qquad
  k_\perp^2 = -\vec k_\perp^2
  \,.
\end{align}
In terms of these light-cone coordinates, scalar products are given by
\begin{align}
  \label{eq:lightconescalar}
  k\cdot\ell = \frac{k^+ \ell^- + k^- \ell^+}{2}
    - \vec k_\perp \cdot \vec\ell_\perp
    \,,\qquad
  k^2 = k^+ k^- - \vec k_\perp^2
  \,,
\end{align}
the propagators read
\begin{align}
  I_1 &= \frac{1}{(k^+ k^- - \vec k_\perp^2 + Q k^+)^{n_1}}
  \,,\qquad
  I_2 = \frac{1}{(k^+ k^- - \vec k_\perp^2 + Q k^-)^{n_2}}
  \,,
\nonumber \\
  I_3 &= \frac{1}{(k^+ k^- - \vec k_\perp^2 - m^2)^{n_3}}
  \,,
\end{align}
and the integration measure~(\ref{eq:Dk}) factorizes into
\begin{align}
  \int\!\rD k = \frac{\mu^{2\eps} \, e^{\eps\gammaE}}{2i \pi^{d/2}}
    \int\!\rd^{d-2}\vec k_\perp
    \int_{-\infty}^\infty\!\rd k^+ \, \rd k^-
  \,.
\end{align}

When looking for the relevant regions of this integral, let us start with
the known hard and collinear types:
\begin{itemize}
\item the \textbf{hard region \boldmath$(h)$}, characterized by $k^+ \sim
  k^- \sim \vec k_\perp \sim Q$, with the expansion
  \begin{align}
    \label{eq:SudakovTh}
    T^{(h)} I_{1,2} = I_{1,2}
    \,,\qquad
    T^{(h)} I_3 = \sum_{j=0}^\infty \frac{(n_3)_j}{j!} \,
      \frac{(m^2)^j}{(k^+ k^- - \vec k_\perp^2)^{n_3+j}}
    \,,
  \end{align}
  converging absolutely within $D_h = \bigl\{ k \in D : \,
  \vec k_\perp^2 \gg m^2 \bigr\}$,
\item the \textbf{1-collinear region \boldmath$(1c)$}, characterized by
  $k^+ \sim m^2/Q$, $k^- \sim Q$ and $\vec k_\perp \sim m$, with the
  expansion
  \begin{align}
    \label{eq:SudakovT1c}
    T^{(1c)} I_{1,3} = I_{1,3}
    \,,\qquad
    T^{(1c)} I_2 = \sum_{j_1,j_2=0}^\infty
      \frac{(n_2)_{j_{12}}}{j_1! \, j_2!} \,
      \frac{(-k^+ k^-)^{j_1} \, (\vec k_\perp^2)^{j_2}}{
        (Q k^-)^{n_2+j_{12}}}
    \,,
  \end{align}
  converging absolutely within
  \begin{displaymath}
    D_{1c} = \bigl\{ k \in D : \,
      |k^+| \ll Q \,\wedge\, \vec k_\perp^2 \ll Q |k^-| \,\wedge\,
      \vec k_\perp^2 \lesssim m^2 \bigr\}
    \,,
  \end{displaymath}
\item and the \textbf{2-collinear region \boldmath$(2c)$}, characterized by
  $k^+ \sim Q$, $k^- \sim m^2/Q$ and $\vec k_\perp \sim m$, with the
  expansion
  \begin{align}
    \label{eq:SudakovT2c}
    T^{(2c)} I_1 = \sum_{j_1,j_2=0}^\infty
      \frac{(n_1)_{j_{12}}}{j_1! \, j_2!} \,
      \frac{(-k^+ k^-)^{j_1} \, (\vec k_\perp^2)^{j_2}}{
        (Q k^+)^{n_1+j_{12}}}
    \,,\qquad
    T^{(2c)} I_{2,3} = I_{2,3}
    \,,
  \end{align}
  converging absolutely within
  \begin{displaymath}
    D_{2c} = \bigl\{ k \in D \setminus D_{1c} : \,
      |k^-| \ll Q \,\wedge\, \vec k_\perp^2 \ll Q |k^+| \,\wedge\,
      \vec k_\perp^2 \lesssim m^2 \bigr\}
    \,,
  \end{displaymath}
\end{itemize}
where $D = \Rb^d$ is the complete integration domain.
The domains $D_h, D_{1c}, D_{2c}$ have been chosen to cover as much of each
expansion's convergence domain as possible, but in such a way that they are
non-intersecting. In particular, in the domain with $\vec k_\perp^2
\lesssim m^2$ and $\vec k_\perp^2/Q \ll |k^\pm| \ll Q$, both collinear
expansions $T^{(1c)}$ and $T^{(2c)}$ converge. Here this domain is
attributed to~$D_{1c}$ and therefore explicitly excluded from $D_{2c}$, but
this is an arbitrary choice, and the boundary between $D_{1c}$ and $D_{2c}$
can be chosen differently.

For the convergence of~$T^{(h)}$ we assume that $|k^+ k^- - \vec k_\perp^2|
\gg m^2$ can be obtained whenever $\vec k_\perp^2 \gg m^2$, through an
appropriate bending of the integration contours away from the real axes
where necessary to avoid cancellations between the terms $k^+ k^-$ and
$\vec k_\perp^2$. Such convergence issues have been checked explicitly for
the threshold expansion in appendix~\ref{app:thresholdconv}.\footnote{%
  Based on the same reasoning, we could have included into~$D_h$ the domain
  with $|k^+ k^-| \gg m^2$ and $\vec k_\perp^2 \lesssim m^2$. But this only
  works as long as the term~$k^+ k^-$ is present in the denominator of the
  third propagator. We also need a region where $k^+$ and $k^-$ are both
  small (see below), expanding~$I_3$ by eliminating $k^+ k^-$ from its
  denominator. A second expansion according to the hard region is then
  impossible within this domain where $\vec k_\perp^2 \lesssim m^2$. So we
  choose to exclude this domain from~$D_h$ and rather establish an
  additional region especially dedicated to large $k^\pm$.}

Before we search for further relevant regions, let us state why we do not
need a soft region here. The soft scaling is $k^+ \sim k^- \sim \vec
k_\perp \sim m$, and the corresponding expansion is equivalent to the
double collinear expansion, $T^{(1c,2c)}$, expanding both propagators $I_1$
and~$I_2$ according to (\ref{eq:SudakovT1c}) and~(\ref{eq:SudakovT2c}).%
\footnote{See the expanded integral~(\ref{eq:SudakovF1c2c}) in
  appendix~\ref{app:Sudakovcalcscaleless} for details.}
So a soft, i.e.\ double collinear expansion is involved anyway, even if we
do not add a soft region. Adding a soft region would not change the result
within our formalism. As $T^{(s)} = T^{(1c,2c)} = T^{(1c,s)} = T^{(2c,s)} =
T^{(1c,2c,s)}$, all additional terms would cancel each other:
\begin{align}
  (-1)^n \, \Bigl(
    F^{(s,x_1,\ldots,x_n)} - F^{(1c,s,x_1,\ldots,x_n)}
    - F^{(2c,s,x_1,\ldots,x_n)} + F^{(1c,2c,s,x_1,\ldots,x_n)}
    \Bigr)
  = 0 \,,
\end{align}
for any set of other regions with $x_i \not\in \{1c,2c,s\}$.

\begin{figure}[t]
  \centering
  \includegraphics[scale=\gscaler]{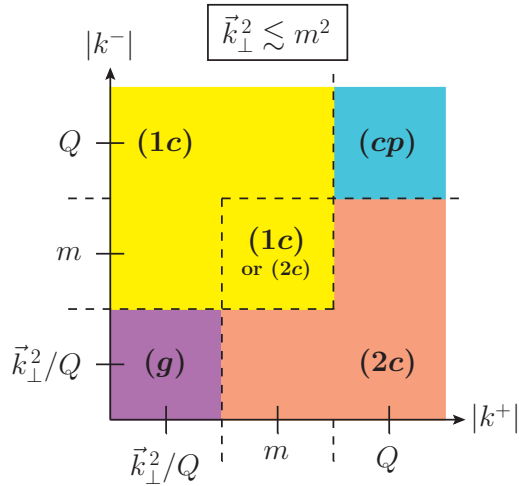}
  \caption{Convergence domains of the regions for the Sudakov form factor
    in the case $\vec k_\perp^2 \lesssim m^2$.}
  \label{fig:Sudakovregions}
\end{figure}

To determine which further regions we need, we take a look at which parts
of the integration domain~$D = \Rb^d$ are already covered by $D_h$,
$D_{1c}$ and $D_{2c}$. The domain with $\vec k_\perp^2 \gg m^2$ is
identical to~$D_h$. For $\vec k_\perp^2 \lesssim m^2$, the partitioning of
the $|k^+|$--$|k^-|$-plane is schematically shown in
figure~\ref{fig:Sudakovregions}.
Two corners of this plane are not covered by $D_{1c}$ and $D_{2c}$: The
domain where both $|k^+|$ and $|k^-|$ are small, of the order $\vec
k_\perp^2/Q$. And the opposite corner where both~$|k^\pm|$ are large, of
the order~$Q$. We associate these two missing domains to convergence
domains of additional regions and establish the corresponding
expansions. To the list of regions we add
\begin{itemize}
\item the \textbf{Glauber region \boldmath$(g)$}, characterized by $k^+
  \sim k^- \sim m^2/Q$ and $\vec k_\perp \sim m$, with the expansion
  \begin{align}
    \label{eq:SudakovTg}
    T^{(g)} I_{1,2} &= \sum_{j=0}^\infty \frac{(n_{1,2})_j}{j!} \,
      \frac{(-k^+ k^-)^j}{(-\vec k_\perp^2 + Q k^\pm)^{n_{1,2}+j}}
    \,,
  \nonumber \\*
    T^{(g)} I_3 &= \sum_{j=0}^\infty \frac{(n_3)_j}{j!} \,
      \frac{(-k^+ k^-)^j}{(-\vec k_\perp^2 - m^2)^{n_3+j}}
    \,,
  \end{align}
  converging absolutely within $D_g = \bigl\{ k \in D : \,
  Q |k^\pm| \lesssim \vec k_\perp^2 \lesssim m^2 \bigr\}$,
\item and the \textbf{collinear-plane region \boldmath$(cp)$},
  characterized by $k^+ \sim k^- \sim Q$ and $\vec k_\perp \sim m$, with
  the expansion
  \begin{align}
    \label{eq:SudakovTcp}
    T^{(cp)} I_{1,2} &= \sum_{j=0}^\infty \frac{(n_{1,2})_j}{j!} \,
      \frac{(\vec k_\perp^2)^j}{(k^+ k^- + Q k^\pm)^{n_{1,2}+j}}
    \,,
  \nonumber \\*
    T^{(cp)} I_3 &=
      \sum_{j_1,j_2=0}^\infty \frac{(n_3)_{j_{12}}}{j_1! \, j_2!} \,
      \frac{(\vec k_\perp^2)^{j_1} \, (m^2)^{j_2}}{(k^+ k^-)^{n_3+j_{12}}}
    \,,
  \end{align}
  converging absolutely within $D_{cp} = \bigl\{ k \in D : \,
  |k^\pm| \gtrsim Q \,\wedge\, \vec k_\perp^2 \lesssim m^2 \bigr\}$.
\end{itemize}
While the Glauber region has some physical meaning and can yield relevant
contributions for other integrals or with other regularization schemes, the
collinear-plane region is completely artificial and just needed to cover
the integration domain with convergence domains.

We now have five regions with non-intersecting convergence domains which
fulfill
\begin{align}
  D_h \cup D_{1c} \cup D_{2c} \cup D_g \cup D_{cp} = D
  \,.
\end{align}
Most of the corresponding expansions commute with each other, as one can
easily verify, with one exception: The Glauber and collinear-plane
expansions do not commute, their double expansions read
\begin{align}
  T^{(cp \leftarrow g)} I &= \sum_{j_1,\ldots,j_5=0}^\infty
    \frac{(n_1)_{j_{12}} \, (n_2)_{j_{34}} \, (n_3)_{j_5}}{
      j_1! \cdots j_5!} \,
    \frac{(-k^+ k^-)^{j_{135}} \, (\vec k_\perp^2)^{j_{24}}}{
      (Q k^+)^{n_1+j_{12}} \, (Q k^-)^{n_2+j_{34}} \,
      (-\vec k_\perp^2 - m^2)^{n_3+j_5}}
    \,,
\nonumber \\*
  T^{(g \leftarrow cp)} I &= \sum_{j_1,\ldots,j_6=0}^\infty
    \frac{(n_1)_{j_{12}} \, (n_2)_{j_{34}} \, (n_3)_{j_{56}}}{
      j_1! \cdots j_6!} \,
    \frac{(-k^+ k^-)^{j_{13}} \, (\vec k_\perp^2)^{j_{245}} \, (m^2)^{j_6}}{
      (Q k^+)^{n_1+j_{12}} \, (Q k^-)^{n_2+j_{34}} \,
      (k^+ k^-)^{n_3+j_{56}}}
    \,.
\end{align}
So we have a situation where the formalism for non-commuting expansions
developed in section~\ref{sec:formalismnc} is needed. The sets of commuting
and non-commuting expansions are $\Rc = \{h,1c,2c\}$ and $\Rnc = \{g,cp\}$,
respectively. Condition~\ref{item:EbRcondnc} (p.~\pageref{item:EbRcondnc})
is easily satisfied, there are always several regions whose expansions do
not further change the doubly expanded integrands:
\begin{gather}
  T^{(cp \leftarrow g,1c)} = T^{(cp \leftarrow g,2c)} =
    T^{(cp \leftarrow g)}
    \,,
\nonumber \\*
  T^{(g \leftarrow cp,h)} = T^{(g \leftarrow cp,1c)} =
    T^{(g \leftarrow cp,2c)} = T^{(g \leftarrow cp)}
    \,.
\end{gather}
Therefore the extra terms in the identity~(\ref{eq:formalismncidentity})
vanish:
\begin{align}
  &-F^{(cp \leftarrow g)}_{[cp]}
    + F^{(cp \leftarrow g,h)}_{[cp]}
    + F^{(cp \leftarrow g,1c)}_{[cp]} + F^{(cp \leftarrow g,2c)}_{[cp]}
\nonumber \\* &\qquad\qquad{}
    - F^{(cp \leftarrow g,h,1c)}_{[cp]} - F^{(cp \leftarrow g,h,2c)}_{[cp]}
    - F^{(cp \leftarrow g,1c,2c)}_{[cp]}
    + F^{(cp \leftarrow g,h,1c,2c)}_{[cp]}
\nonumber \\* &\qquad
    = (-1+2-1) \, F^{(cp \leftarrow g)}_{[cp]}
      + (1-2+1) \, F^{(cp \leftarrow g,h)}_{[cp]}
    = 0
\end{align}
and
\begin{align}
  &-F^{(g \leftarrow cp)}_{[g]}
    + F^{(g \leftarrow cp,h)}_{[g]}
    + F^{(g \leftarrow cp,1c)}_{[g]} + F^{(g \leftarrow cp,2c)}_{[g]}
\nonumber \\* &\qquad\qquad{}
    - F^{(g \leftarrow cp,h,1c)}_{[g]} - F^{(g \leftarrow cp,h,2c)}_{[g]}
    - F^{(g \leftarrow cp,1c,2c)}_{[g]}
    + F^{(g \leftarrow cp,h,1c,2c)}_{[g]}
\nonumber \\* &\qquad
    = (-1+3-3+1) \, F^{(g \leftarrow cp)}_{[g]}
    = 0
  \,.
\end{align}
We are left with the terms of the identity~(\ref{eq:formalismncidred}),
\begin{align}
  \label{eq:Sudakovidred}
  F &= F^{(h)} + F^{(1c)} + F^{(2c)} + F^{(g)} + F^{(cp)}
    - \Bigl(
      F^{(h,1c)} + F^{(h,2c)} + F^{(h,g)} + F^{(h,cp)}
\nonumber \\* &\quad{}
      + F^{(1c,2c)} + F^{(1c,g)} + F^{(1c,cp)}
      + F^{(2c,g)} + F^{(2c,cp)}
      \Bigr)
\nonumber \\ &\quad{}
    + F^{(h,1c,2c)} + F^{(h,1c,g)} + F^{(h,1c,cp)} + F^{(h,2c,g)}
      + F^{(h,2c,cp)}
      + F^{(1c,2c,g)} + F^{(1c,2c,cp)}
\nonumber \\ &\quad{}
    - \Bigl(
      F^{(h,1c,2c,g)} + F^{(h,1c,2c,cp)}
      \Bigr)
    \,,
\end{align}
summing over all combinations of regions where the expansions commute with
each other. The evaluation of these contributions is described in
appendix~\ref{app:Sudakovcalc}. The first subsection,
appendix~\ref{app:Sudakovcalcscaleless}, shows that the contributions
$F^{(g)}$ and $F^{(cp)}$ from the Glauber and collinear-plane regions are
scaleless, and the same holds for all overlap contributions with multiple
expansions. Omitting these scaleless contributions, the integral~$F$ can be
expressed as follows:
\begin{align}
  \label{eq:Sudakovidred2}
  F = F^{(h)} + F^{(1c)} + F^{(2c)}
  \,.
\end{align}
Remember that we have used analytic regularization, and we actually need it
for having all scaleless contributions well-defined.

If we had introduced a soft region and added~$F^{(s)}$ to the
result~(\ref{eq:Sudakovidred2}), according to the recipe in
section~\ref{sec:intro} for the expansion by regions, then nothing would
have changed because $F^{(s)} = F^{(1c,2c)} = 0$.

We also note that even if the collinear-plane region did not yield
scaleless integrals, all contributions involving the collinear-plane
expansion in~(\ref{eq:Sudakovidred}) would cancel each other. This is
because adding the hard expansion does not change the collinear-plane
expansion, $T^{(h,cp)} = T^{(cp)}$ (unless applied after $T^{(g)}$), which
can be seen in the expressions~(\ref{eq:Sudakovotherscaleless}) in
appendix~\ref{app:Sudakovcalcscaleless}. So the contributions
involving~$T^{(cp)}$ can be grouped into pairs
\begin{align}
  \label{eq:Sudakovcpcancel}
  (-1)^n \, \Bigl( F^{(cp,x_1,\ldots,x_n)} - F^{(h,cp,x_1,\ldots,x_n)} \Bigr)
  = 0 \,,\quad
  \text{with } \{x_1,\ldots,x_n\} \subset \{1c,2c\}
  \,,
\end{align}
which cancel each other.

The contributions remaining in the identity~(\ref{eq:Sudakovidred2}) read
\begin{align}
  \label{eq:Sudakovints}
  F^{(h)} &= \sum_{j=0}^\infty
    \frac{(n_3)_j}{j!} \, (m^2)^j
    \int\! \frac{\rD k}{
      (k^2+2p_1\cdot k)^{n_1} \, (k^2+2p_2\cdot k)^{n_2} \,
      (k^2)^{n_3+j}}
    \,,
\nonumber \\
  F^{(1c)} &= \sum_{j=0}^\infty
    \frac{(n_2)_j}{j!}
    \int\! \frac{\rD k \, (-k^2)^j}{
      (k^2+2p_1\cdot k)^{n_1} \, (2p_2\cdot k)^{n_2+j} \,
      (k^2-m^2)^{n_3}}
    \,,
\nonumber \\
  F^{(2c)} &= \sum_{j=0}^\infty
    \frac{(n_1)_j}{j!}
    \int\! \frac{\rD k \, (-k^2)^j}{
      (2p_1\cdot k)^{n_1+j} \, (k^2+2p_2\cdot k)^{n_2} \,
      (k^2-m^2)^{n_3}}
    \,,
\end{align}
based on the expansions $T^{(h)}$~(\ref{eq:SudakovTh}),
$T^{(1c)}$~(\ref{eq:SudakovT1c}) and $T^{(2c)}$~(\ref{eq:SudakovT2c}), but
rewriting the expressions in a Lorentz-invariant way, for the collinear
regions using~(\ref{eq:SudakovT1cLI}).
The evaluation of the hard contribution~$F^{(h)}$ is straightforward with
Feynman parameters (see appendix~\ref{app:Sudakovhcalc}). The collinear
contributions are calculated in appendix~\ref{app:Sudakovccalc}. The
integrals yield
\begin{align}
  \label{eq:Sudakovresgen}
  F^{(h)} &=
    \frac{\mu^{2\eps} \, e^{\eps\gammaE} \, e^{-i\pi n_{123}}}{
      \Gamma(n_1) \, \Gamma(n_2) \, \Gamma(n_3)} \,
    (Q^2)^{2-n_{123}-\eps}
    \sum_{j=0}^\infty \left(-\frac{m^2}{Q^2}\right)^j \,
    \frac{\Gamma(n_3+j) \, \Gamma(n_{123}-2+\eps+j)}{
      j! \, \Gamma(4-n_{123}-2\eps-j)}
\nonumber \\* &\qquad {}\times
    \Gamma(2-n_{13}-\eps-j) \, \Gamma(2-n_{23}-\eps-j)
    \,,
\nonumber \\
  F^{(1c)} &=
    \frac{\mu^{2\eps} \, e^{\eps\gammaE} \, e^{-i\pi n_{123}}}{
      \Gamma(n_1) \, \Gamma(n_2) \, \Gamma(n_3)} \,
    (m^2)^{2-n_{13}-\eps} \, (Q^2)^{-n_2}
    \sum_{j=0}^\infty \left(-\frac{m^2}{Q^2}\right)^j
\nonumber \\* &\qquad {}\times
    \frac{\Gamma(n_2+j) \, \Gamma(n_1-n_2-j) \, \Gamma(n_{13}-2+\eps-j) \,
        \Gamma(2-n_1-\eps+j)}{
      j! \, \Gamma(2-n_2-\eps-j)}
    \,,
\nonumber \\
  F^{(2c)} &=
    \frac{\mu^{2\eps} \, e^{\eps\gammaE} \, e^{-i\pi n_{123}}}{
      \Gamma(n_1) \, \Gamma(n_2) \, \Gamma(n_3)} \,
    (m^2)^{2-n_{23}-\eps} \, (Q^2)^{-n_1}
    \sum_{j=0}^\infty \left(-\frac{m^2}{Q^2}\right)^j
\nonumber \\* &\qquad {}\times
    \frac{\Gamma(n_1+j) \, \Gamma(n_2-n_1-j) \, \Gamma(n_{23}-2+\eps-j) \,
        \Gamma(2-n_2-\eps+j)}{
      j! \, \Gamma(2-n_1-\eps-j)}
    \,.
\end{align}
Two remarks on these results are in order. First, all terms in the series
expansions in~(\ref{eq:Sudakovresgen}) are homogeneous functions of $m^2$
and~$Q^2$, where the leading-order terms scale as
\begin{align}
  F^{(h)}_0 &\propto (Q^2)^{\frac{d}{2}-n_{123}} \,,
\nonumber \\*
  F^{(1c)}_0 &\propto (m^2)^{\frac{d}{2}-n_{13}} \, (Q^2)^{-n_2} \,,
\nonumber \\*
  F^{(2c)}_0 &\propto (m^2)^{\frac{d}{2}-n_{23}} \, (Q^2)^{-n_1} \,.
\end{align}
These scalings could have been obtained directly from the
integrals~(\ref{eq:Sudakovints}), taking into account in each region the
scaling of the integration measure and of each propagator with $m^2$
and~$Q^2$. Using such scaling arguments, the contributions from the Glauber
and collinear-plane regions would scale as
\begin{align}
  F^{(g)}_0 &\propto (m^2)^{\frac{d}{2}+1-n_{123}} \, (Q^2)^{-1} \,,
\nonumber \\
  F^{(cp)}_0 &\propto (m^2)^{\frac{d}{2}-1} \, (Q^2)^{1-n_{123}} \,,
\end{align}
if they were not scaleless. Each region's scaling has a unique dependence
on the regularization parameters $d$ and $n_1,n_2,n_3$ by which it differs
from the other regions. The overlap contributions originate from multiple
expansions with different scalings, so we could have expected all overlap
contributions to be scaleless even without calculating them.

Second, the two collinear contributions $F^{(1c)}$ and~$F^{(2c)}$
in~(\ref{eq:Sudakovresgen}) are individually divergent when $n_1$ and~$n_2$
tend to integer numbers, in particular for the case $n_1=n_2=n_3=1$ we are
interested in. The Gamma functions $\Gamma(n_1-n_2-j)$ and
$\Gamma(n_2-n_1-j)$, respectively, provide single poles in the variable
$(n_1-n_2)$ once the expansion order~$j$ is large enough (starting at $j=0$
for $n_1=n_2$). So here analytic regularization is not only needed to make
scaleless contributions well-defined, but also to prevent contributing
regions from yielding ill-defined integrals.

Let us have a look at the particular case $n_1=n_2=n_3=1$. We cannot simply
set all propagator powers to integer values, but we have to understand this
as a limit. We use the symmetric values $n_1=1+\delta$ and $n_2=1-\delta$
and expand both collinear contributions in a Laurent expansion about
$\delta=0$ up to the finite order~$\delta^0$. This choice of the limit has
the advantage that the overall dimension of the integral is not changed,
but any other way to approach the point $n_1=n_2=1$ would yield the same
result for the sum $F^{(1c)} + F^{(2c)}$, though not for the individual
terms.

The hard contribution is not singular in the limit $\delta \to 0$, so we
can directly set $n_1=n_2=n_3=1$ here. According to
(\ref{eq:Sudakovhcalcres111}) and~(\ref{eq:SudakovhcalcF21}) in
appendix~\ref{app:Sudakovhcalc},
\begin{align}
  \label{eq:SudakovFhres111}
  F^{(h)}
  &= -\frac{1}{Q^2} \left(\frac{\mu^2}{Q^2}\right)^\eps \,
    \frac{e^{\eps\gammaE} \, \Gamma(1+\eps) \, \Gamma^2(-\eps)}{
      \Gamma(1-2\eps)} \,
    \hyperF21\!\left(2\eps,1;1+\eps; \frac{m^2}{Q^2}\right)
\nonumber \\
  &= -\frac{1}{Q^2} \left(\frac{\mu^2}{Q^2}\right)^\eps
    \left[
      \frac{1}{\eps^2}
      - \frac{2}{\eps} \, \ln\!\left(1-\frac{m^2}{Q^2}\right)
      + \ln^2\!\left(1-\frac{m^2}{Q^2}\right)
      - 2\,\Li2\!\left(\frac{m^2}{Q^2}\right)
      - \frac{\pi^2}{12}
    \right]
\nonumber \\* &\qquad{}
    + \Oc(\eps)
  \,.
\end{align}
The original integral $F$~(\ref{eq:Sudakov}) is finite in four dimensions
(i.e.\ for $\eps=0$) when $n_1=n_2=n_3=1$, because there are no ultraviolet
divergences and the infrared (soft and collinear) divergences are absent
due to the finite mass~$m$ of the exchanged particle. The hard
contribution~$F^{(h)}$, however, involves infrared singularities (single
and double poles in~$\eps$) because the expansion, assuming large loop
momenta, yields massless integrals. These singularities
in~(\ref{eq:SudakovFhres111}) have to be cancelled by the contributions
from the collinear regions.

The collinear contributions are evaluated for $n_1=1+\delta$ and
$n_2=1-\delta$. The results obtained in (\ref{eq:Sudakovccalcres111})
and~(\ref{eq:Sudakovccalcres111exp}) in appendix~\ref{app:Sudakovccalc}
read
\begin{align}
  \label{eq:SudakovF1cres111}
  F^{(1c)}
  &= -\frac{1}{Q^2} \left(\frac{\mu^2}{m^2}\right)^\eps \,
    \left(\frac{Q^2}{m^2}\right)^\delta \,
    \frac{e^{\eps\gammaE} \, \Gamma(\eps+\delta) \, \Gamma(1-\eps-\delta)}{
      \Gamma(1-\eps+\delta)} \,
    \frac{\Gamma(2\delta)}{\Gamma(1+\delta)}
\nonumber \\* &\qquad {}\times
    \hyperF21\!\left(\eps-\delta,1-\delta; 1-2\delta; \frac{m^2}{Q^2}\right)
\nonumber \\
  &= -\frac{1}{2 Q^2} \left(\frac{\mu^2}{Q^2}\right)^\eps \,
    \Biggl(
      \frac{1}{\delta} \left[
        \frac{1}{\eps} + \ln\!\left(\frac{Q^2}{m^2}\right)
        - \ln\!\left(1-\frac{m^2}{Q^2}\right) \right]
      - \frac{1}{\eps^2}
      + \frac{2}{\eps} \, \ln\!\left(1-\frac{m^2}{Q^2}\right)
\nonumber \\* &\qquad{}
      + \frac{1}{2} \, \ln^2\!\left(\frac{Q^2}{m^2}\right)
      + \ln\!\left(\frac{Q^2}{m^2}\right)
        \ln\!\left(1-\frac{m^2}{Q^2}\right)
      - \ln^2\!\left(1-\frac{m^2}{Q^2}\right)
      + \Li2\!\left(\frac{m^2}{Q^2}\right)
\nonumber \\* &\qquad{}
      + \frac{5\pi^2}{12}
    \Biggr)
    + \Oc(\delta) + \Oc(\eps)
  \,.
\end{align}
The 2-collinear contribution is obtained by replacing $\delta \to -\delta$:
\begin{align}
  \label{eq:SudakovF2cres111}
  F^{(2c)} &= -\frac{1}{2 Q^2} \left(\frac{\mu^2}{Q^2}\right)^\eps \,
    \Biggl(
      -\frac{1}{\delta} \left[
        \frac{1}{\eps} + \ln\!\left(\frac{Q^2}{m^2}\right)
        - \ln\!\left(1-\frac{m^2}{Q^2}\right) \right]
      - \frac{1}{\eps^2}
      + \frac{2}{\eps} \, \ln\!\left(1-\frac{m^2}{Q^2}\right)
\nonumber \\* &\qquad{}
      + \frac{1}{2} \, \ln^2\!\left(\frac{Q^2}{m^2}\right)
      + \ln\!\left(\frac{Q^2}{m^2}\right)
        \ln\!\left(1-\frac{m^2}{Q^2}\right)
      - \ln^2\!\left(1-\frac{m^2}{Q^2}\right)
      + \Li2\!\left(\frac{m^2}{Q^2}\right)
\nonumber \\* &\qquad{}
      + \frac{5\pi^2}{12}
    \Biggr)
    + \Oc(\delta) + \Oc(\eps)
  \,.
\end{align}
When the two collinear contributions are added, the
$1/\delta$-singularities drop out. The total collinear contribution is
finite in the limit $\delta \to 0$, i.e.\ in the case $n_1=n_2=n_3=1$:
\begin{align}
  \label{eq:SudakovFcres111}
  F^{(1c)} + F^{(2c)} &= -\frac{1}{Q^2} \left(\frac{\mu^2}{Q^2}\right)^\eps \,
    \biggl[
      - \frac{1}{\eps^2}
      + \frac{2}{\eps} \, \ln\!\left(1-\frac{m^2}{Q^2}\right)
      + \frac{1}{2} \, \ln^2\!\left(\frac{Q^2}{m^2}\right)
\nonumber \\* &\qquad\quad{}
      + \ln\!\left(\frac{Q^2}{m^2}\right)
        \ln\!\left(1-\frac{m^2}{Q^2}\right)
      - \ln^2\!\left(1-\frac{m^2}{Q^2}\right)
      + \Li2\!\left(\frac{m^2}{Q^2}\right)
      + \frac{5\pi^2}{12}
    \biggr]
\nonumber \\* &\qquad{}
    + \Oc(\eps)
  \,.
\end{align}
The total collinear contribution exhibits the same single and double poles
in~$\eps$ as the hard contribution $F^{(h)}$~(\ref{eq:SudakovFhres111}),
just with opposite signs. So the sum of hard and collinear contributions is
finite in the limit $\eps \to 0$:
\begin{align}
  \label{eq:Sudakovres111}
  F &= F^{(h)} + F^{(1c)} + F^{(2c)}
\nonumber \\*
  &= -\frac{1}{Q^2}
    \left[
      \frac{1}{2} \, \ln^2\!\left(\frac{Q^2}{m^2}\right)
      + \ln\!\left(\frac{Q^2}{m^2}\right)
        \ln\!\left(1-\frac{m^2}{Q^2}\right)
      - \Li2\!\left(\frac{m^2}{Q^2}\right)
      + \frac{\pi^2}{3}
    \right]
    + \Oc(\eps)
  \,.
\end{align}
This result agrees with a direct evaluation of the original integral
$F$~(\ref{eq:Sudakov}) for $n_1=n_2=n_3=1$ via Feynman parameters without
expanding in $m^2/Q^2$.

For general propagator powers $n_{1,2,3}$ the following Mellin--Barnes
representation of the full integral $F$~(\ref{eq:Sudakov}) is found:
\begin{multline}
  \label{eq:SudakovMB}
  F = \frac{\mu^{2\eps} \, e^{\eps\gammaE} \, e^{-i\pi n_{123}}}{
      \Gamma(n_1) \, \Gamma(n_2) \, \Gamma(n_3)} \,
    (Q^2)^{2-n_{123}-\eps}
    \int_{-i\infty}^{i\infty}\! \frac{\rd z}{2i\pi}
    \left(\frac{m^2}{Q^2}\right)^z \,
    \Gamma(n_3+z) \, \Gamma(n_{123}-2+\eps+z)
\\* {}\times
    \frac{\Gamma(-z) \, \Gamma(2-n_{13}-\eps-z) \, \Gamma(2-n_{23}-\eps-z)}{
      \Gamma(4-n_{123}-2\eps-z)}
  \,.
\end{multline}
An asymptotic expansion for $m^2 \ll Q^2$ is obtained by closing the
Mellin--Barnes integral to the right and picking up the residues of the
poles from the three $\Gamma(\ldots - z)$ functions in the numerator. If
the difference $(n_1-n_2)$ is non-integer, then each of these three Gamma
functions provides a series of single poles, and the three contributions
$F^{(h)}$, $F^{(1c)}$ and $F^{(2c)}$ in~(\ref{eq:Sudakovresgen}) are
exactly reproduced. Otherwise, especially in the case $n_1=n_2=1$, the
poles of two of the Gamma functions coincide, providing a series of double
poles. For this reason the contributions $F^{(1c)}$ and $F^{(2c)}$ cannot
be individually finite in the case $n_1=n_2=1$, but their finite sum,
$F^{(1c)} + F^{(2c)}$, corresponds to the series of residues from the
double poles.
If we go one step further and set $\eps = 0$ in~(\ref{eq:SudakovMB}) in
addition to $n_1=n_2=n_3=1$, then a series of triple poles arises which
yields the finite result~(\ref{eq:Sudakovres111}).

The residues of the Mellin--Barnes integrand~(\ref{eq:SudakovMB}) are in
one-to-one correspondence with the contributions from the $(h)$, $(1c)$ and
$(2c)$ regions. This correspondence links singularities arising in
contributions from individual regions to coinciding poles of the
Mellin--Barnes integrand.

\section{Example: forward scattering
  with small momentum exchange}
\label{sec:forward}

Let us look at a last example which permits us to study the dependence of
individual contributions on different regularization schemes, in particular
with and without analytic regularization. It also is an example where
overlap contributions turn out to be relevant under certain
circumstances. We evaluate a set of one-loop corrections to the
forward-scattering amplitude of two light-like particles with a large
centre-of-mass energy, but small momentum exchange between them.

\begin{figure}[t]
  \centering
  \includegraphics[scale=\gscale]{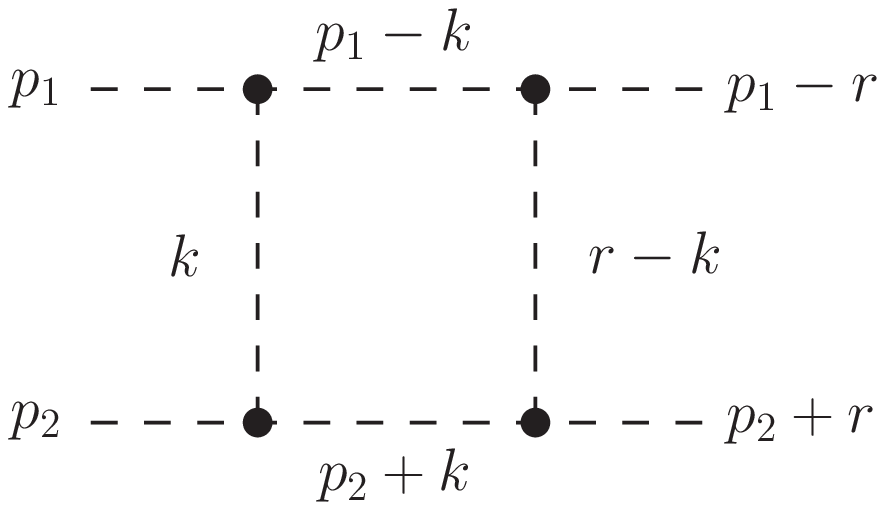}
  \includegraphics[scale=\gscale]{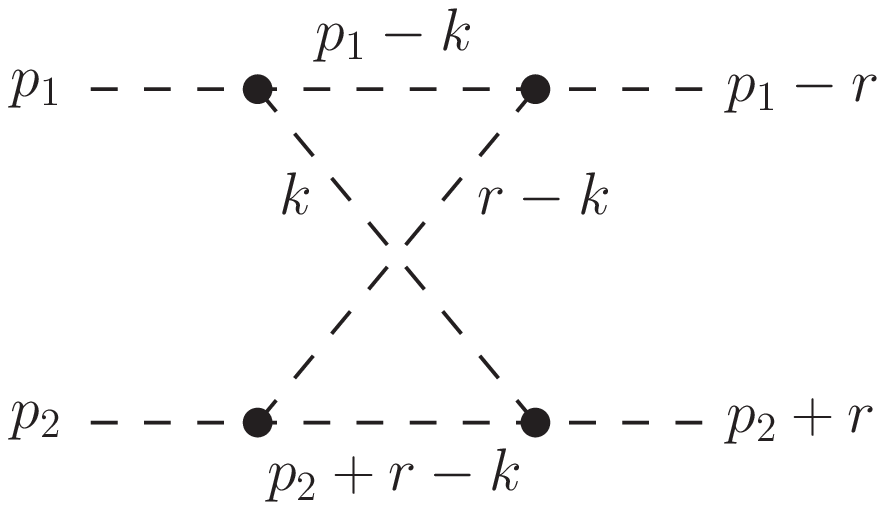}
  \caption{One-loop corrections to the forward-scattering amplitude with
    small momentum exchange~$r$.}
  \label{fig:forward}
\end{figure}

The two contributing diagrams are displayed in figure~\ref{fig:forward}.
For reasons to be seen later we symmetrize the loop integral under
commutation of the momenta of the exchanged particles, $k \leftrightarrow
(r-k)$, before specifying general propagator powers~$n_{1,2,3,4}$.
The loop integral reads
\begin{multline}
  \label{eq:forward}
  F = \frac{1}{2} \int\! \frac{\rD k}{
        (k^2)^{n_1} \, \bigl((r-k)^2\bigr)^{n_2}}
      \left( \frac{1}{\bigl((p_1-k)^2\bigr)^{n_3}}
        + \frac{1}{\bigl((p_1-r+k)^2\bigr)^{n_3}} \right)
\\* {}\times
      \left( \frac{1}{\bigl((p_2+k)^2\bigr)^{n_4}}
        + \frac{1}{\bigl((p_2+r-k)^2\bigr)^{n_4}} \right)
  .
\end{multline}
We are interested in the case $n_1=n_2=n_3=n_4=1$, but we will partly use
the propagator powers as analytic regulators.

All internal and external lines are massless. The on-shell conditions for
the external particles, $p_1^2 = (p_1-r)^2 = p_2^2 = (p_2+r)^2 = 0$, imply
$2p_1\cdot r = -2p_2\cdot r = r^2$. The centre-of-mass energy shall be much
larger than the momentum exchange, $Q^2 = (p_1+p_2)^2 = 2p_1\cdot p_2 \gg
|r^2|$.
We use the light-cone coordinates specified in (\ref{eq:lightconecoords})
and~(\ref{eq:lightconedecomp}) such that $r^\pm = 2p_{1,2}\cdot r/Q = \pm
r^2/Q$ and $r^2 = r^+ r^- - \vec r_\perp^{\,2}$. The only independent
kinematical parameters are $Q^2$ and~$\vec r_\perp^{\,2}$ with
\begin{align}
  Q^2 \gg \vec r_\perp^{\,2}
  \,.
\end{align}
The dependent parameters~$r^\pm$ are determined by
\begin{align}
  \label{eq:forwardrpm}
  r^\pm = \mp \frac{\vec r_\perp^{\,2} - r^+ r^-}{Q} 
        = \mp \frac{\vec r_\perp^{\,2}}{Q}
          + \Oc\!\left(\frac{(\vec r_\perp^{\,2})^2}{Q^3}\right)
  .
\end{align}
In terms of the expansion parameter~$\vec r_\perp^{\,2}/Q^2$, the exchanged
momentum~$r$ thus has the scaling of the Glauber region. The loop integral
is written using light-cone coordinates:
\begin{align}
  \label{eq:forwardlc}
  F &= \frac{\mu^{2\eps} \, e^{\eps\gammaE}}{2i\pi^{d/2}}
    \int\!\rd^{d-2}\vec k_\perp \int_{-\infty}^\infty\!\rd k^+ \, \rd k^- \,
    I
    \,,
\nonumber \\
  I &= \frac{1}{2} \,
    \frac{1}{(k^+ k^- - \vec k_\perp^2)^{n_1}} \,
    \frac{1}{\bigl( (k^+-r^+)(k^--r^-)
      - (\vec k_\perp - \vec r_\perp)^2 \bigr)^{n_2}}
\nonumber \\* &\quad {}\times
    \left(
      \frac{1}{\bigl( k^+ (k^--Q) - \vec k_\perp^2 \bigr)^{n_3}}
      + \frac{1}{\bigl( k^+ (k^--r^-+Q) - r^+ k^- - \vec k_\perp^2
          + 2\vec r_\perp\cdot\vec k_\perp \bigr)^{n_3}}
    \right)
\nonumber \\* &\quad {}\times
    \left(
      \frac{1}{\bigl( (k^++Q) k^- - \vec k_\perp^2 \bigr)^{n_4}}
      + \frac{1}{\bigl( (k^+-r^+-Q) k^- - k^+ r^- - \vec k_\perp^2
          + 2\vec r_\perp\cdot\vec k_\perp \bigr)^{n_4}}
    \right)
  ,
\end{align}
where relation~(\ref{eq:forwardrpm}) has been used to cancel certain terms
in the denominators.

\subsection{Regions for the forward-scattering integral}

\begin{figure}[t]
  \centering
  \includegraphics[scale=\gscaler]{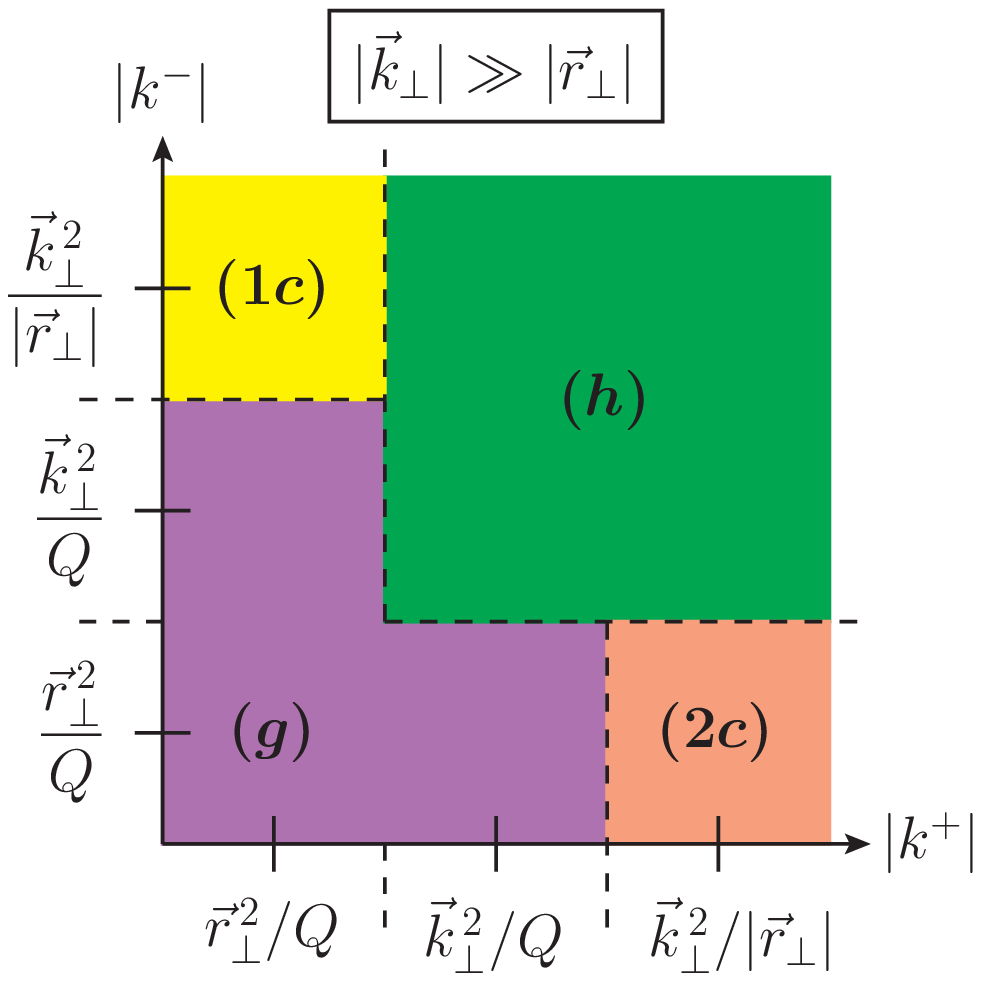}
  \includegraphics[scale=\gscaler]{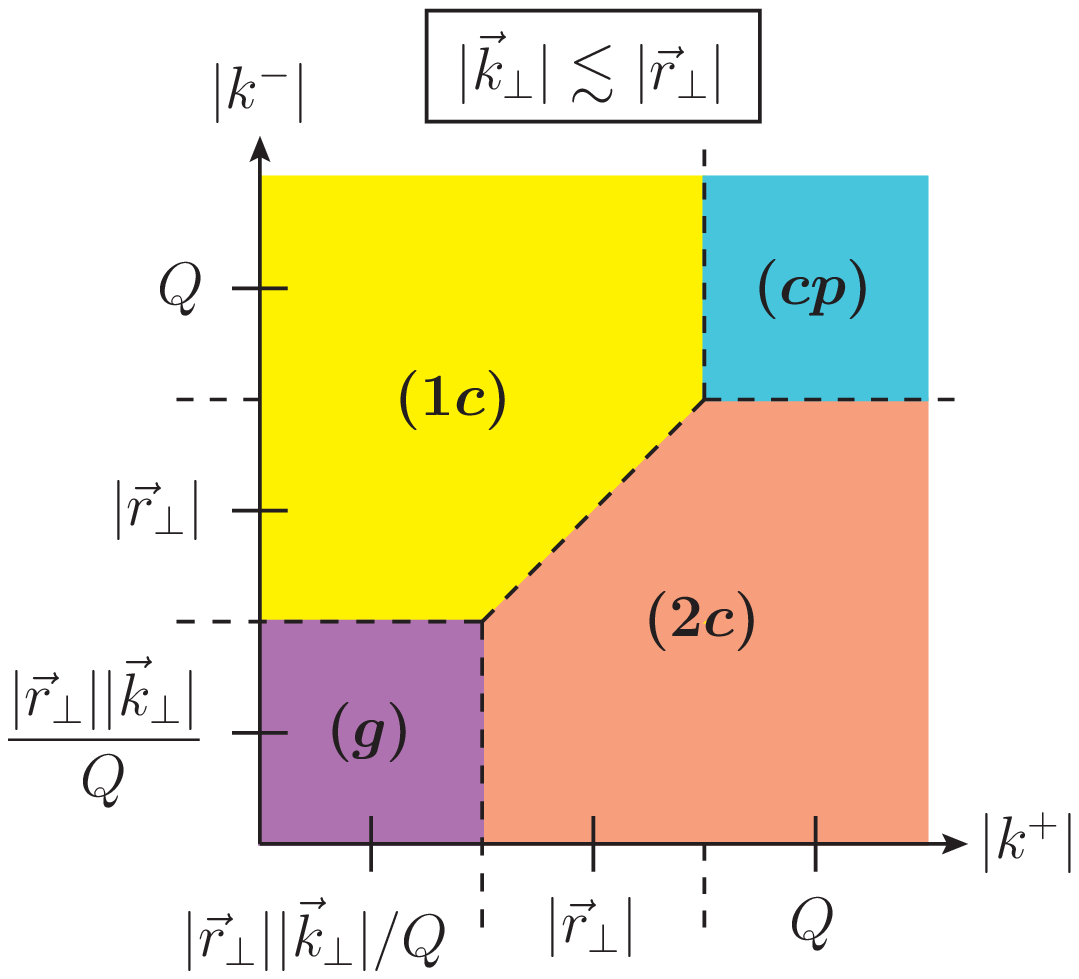}
  \caption{Convergence domains of the regions for the forward scattering
    with small momentum exchange.}
  \label{fig:forwardregions}
\end{figure}

One finds that the same regions as in the previous example of
section~\ref{sec:Sudakov} are relevant here, with $m$ replaced by $|\vec
r_\perp|$ for the scaling prescriptions. The convergence domains~$D_x$
($x=h,1c,2c,g,cp$) are similar, but slightly more involved. They are
illustrated in figure~\ref{fig:forwardregions}, where the partitioning of
the $|k^+|$--$|k^-|$-plane is shown for the cases $|\vec k_\perp| \gg |\vec
r_\perp|$ (left diagram) and $|\vec k_\perp| \lesssim |\vec r_\perp|$
(right diagram).

In the present example we will study all expansions only to leading order
in the expansion parameter~$\vec r_\perp^{\,2}/Q^2$. In detail the regions
we need are
\begin{itemize}
\item the \textbf{hard region \boldmath$(h)$}, characterized by $k^+ \sim
  k^- \sim \vec k_\perp \sim Q$, with the (leading-order) expansion
  \begin{align}
    \label{eq:forwardTh}
    T^{(h)}_0 I &= \frac{1}{2} \,
      \frac{1}{(k^+ k^- - \vec k_\perp^2)^{n_1+n_2}}
  \nonumber \\* &\qquad {}\times
      \left(
        \frac{1}{\bigl( k^+ (k^--Q) - \vec k_\perp^2 \bigr)^{n_3}}
        + \frac{1}{\bigl( k^+ (k^-+Q) - \vec k_\perp^2 \bigr)^{n_3}}
      \right)
  \nonumber \\* &\qquad {}\times
      \left(
        \frac{1}{\bigl( (k^++Q) k^- - \vec k_\perp^2 \bigr)^{n_4}}
        + \frac{1}{\bigl( (k^+-Q) k^- - \vec k_\perp^2 \bigr)^{n_4}}
      \right)
    ,
  \end{align}
  where the all-order expansion converges absolutely within
  \begin{align}
    D_h = \left\{ k \in D : \, |\vec k_\perp| \gg |\vec r_\perp|
      \,\wedge\, |k^\pm| \gg \frac{\vec r_\perp^{\,2}}{Q} \right\}
    ,
  \end{align}
\item the \textbf{1-collinear region \boldmath$(1c)$}, characterized by
  $k^+ \sim \vec r_\perp^{\,2}/Q$, $k^- \sim Q$ and $\vec k_\perp \sim \vec
  r_\perp$, with the expansion
  \begin{align}
    \label{eq:forwardT1c}
    T^{(1c)}_0 I &= \frac{1}{2} \,
      \frac{1}{(k^+ k^- - \vec k_\perp^2)^{n_1}} \,
      \frac{1}{\bigl( (k^+ - r^+_0) k^-
        - (\vec k_\perp - \vec r_\perp)^2 \bigr)^{n_2}}
  \nonumber \\* &\qquad {}\times
      \left(
        \frac{1}{\bigl( k^+ (k^--Q) - \vec k_\perp^2 \bigr)^{n_3}}
        + \frac{1}{\bigl( k^+ (k^-+Q) - r^+_0 k^- - \vec k_\perp^2
          + 2\vec r_\perp\cdot\vec k_\perp \bigr)^{n_3}}
      \right)
  \nonumber \\* &\qquad {}\times
      \left(
        \frac{1}{(Q k^-)^{n_4}} + \frac{1}{(-Q k^-)^{n_4}}
      \right)
    ,
  \end{align}
  converging absolutely within
  \begin{multline}
    D_{1c} = \Biggl\{ k \in D :
      \left( |\vec k_\perp| \gg |\vec r_\perp| \,\wedge\,
        |k^+| \lesssim \frac{\vec r_\perp^{\,2}}{Q} \,\wedge\,
        |k^-| \gg \frac{\vec k_\perp^2}{Q} \right)
  \\*
      \vee
      \left( |\vec k_\perp| \lesssim |\vec r_\perp| \,\wedge\,
        |k^+| \ll Q \,\wedge\,
        |k^-| \gg \frac{|\vec r_\perp| \, |\vec k_\perp|}{Q} \,\wedge\,
        |k^+| \le |k^-| \right)
      \Biggr\}
    \,,
  \end{multline}
\item the \textbf{2-collinear region \boldmath$(2c)$}, characterized by
  $k^+ \sim Q$, $k^- \sim \vec r_\perp^{\,2}/Q$ and $\vec k_\perp \sim \vec
  r_\perp$, with the expansion
  \begin{align}
    \label{eq:forwardT2c}
    T^{(2c)}_0 I &= \frac{1}{2} \,
      \frac{1}{(k^+ k^- - \vec k_\perp^2)^{n_1}} \,
      \frac{1}{\bigl( k^+ (k^- - r^-_0)
        - (\vec k_\perp - \vec r_\perp)^2 \bigr)^{n_2}}
  \nonumber \\* &\quad {}\times
      \left(
        \frac{1}{(-Q k^+)^{n_3}} + \frac{1}{(Q k^+)^{n_3}}
      \right)
  \nonumber \\* &\quad {}\times
      \left(
        \frac{1}{\bigl( (k^++Q) k^- - \vec k_\perp^2 \bigr)^{n_4}}
        + \frac{1}{\bigl( (k^+-Q) k^- - k^+ r^-_0 - \vec k_\perp^2
          + 2\vec r_\perp\cdot\vec k_\perp \bigr)^{n_4}}
      \right)
    ,
  \end{align}
  converging absolutely within
  \begin{multline}
    D_{2c} = \Biggl\{ k \in D :
      \left( |\vec k_\perp| \gg |\vec r_\perp| \,\wedge\,
        |k^-| \lesssim \frac{\vec r_\perp^{\,2}}{Q} \,\wedge\,
        |k^+| \gg \frac{\vec k_\perp^2}{Q} \right)
  \\*
      \vee
      \left( |\vec k_\perp| \lesssim |\vec r_\perp| \,\wedge\,
        |k^-| \ll Q \,\wedge\,
        |k^+| \gg \frac{|\vec r_\perp| \, |\vec k_\perp|}{Q} \,\wedge\,
        |k^+| > |k^-| \right)
      \Biggr\}
    \,,
  \end{multline}
\item the \textbf{Glauber region \boldmath$(g)$}, characterized by $k^+
  \sim k^- \sim \vec r_\perp^{\,2}/Q$ and $\vec k_\perp \sim \vec r_\perp$,
  with the expansion
  \begin{align}
    \label{eq:forwardTg}
    T^{(g)}_0 I &= \frac{1}{2} \,
      \frac{1}{(-\vec k_\perp^2)^{n_1}} \,
      \frac{1}{\bigl( -(\vec k_\perp - \vec r_\perp)^2 \bigr)^{n_2}}
  \nonumber \\* &\qquad {}\times
      \left(
        \frac{1}{\bigl( -Q k^+ - \vec k_\perp^2 \bigr)^{n_3}}
        + \frac{1}{\bigl( Q k^+ - \vec k_\perp^2
            + 2\vec r_\perp\cdot\vec k_\perp \bigr)^{n_3}}
      \right)
  \nonumber \\* &\qquad {}\times
      \left(
        \frac{1}{\bigl( Q k^- - \vec k_\perp^2 \bigr)^{n_4}}
        + \frac{1}{\bigl( -Q k^- - \vec k_\perp^2
            + 2\vec r_\perp\cdot\vec k_\perp \bigr)^{n_4}}
      \right)
    ,
  \end{align}
  converging absolutely within
  \begin{align}
    D_{g} = \Biggl\{ k \in D :
      &\left( |\vec k_\perp| \gg |\vec r_\perp| \,\wedge\,
        |k^+| \lesssim \frac{\vec k_\perp^2}{Q} \,\wedge\,
        |k^-| \lesssim \frac{\vec r_\perp^{\,2}}{Q} \right)
  \nonumber \\*
      \vee
      &\left( |\vec k_\perp| \gg |\vec r_\perp| \,\wedge\,
        |k^-| \lesssim \frac{\vec k_\perp^2}{Q} \,\wedge\,
        |k^+| \lesssim \frac{\vec r_\perp^{\,2}}{Q} \right)
  \nonumber \\*
      \vee
      &\left( |\vec k_\perp| \lesssim |\vec r_\perp| \,\wedge\,
        |k^\pm| \lesssim \frac{|\vec r_\perp| \, |\vec k_\perp|}{Q} \right)
      \Biggr\}
    \,,
  \end{align}
\item and the \textbf{collinear-plane region \boldmath$(cp)$},
  characterized by $k^+ \sim k^- \sim Q$ and $\vec k_\perp \sim \vec
  r_\perp$, with the expansion
  \begin{align}
    \label{eq:forwardTcp}
    T^{(cp)}_0 I &= \frac{1}{2} \,
      \frac{1}{(k^+ k^-)^{n_1+n_2}}
      \left(
        \frac{1}{\bigl( k^+ (k^--Q) \bigr)^{n_3}}
        + \frac{1}{\bigl( k^+ (k^-+Q) \bigr)^{n_3}}
      \right)
  \nonumber \\* &\qquad {}\times
      \left(
        \frac{1}{\bigl( (k^++Q) k^- \bigr)^{n_4}}
        + \frac{1}{\bigl( (k^+-Q) k^- \bigr)^{n_4}}
      \right)
    ,
  \end{align}
  converging absolutely within
  \begin{align}
    D_{cp} = \Bigl\{ k \in D :
      |\vec k_\perp| \lesssim |\vec r_\perp| \,\wedge\,
      |k^\pm| \gtrsim Q \Bigr\}
    \,,
  \end{align}
\end{itemize}
where $D = \Rb^d$ is the complete integration domain and $r^\pm_0 = \mp\vec
r_\perp^{\,2}/Q$ is the leading-order approximation of~$r^\pm$ according
to~(\ref{eq:forwardrpm}).

Note that without analytic regularization, i.e.\ setting
$n_1=n_2=n_3=n_4=1$, the Glauber contribution $F^{(g)}_0 = \int\!\rD k \,
T^{(g)}_0 I$ from~(\ref{eq:forwardTg}) is only convergent because the
integral has been symmetrized. Integrals over individual terms are
divergent for $|k^+| \to \infty$ or $|k^-| \to \infty$ like $\rd
k^\pm/k^\pm$, but the leading term in each of the round brackets
of~(\ref{eq:forwardTg}) is cancelled, making the integrals convergent at
infinity like $\rd k^\pm/(k^\pm)^2$.

The convergence domains~$D_x$ ($x=h,1c,2c,g,cp$) are non-intersecting and
cover the complete integration domain,
\begin{align}
  D_h \cup D_{1c} \cup D_{2c} \cup D_g \cup D_{cp} = D
  \,.
\end{align}
It might look strange that the domains of the collinear and Glauber regions
extend into the zone with $|\vec k_\perp| \gg |\vec r_\perp|$ (cf.\ left
diagram of figure~\ref{fig:forwardregions}). But this is necessary if we do
not want to invent new artificial regions, because the hard convergence
domain cannot cover the complete $|k^+|$--$|k^-|$-plane with $|\vec
k_\perp| \gg |\vec r_\perp|$. The definitions of the convergence domains
given above ensure that for every region~$x$ and every point $k \in D_x$
there is at least one large term left by~$T^{(x)}$ in the denominator
against which the smaller terms according to the scaling of the region~$x$
can be expanded.

The multiple expansions are determined from the scaling prescriptions for
the regions given above; they are listed in
appendix~\ref{app:forwardanalytic}. All expansions commute with each other
with the exception of $T^{(g)}$ and~$T^{(cp)}$:
\begin{align}
  \label{eq:forwardTcpg}
  T^{(cp \leftarrow g)}_0 I &= \frac{1}{2} \,
    \frac{1}{(-\vec k_\perp^2)^{n_1}} \,
    \frac{1}{\bigl( -(\vec k_\perp - \vec r_\perp)^2 \bigr)^{n_2}}
    \left(
      \frac{1}{(-Q k^+)^{n_3}} + \frac{1}{(Q k^+)^{n_3}}
    \right)
\nonumber \\* &\qquad {}\times
    \left(
      \frac{1}{(Q k^-)^{n_4}} + \frac{1}{(-Q k^-)^{n_4}}
    \right)
    ,
\\
  \label{eq:forwardTgcp}
  T^{(g \leftarrow cp)}_0 I &= T^{(cp)}_0 I
    = \frac{1}{2} \,
    \frac{1}{(k^+ k^-)^{n_1+n_2}}
    \left(
      \frac{1}{\bigl( k^+ (k^--Q) \bigr)^{n_3}}
      + \frac{1}{\bigl( k^+ (k^-+Q) \bigr)^{n_3}}
    \right)
\nonumber \\* &\qquad\qquad\quad {}\times
    \left(
      \frac{1}{\bigl( (k^++Q) k^- \bigr)^{n_4}}
      + \frac{1}{\bigl( (k^+-Q) k^- \bigr)^{n_4}}
    \right)
    .
\end{align}
The expansion $T^{(g)}$ does not alter the integrand when it is applied
after $T^{(cp)}$, because the Glauber domain~$D_g$ also contains a part
with $|\vec k_\perp| \gg |\vec r_\perp|$ (so possibly $|\vec k_\perp|
\gtrsim Q$) and $|k^+| \sim \vec k_\perp^2/Q$ or $|k^-| \sim \vec
k_\perp^2/Q$. Thus we cannot rely on $|k^\pm| \ll Q$ throughout~$D_g$ and
have to leave $(k^\pm \pm Q)$ unexpanded in $T^{(g \leftarrow
  cp)}$~(\ref{eq:forwardTgcp}). When, however, $T^{(g)}$ is applied to the
original integrand~(\ref{eq:forwardlc}) or to the integrand expanded
according to any other region except~$(cp)$, the term~$k^+ k^-$ in the
denominators is always accompanied by~$\vec k_\perp^2$. So we can use $|k^+
k^-| \ll \vec k_\perp^2$, which always holds within~$D_g$, in order to
perform the expansion~$T^{(g)}$ as stated in~(\ref{eq:forwardTg}).
All other multiple expansions converge within the convergence domain of the
last expansion, whatever expansions have been applied before.

With the notations from section~\ref{sec:formalismnc}, the sets of regions
with commuting and non-commuting expansions are $\Rc = \{h, 1c, 2c\}$ and
$\Rnc = \{g, cp\}$, respectively, as in the previous example of
section~\ref{sec:Sudakov}. Condition~\ref{item:EbRcondnc} from
p.~\pageref{item:EbRcondnc} holds because the doubly expanded
integrand~(\ref{eq:forwardTcpg}) is unchanged from further applications of
$T^{(1c)}$ or~$T^{(2c)}$, and the expression~(\ref{eq:forwardTgcp}) is
invariant under the hard expansion~$T^{(h)}$:
\begin{gather}
  T^{(cp \leftarrow g,1c)} = T^{(cp \leftarrow g,2c)}
    = T^{(cp \leftarrow g)}
    \,,
\nonumber \\
  T^{(g \leftarrow cp,h)} = T^{(g \leftarrow cp)}
    \,.
\end{gather}
So the identity~(\ref{eq:formalismncidred}) holds and has the same form as
the identity~(\ref{eq:Sudakovidred}) of the previous example. Here we
restrict ourselves to the leading-order contributions:
\begin{align}
  \label{eq:forwardidred}
  F_0 &= F^{(h)}_0 + F^{(1c)}_0 + F^{(2c)}_0 + F^{(g)}_0 + F^{(cp)}_0
    - \Bigl(
      F^{(h,1c)}_0 + F^{(h,2c)}_0 + F^{(h,g)}_0 + F^{(h,cp)}_0
\nonumber \\* &\quad{}
      + F^{(1c,2c)}_0 + F^{(1c,g)}_0 + F^{(1c,cp)}_0
      + F^{(2c,g)}_0 + F^{(2c,cp)}_0
      \Bigr)
\nonumber \\ &\quad{}
    + F^{(h,1c,2c)}_0 + F^{(h,1c,g)}_0 + F^{(h,1c,cp)}_0 + F^{(h,2c,g)}_0
      + F^{(h,2c,cp)}_0
      + F^{(1c,2c,g)}_0 + F^{(1c,2c,cp)}_0
\nonumber \\ &\quad{}
    - \Bigl(
      F^{(h,1c,2c,g)}_0 + F^{(h,1c,2c,cp)}_0
      \Bigr)
    \,.
\end{align}
Then, as explained at the end of section~\ref{sec:formalism}, we just have
to ensure that these leading-order integrals are properly regularized
(condition~\ref{item:EbRcondreg} of p.~\pageref{item:EbRcondreg}). And
condition~\ref{item:EbRcondsums}, concerning the convergence of the
expansions outside their corresponding domains, is irrelevant.

For the same reasons as in~(\ref{eq:Sudakovcpcancel}) we see that all
contributions involving the collinear-plane expansion~$T^{(cp)}$ cancel
each other.

Using scaling arguments, i.e.\ considering the scaling of the integration
measure and the propagators in each region, we predict how the
leading-order contributions of the individual regions depend on $\vec
r_\perp^{\,2}$ and~$Q^2$ (this will be confirmed by explicit evaluations later):
\begin{align}
  \label{eq:forwardscaling}
  F^{(h)}_0 &\propto (Q^2)^{\frac{d}{2}-n_{1234}} \,,
\nonumber \\
  F^{(1c)}_0 &\propto (\vec r_\perp^{\,2})^{\frac{d}{2}-n_{123}} \,
    (Q^2)^{-n_4} \,,
\nonumber \\
  F^{(2c)}_0 &\propto (\vec r_\perp^{\,2})^{\frac{d}{2}-n_{124}} \,
    (Q^2)^{-n_3} \,,
\nonumber \\
  F^{(g)}_0 &\propto (\vec r_\perp^{\,2})^{\frac{d}{2}+1-n_{1234}} \,
    (Q^2)^{-1} \,,
\nonumber \\
  F^{(cp)}_0 &\propto (\vec r_\perp^{\,2})^{\frac{d}{2}-1} \,
    (Q^2)^{1-n_{1234}} \,.
\end{align}
When the propagator powers~$n_i$ are used as analytic regulators, each
region exhibits a unique dependence on $\vec r_\perp^{\,2}$ and~$Q^2$, so
we expect that all overlap contributions must be scaleless. Without
analytic regulators, however, i.e.\ for $n_1=n_2=n_3=n_4=1$, the scalings
read
\begin{align}
  \label{eq:forwardscaling1111}
  F^{(h)}_0 &\propto (Q^2)^{-2-\eps} \,,
\nonumber \\
  F^{(1c)}_0 &\propto (\vec r_\perp^{\,2})^{-1-\eps} \,
    (Q^2)^{-1} \,,
\nonumber \\
  F^{(2c)}_0 &\propto (\vec r_\perp^{\,2})^{-1-\eps} \,
    (Q^2)^{-1} \,,
\nonumber \\
  F^{(g)}_0 &\propto (\vec r_\perp^{\,2})^{-1-\eps} \,
    (Q^2)^{-1} \,,
\nonumber \\*
  F^{(cp)}_0 &\propto (\vec r_\perp^{\,2})^{1-\eps} \,
    (Q^2)^{-3} \,.
\end{align}
We notice that the hard and collinear-plane contributions, $F^{(h)}_0$
and~$F^{(cp)}_0$, are suppressed with respect to the ``true'' leading-order
contributions $F^{(1c)}_0$, $F^{(2c)}_0$ and~$F^{(g)}_0$. And the leading
collinear and Glauber contributions share exactly the same dependence on
$\vec r_\perp^{\,2}$ and~$Q^2$. So there is no reason why the overlap
contributions $F^{(1c,2c)}_0$, $F^{(1c,g)}_0$, $F^{(2c,g)}_0$ and
$F^{(1c,2c,g)}_0$ should be scaleless; we must be prepared for relevant
contributions from these multiple expansions.

Altogether we expect that the leading-order result for the complete
integral $F$~(\ref{eq:forwardlc}), with or without analytic regulators (in
any case for $n_i \approx 1 \, \forall i$ and $\eps \approx 0$) is given by
\begin{align}
  \label{eq:forwardidred2}
  F_0 &= F^{(1c)}_0 + F^{(2c)}_0 + F^{(g)}_0
    - \Bigl(
      F^{(1c,2c)}_0 + F^{(1c,g)}_0 + F^{(2c,g)}_0
      \Bigr)
    + F^{(1c,2c,g)}_0
  \,,
\end{align}
where omitted contributions are either suppressed or scaleless, which is
checked explicitly in appendix~\ref{app:forward}.

\subsection{Evaluation with analytic regulators}
\label{sec:forwardanalytic}

The integrals contributing to~(\ref{eq:forwardidred}) are evaluated in
appendix~\ref{app:forwardanalytic} using the propagator
powers~$n_{1,2,3,4}$ as analytic regulators. It is shown there that the
Glauber contribution~$F^{(g)}_0$, the collinear-plane
contribution~$F^{(cp)}_0$ and all overlap contributions are scaleless. For
the non-vanishing contributions $F^{(h)}_0$, $F^{(1c)}_0$ and $F^{(2c)}_0$,
the scaling with $\vec r_\perp^{\,2}$ and~$Q^2$ predicted
in~(\ref{eq:forwardscaling}) is confirmed. In the limit $n_i \to 1 \,
\forall i$, i.e.\ when all propagator powers tend to~$1$, and close to
$d=4$ dimensions ($\eps \approx 0$), the hard contribution~$F^{(h)}_0$ is
suppressed with respect to the collinear contributions by one power of
$\vec r_\perp^{\,2}/Q^2$. So the general
expression~(\ref{eq:forwardidred2}) for the leading-order result is
confirmed. When analytic regularization is employed, all but the first two
terms are scaleless, and the leading-order result reads
\begin{align}
  \label{eq:forwardidredanalytic}
  F_0 &= F^{(1c)}_0 + F^{(2c)}_0
  \,.
\end{align}
These two collinear contributions are given in
(\ref{eq:forwardcalccF01cres}) and~(\ref{eq:forwardcalccF02cres}) of
appendix~\ref{app:forwardanalytic}. They both involve a factor which is
singular for $n_3 = n_4$, but the limits $n_1 \to 1$ and $n_2 \to 1$ are
well-defined. Let us evaluate the contributions for $n_1=n_2=1$,
$n_3=1+\delta_3$ and $n_4=1+\delta_4$:
\begin{align}
  F^{(1c)}_0 &=
    \frac{1}{\vec r_\perp^{\,2} \, Q^2}
    \left(\frac{\mu^2}{\vec r_\perp^{\,2}}\right)^\eps \,
    (\vec r_\perp^{\,2})^{-\delta_3} \, (Q^2)^{-\delta_4} \,
    e^{-i\pi\delta_3} \, (e^{-i\pi\delta_4} - 1) \,
    \Gamma(\delta_3-\delta_4)
\nonumber \\* &\qquad {}\times
    \frac{e^{\eps\gammaE} \,
        \Gamma(1+\eps+\delta_3) \, \Gamma^2(-\eps-\delta_3)}{
      \Gamma(1+\delta_3) \, \Gamma(-2\eps-\delta_3-\delta_4)}
    \,,
\nonumber \\
  F^{(2c)}_0 &=
    \frac{1}{\vec r_\perp^{\,2} \, Q^2}
    \left(\frac{\mu^2}{\vec r_\perp^{\,2}}\right)^\eps \,
    (\vec r_\perp^{\,2})^{-\delta_4} \, (Q^2)^{-\delta_3} \,
    e^{-i\pi\delta_4} \, (e^{-i\pi\delta_3} - 1) \,
    \Gamma(\delta_4-\delta_3)
\nonumber \\* &\qquad {}\times
    \frac{e^{\eps\gammaE} \,
        \Gamma(1+\eps+\delta_4) \, \Gamma^2(-\eps-\delta_4)}{
      \Gamma(1+\delta_4) \, \Gamma(-2\eps-\delta_3-\delta_4)}
    \,.
\end{align}
We are interested in the case $n_1=n_2=n_3=n_4=1$, i.e.\ in the point
$\delta_3=\delta_4=0$. The two collinear contributions individually depend
on the direction in the $\delta_3$--$\delta_4$-plane by which we approach
this point. One possible choice is the antisymmetric case of approaching
the desired limit on the line~$\delta_3 = -\delta_4 \equiv \delta$:
\begin{align}
  F^{(1c)}_0 \big|_{\delta_{3,4}=\pm\delta} &=
      \frac{1}{\vec r_\perp^{\,2} \, Q^2}
      \left(\frac{\mu^2}{\vec r_\perp^{\,2}}\right)^\eps
      \left(\frac{Q^2}{\vec r_\perp^{\,2}}\right)^\delta \,
      (1 - e^{-i\pi\delta}) \, \Gamma(2\delta) \,
\nonumber \\* &\qquad {}\times
      \frac{e^{\eps\gammaE} \,
          \Gamma(1+\eps+\delta) \, \Gamma^2(-\eps-\delta)}{
        \Gamma(1+\delta) \, \Gamma(-2\eps)}
\nonumber \\*
    &= \frac{1}{2} \, \frac{i\pi}{\vec r_\perp^{\,2} \, Q^2}
      \left(\frac{\mu^2}{\vec r_\perp^{\,2}}\right)^\eps \,
      \frac{e^{\eps\gammaE} \, \Gamma(1+\eps) \, \Gamma^2(-\eps)}{
        \Gamma(-2\eps)}
      + \Oc(\delta)
    \,,
\nonumber \\
  F^{(2c)}_0 \big|_{\delta_{3,4}=\pm\delta} &=
      \frac{1}{\vec r_\perp^{\,2} \, Q^2}
      \left(\frac{\mu^2}{\vec r_\perp^{\,2}}\right)^\eps
      \left(\frac{Q^2}{\vec r_\perp^{\,2}}\right)^{-\delta} \,
      (1 - e^{i\pi\delta}) \, \Gamma(-2\delta) \,
\nonumber \\* &\qquad {}\times
      \frac{e^{\eps\gammaE} \,
          \Gamma(1+\eps-\delta) \, \Gamma^2(-\eps+\delta)}{
        \Gamma(1-\delta) \, \Gamma(-2\eps)}
\nonumber \\*
    &= \frac{1}{2} \, \frac{i\pi}{\vec r_\perp^{\,2} \, Q^2}
      \left(\frac{\mu^2}{\vec r_\perp^{\,2}}\right)^\eps \,
      \frac{e^{\eps\gammaE} \, \Gamma(1+\eps) \, \Gamma^2(-\eps)}{
        \Gamma(-2\eps)}
      + \Oc(\delta)
    \,.
\end{align}
In the limit $\delta \to 0$ both collinear contributions are equal. Note
that each collinear contribution is individually finite, which is due to a
cancellation of $1/\delta$ singularities between the two diagrams shown in
figure~\ref{fig:forward}, such that the factor $(1-e^{\mp i\pi\delta})$
cancels the singularity from $\Gamma(\pm 2\delta)$. The total result with
$n_1=n_2=n_3=n_4=1$ reads
\begin{align}
  \label{eq:forwardF0ressym}
  F_0 \big|_{n_i=1}
  = \lim_{\delta \to 0} F^{(1c)}_0 \big|_{\delta_{3,4}=\pm\delta}
      + \lim_{\delta \to 0} F^{(2c)}_0 \big|_{\delta_{3,4}=\pm\delta}
  = \frac{i\pi}{\vec r_\perp^{\,2} \, Q^2}
      \left(\frac{\mu^2}{\vec r_\perp^{\,2}}\right)^\eps \,
      \frac{e^{\eps\gammaE} \, \Gamma(1+\eps) \, \Gamma^2(-\eps)}{
        \Gamma(-2\eps)}
  \,.
\end{align}
Another choice for the analytic regularization consists in taking the limit
$\delta_3 \to 0$ first:
\begin{align}
  F^{(1c)}_0 \big|_{\delta_3=0} &=
    \frac{1}{\vec r_\perp^{\,2} \, Q^2}
    \left(\frac{\mu^2}{\vec r_\perp^{\,2}}\right)^\eps \,
    (Q^2)^{-\delta_4} \,
    (e^{-i\pi\delta_4} - 1) \,
    \Gamma(-\delta_4) \,
    \frac{e^{\eps\gammaE} \,
        \Gamma(1+\eps) \, \Gamma^2(-\eps)}{
      \Gamma(-2\eps-\delta_4)}
    \,,
\nonumber \\
  F^{(2c)}_0 \big|_{\delta_3=0} &= 0
    \,.
\end{align}
The total result is then reproduced entirely by the 1-collinear
contribution:
\begin{align}
  \label{eq:forwardF0res31}
  F_0 \big|_{n_i=1} = \lim_{\delta_4 \to 0} F^{(1c)}_0 \big|_{\delta_3=0}
  = \frac{i\pi}{\vec r_\perp^{\,2} \, Q^2}
      \left(\frac{\mu^2}{\vec r_\perp^{\,2}}\right)^\eps \,
      \frac{e^{\eps\gammaE} \, \Gamma(1+\eps) \, \Gamma^2(-\eps)}{
        \Gamma(-2\eps)}
  \,.
\end{align}
Other choices are possible, but they all lead to the same total result. In
fact, the analytic regularization parameters can be expressed through
$\delta_{3,4} = \Delta \pm \delta$, isolating the singularity in the
variable~$\delta$. The total result remains invariant upon subtracting the
$1/\delta$ terms from each collinear contribution because these subtraction
terms cancel each other:
\begin{align}
  \lefteqn{
    \Bigl[ F^{(1c)}_0 + F^{(2c)}_0 \Bigr]_{\delta_{3,4} = \Delta \pm \delta}
    = \frac{1}{\vec r_\perp^{\,2} \, Q^2}
      \left(\frac{\mu^2}{\vec r_\perp^{\,2}}\right)^\eps \,
      (\vec r_\perp^{\,2} \, Q^2)^{-\Delta} \,
      \frac{e^{\eps\gammaE} \, e^{-i\pi\Delta}}{\Gamma(-2\eps-2\Delta)}
  } \qquad
\nonumber \\* &{}\times
    \Biggl[
      \left(\frac{Q^2}{\vec r_\perp^{\,2}}\right)^\delta \,
      (e^{-i\pi\Delta} - e^{-i\pi\delta}) \,
      \Gamma(2\delta) \,
      \frac{\Gamma(1+\eps+\Delta+\delta) \, \Gamma^2(-\eps-\Delta-\delta)}{
        \Gamma(1+\Delta+\delta)}
\nonumber \\* &\qquad{}
    - \frac{e^{-i\pi\Delta} - 1}{2\delta} \,
      \frac{\Gamma(1+\eps+\Delta) \, \Gamma^2(-\eps-\Delta)}{
        \Gamma(1+\Delta)}
    \Biggr]
    + (\delta \to -\delta)
  \,.
\end{align}
Here the expression in the square brackets has a well-defined finite limit
for $\delta \to 0$ and $\Delta \to 0$, independent of the way in which this
limit is approached, which reproduces the known result:
\begin{align}
  \label{eq:forwardF0resgen}
  F_0 \big|_{n_i=1} = \lim_{\delta,\Delta \to 0}
    \Bigl[ F^{(1c)}_0 + F^{(2c)}_0 \Bigr]_{\delta_{3,4} = \Delta \pm \delta}
  = \frac{i\pi}{\vec r_\perp^{\,2} \, Q^2}
      \left(\frac{\mu^2}{\vec r_\perp^{\,2}}\right)^\eps \,
      \frac{e^{\eps\gammaE} \, \Gamma(1+\eps) \, \Gamma^2(-\eps)}{
        \Gamma(-2\eps)}
  \,.
\end{align}
It is evident that $\lim_{n_i \to 1} F_0$ must be independent of the order
in which the propagator powers~$n_{1,2,3,4}$ are sent to~$1$, because
expression~(\ref{eq:forwardidredanalytic}) is derived from the
identity~(\ref{eq:forwardidred}) which is valid for any regularization
scheme (as long as the conditions of the formalism presented in sections
\ref{sec:formalism} and~\ref{sec:formalismnc} hold).

The result obtained with the expansion by regions can be checked by
evaluating the original integral $F$~(\ref{eq:forward}) with the help of a
Mellin--Barnes representation:
\begin{align}
  \label{eq:forwardMB}
  F &= \frac{\mu^{2\eps} \, e^{\eps\gammaE} \, e^{-i\pi n_{1234}}}{
      \Gamma(n_1) \, \Gamma(n_2) \, \Gamma(n_3) \, \Gamma(n_4) \,
      \Gamma(4-n_{1234}-2\eps)}
\nonumber \\* &\qquad {}\times
    \int_{-i\infty}^{i\infty}\!\frac{\rd z}{2i\pi} \, (-r^2)^z \,
    \Bigl[ (-Q^2-i0)^{2-n_{1234}-\eps-z}
      + (Q^2+r^2)^{2-n_{1234}-\eps-z} \Bigr]
\nonumber \\* &\qquad {}\times
    \Gamma(n_1+z) \, \Gamma(n_2+z) \, \Gamma(n_{1234}-2+\eps+z)
\nonumber \\* &\qquad {}\times
    \Gamma(-z) \, \Gamma(2-n_{123}-\eps-z) \, \Gamma(2-n_{124}-\eps-z)
  \,.
\end{align}
The asymptotic expansion for $|r^2| \ll Q^2$ is obtained by closing the
Mellin--Barnes integral to the right and extracting the residues of the
three Gamma functions in the last line of~(\ref{eq:forwardMB}). The
(possibly) leading terms originate from the residues at $z=0$,
$z=2-n_{123}-\eps$ and $z=2-n_{124}-\eps$. Using $-r^2 \approx \vec
r_\perp^{\,2}$ and $Q^2 + r^2 \approx Q^2$, they reproduce the
contributions from the hard~(\ref{eq:forwardcalchres}),
1-collinear~(\ref{eq:forwardcalccF01cres}) and
\mbox{2-collinear}~(\ref{eq:forwardcalccF02cres}) regions, respectively.

\subsection{Evaluation without analytic regulators}
\label{sec:forward1111}

Now we would like to see how the contributions from the previous
section~\ref{sec:forwardanalytic} change when we evaluate them without
using analytic regulators. This means that we set all propagator powers to
the fixed values $n_1=n_2=n_3=n_4=1$ before performing the integrals.

The evaluation without analytic regularization of the integrals
contributing to the complete expression~(\ref{eq:forwardidred}) is
described in appendix~\ref{app:forward1111}. The scaling of the
contributions $F^{(h)}_0$~(\ref{eq:forward1111calcF0hres}),
$F^{(1c)}_0$~(\ref{eq:forward1111calcF01cres}),
$F^{(2c)}_0$~(\ref{eq:forward1111calcF02cres}) and
$F^{(g)}_0$~(\ref{eq:forward1111calcF0gres}) with the parameters $\vec
r_\perp^{\,2}$ and~$Q^2$ as predicted in~(\ref{eq:forwardscaling1111}) is
confirmed by explicit calculation. This implies that the hard
contribution~$F^{(h)}_0$ is suppressed by one power of $\vec
r_\perp^{\,2}/Q^2$ with respect to the collinear and Glauber contributions
and does not contribute to the leading-order result~$F_0$.

The collinear-plane contribution
$F^{(cp)}_0$~(\ref{eq:forward1111calcF0cp}) and all overlap contributions
involving the collinear-plane region are found to be scaleless via
dimensional regularization, even without analytic regulators.\footnote{%
  A subtle point concerning the regularization of poles pinching the
  integration contours at $k^\pm=0$ is treated in
  appendix~\ref{app:forward1111}.}
Also the overlap contributions involving the hard region yield scaleless
integrals~(\ref{eq:forward1111calchoverlapres}).

These results confirm the identity~(\ref{eq:forwardidred2}) for the
complete leading-order result~$F_0$, made up of the contributions from the
1-collinear, 2-collinear and Glauber regions including all their overlap
contributions. The corresponding integrals are evaluated in
(\ref{eq:forward1111calcF01cres}), (\ref{eq:forward1111calcF02cres}),
(\ref{eq:forward1111calcF0gres}), (\ref{eq:forward1111calcF01c2cres}) and
(\ref{eq:forward1111calcoverlapres}). They all yield identical results,
\begin{align}
  \label{eq:forward1111contres}
  F^{(1c)}_0 &= F^{(2c)}_0 = F^{(g)}_0 = F^{(1c,2c)}_0
    = F^{(1c,g)}_0 = F^{(2c,g)}_0 = F^{(1c,2c,g)}_0
\nonumber \\*
  &= \frac{i\pi}{\vec r_\perp^{\,2} \, Q^2}
      \left(\frac{\mu^2}{\vec r_\perp^{\,2}}\right)^\eps \,
      \frac{e^{\eps\gammaE} \, \Gamma(1+\eps) \, \Gamma^2(-\eps)}{
        \Gamma(-2\eps)}
  \,.
\end{align}
This is very different from the contributions with analytic regularization
determined in section~\ref{sec:forwardanalytic}. In particular, the
terms~(\ref{eq:forward1111contres}) involve finite and relevant overlap
contributions.
The complete result, to leading order in~$\vec r_\perp^{\,2}/Q^2$, reads
\begin{align}
  \label{eq:forward1111F0res}
  F_0 &= F^{(1c)}_0 + F^{(2c)}_0 + F^{(g)}_0
      - \Bigl(
        F^{(1c,2c)}_0 + F^{(1c,g)}_0 + F^{(2c,g)}_0
        \Bigr)
      + F^{(1c,2c,g)}_0
\nonumber \\
    &= \Bigl( 1+1+1 - (1+1+1) + 1 \Bigr) \,
      \frac{i\pi}{\vec r_\perp^{\,2} \, Q^2}
      \left(\frac{\mu^2}{\vec r_\perp^{\,2}}\right)^\eps \,
      \frac{e^{\eps\gammaE} \, \Gamma(1+\eps) \, \Gamma^2(-\eps)}{
        \Gamma(-2\eps)}
\nonumber \\
    &= \frac{i\pi}{\vec r_\perp^{\,2} \, Q^2}
      \left(\frac{\mu^2}{\vec r_\perp^{\,2}}\right)^\eps \,
      \frac{e^{\eps\gammaE} \, \Gamma(1+\eps) \, \Gamma^2(-\eps)}{
        \Gamma(-2\eps)}
  \,.
\end{align}
The total result obtained without analytic regularization agrees with the
result from the previous section, expressions (\ref{eq:forwardF0ressym}),
(\ref{eq:forwardF0res31}) or~(\ref{eq:forwardF0resgen}), where analytic
regulators have been used. But the partitioning of the contributions among
the regions is completely different. If we had omitted the overlap
contributions from~(\ref{eq:forward1111F0res}), the result would have been
three times too large.

At first sight it might seem to be just by chance that the cancellations
between the various contributions in~(\ref{eq:forward1111F0res}) leave
exactly once the individual contribution. But this is a general feature
whenever a certain set of regions (with commuting expansions) yields
identical contributions, including all overlap contributions from these
regions. Assume that
\begin{align}
  F^{(x'_1,\ldots,x'_n)}_0
    &\equiv \int\!\rD k \, T^{(x'_1)}_0 \cdots T^{(x'_n)}_0 I
\nonumber \\*
    &= \hat F_0 \quad
    \forall x'_1,\ldots,x'_n \in \hat R \equiv \{x_1,\ldots,x_{\hat N}\}
    \,,\quad
    1 \le n \le \hat N \le N
  \,.
\end{align}
Then the total (leading-order) contribution from these regions, including
the multiple expansions, reads
\begin{align}
  \sum_{n=1}^{\hat N} (-1)^{n-1} \sum_{\{x'_1,\ldots,x'_n\} \subset \hat R}
    F^{(x'_1,\ldots,x'_n)}_0
  &= \int\!\rD k \,
    \Biggl[ 1 -
      \prod_{i=1}^{\hat N} \bigl( 1 - T^{(x_i)}_0 \bigr)
    \Biggr] \, I
\nonumber \\*
  &= \Biggl[ 1 - \prod_{i=1}^{\hat N} (1 - 1) \Biggr] \, \hat F_0
  = \hat F_0
  \,,
\end{align}
where we have used $\int\!\rD k \, (1-1) \, I = (1-1) F = 0 = (1-1) \hat
F_0$. The individual contribution~$\hat F_0$ is left exactly once from all
the cancellations.

The Mellin--Barnes representation~(\ref{eq:forwardMB}) is also valid
without analytic regularization. For $n_1=n_2=n_3=n_4=1$ it reads
\begin{align}
  \label{eq:forward1111MB}
  F &= \frac{\mu^{2\eps} \, e^{\eps\gammaE}}{\Gamma(-2\eps)}
    \int_{-i\infty}^{i\infty}\!\frac{\rd z}{2i\pi} \, (-r^2)^z \,
    \Bigl[ (-Q^2-i0)^{-2-\eps-z} + (Q^2+r^2)^{-2-\eps-z} \Bigr]
\nonumber \\* &\qquad {}\times
    \Gamma^2(1+z) \, \Gamma(2+\eps+z) \,
    \Gamma(-z) \, \Gamma^2(-1-\eps-z)
  \,.
\end{align}
Closing the Mellin--Barnes integral to the right, the leading-order
asymptotic expansion in $r^2/Q^2$ is extracted from the residue at
$z=-1-\eps$. The $z$-integral has a double pole there, and its residue
yields the leading-order result~(\ref{eq:forward1111F0res}), using $-r^2
\approx \vec r_\perp^{\,2}$ and $Q^2+r^2 \approx Q^2$.

\section{Conclusions}
\label{sec:conclusions}

Let us summarize the main statements of the general formalism introduced in
section~\ref{sec:formalism} and generalized in
section~\ref{sec:formalismnc}.

The integral $F = \int\!\rD k \, I$ with integration domain~$D$ shall be
expanded in some limit. We have defined a set~$R$ of $N$~regions,
$R=\{x_1,\ldots,x_N\}$. Each region~$x$ is characterized by an
expansion~$T^{(x)} \equiv \sum_j T^{(x)}_j$ which converges absolutely when
the integration variable is restricted to the corresponding
domain~$D_x$. The conditions for the applicability of the formalism read:
\begin{enumerate}
\item[\ref{item:EbRcondD}.]%
  The convergence domains are non-intersecting, $D_x \cap D_{x'} =
  \emptyset \; \forall x \ne x'$, and cover the complete integration
  domain, $\bigcup_{x \in R} D_x = D$.
\item[\ref{item:EbRcondcommnc}.]%
  All expansions corresponding to regions within some subset $\Rc \subset
  R$ commute with each other and with expansions of any other region
  in~$R$:
  \begin{displaymath}
    T^{(x)} T^{(x')} = T^{(x')} T^{(x)} \; \forall x \in \Rc \,,\ x' \in R
    \,.
  \end{displaymath}
  We write $\Rc = \{x_1,\ldots,x_{\Nc}\}$ with $0 \le \Nc \le N$ and $\Rnc =
  R \setminus \Rc = \{x_{\Nc+1},\ldots,x_N\}$.
\item[\ref{item:EbRcondreg}.]%
  The original integral~$F$ and all integrals over expanded terms,
  \begin{displaymath}
    F^{(x'_1,x'_2,\ldots)}_{j_1,j_2,\ldots}
    = \int\!\rD k \, T^{(x'_1)}_{j_1} T^{(x'_2)}_{j_2} \cdots I
    \,,
  \end{displaymath}
  are well-defined within the chosen regularization scheme, even if these
  integrals are not restricted to any convergence domain~$D_x$.
\item[\ref{item:EbRcondsums}.]%
  The series expansions
  \begin{displaymath}
    F^{(x'_1,x'_2,\ldots)}
    = \sum_{j_1,j_2,\ldots} F^{(x'_1,x'_2,\ldots)}_{j_1,j_2,\ldots}
    \,,
  \end{displaymath}
  converge absolutely, although the expanded terms have been integrated
  over the complete integration domain~$D$.
\item[\ref{item:EbRcondnc}.]%
  For every combination of two non-commuting expansions, there is a
  region from~$\Rc$ whose expansion does not further change the doubly
  expanded integrand:
  \begin{displaymath}
    \forall x'_1,x'_2 \in \Rnc, \, x'_1 \ne x'_2, \:
    \exists \, x \in \Rc : \,
    T^{(x)} T^{(x'_2)} T^{(x'_1)} = T^{(x'_2)} T^{(x'_1)}
    \,.
  \end{displaymath}
  Note that this condition requires $\Nc \ge 1$.
\end{enumerate}
If all these conditions hold, then the original integral is reproduced
exactly, i.e.\ to all orders in the expansion, through the master
identity~(\ref{eq:formalismncidred}):
\begin{align}
  \label{eq:formalismidconc}
    F &=
      \sum_{x \in R} F^{(x)}
      - \sum^{\langle\Rc+1\rangle}_{\{x'_1,x'_2\} \subset R} F^{(x'_1,x'_2)}
      + \ldots
      - (-1)^n \sum^{\langle\Rc+1\rangle}_{\{x'_1,\ldots,x'_n\} \subset R}
        F^{(x'_1,\ldots,x'_n)}
    \nonumber \\* &\qquad{}
      + \ldots
      + (-1)^{\Nc} \sum_{x' \in \Rnc} F^{(x',x_1,\ldots,x_{\Nc})}
    \,,
\end{align}
where the summations with superscript $\langle\Rc+1\rangle$ are defined
in~(\ref{eq:sumRcpone}). They are restricted to such subsets of regions
with at most one region from~$\Rnc$. This means that the
identity~(\ref{eq:formalismidconc}) sums over exactly those
combinations of regions whose expansions commute with each other.

The first term on the right-hand side of~(\ref{eq:formalismidconc})
corresponds to the known recipe of the expansion by regions presented in
section~\ref{sec:intro}. The following terms represent overlap
contributions originating from multiple expansions. These overlap
contributions may be relevant under certain circumstances, see the example
from section~\ref{sec:forward} evaluated without analytic regularization.
In many cases, however, the usual recipe of the expansion by regions is
recovered, especially when dimensional and analytic regularization are
used. If the contributions~$F^{(x)}$ with single expansions yield
homogeneous functions of the expansion parameter in each order of the
expansion and if this dependence on the expansion parameter is unique for
each region, then we expect all overlap contributions to be scaleless, and
only the usual contributions from each region survive.

This means that in situations where the expansion by regions has normally
been employed the known
recipe~\cite{Beneke:1997zp,Smirnov:1998vk,Smirnov:1999bza,Smirnov:2002pj}
remains correct. The original authors of the method have always stressed
how important the homogeneity of the contributions is. Where
non-homogeneous contributions from individual regions appear, the formalism
presented in this paper may still be applied if its conditions hold, and
overlap contributions might arise. But it is usually preferable to change
the choice of regions or regulators in order to obtain only homogeneous
contributions and get rid of overlap contributions. In any case, one can
always check whether overlap contributions are relevant by simply
evaluating them to see if they vanish.

The original recipe of the expansion by
regions~\cite{Beneke:1997zp,Smirnov:1998vk,Smirnov:1999bza,Smirnov:2002pj}
can be understood such that \emph{any} scaleless integral must be set to
zero, whether it is well-defined through regularization or not. However,
this is not the approach of the formalism presented here. Throughout the
whole paper only such scaleless integrals are dropped which are
mathematically well-defined through some regularization (and analytic
continuation) and which can explicitly be shown to vanish (see
appendix~\ref{app:2pointhscalc} and other appendices). Wherever an integral
is ill-defined, a suitable regularization is added. Vanishing scaleless
integrals thus are not a requirement for this formalism, but simply a
well-understood property of dimensional regularization and analytic
continuation which is used here.

The leading-order asymptotic expansion of the integral~$F$ is obtained by
replacing each series expansion in~(\ref{eq:formalismidconc}) with its
leading-order term. Some of the leading-order terms obtained in this way
may be suppressed with respect to others and must be dropped from the
leading-order result. For the leading-order approximation
condition~\ref{item:EbRcondsums} above does not apply, and
condition~\ref{item:EbRcondreg} only needs to hold for the leading-order
integrals.

The examples presented in this paper illustrate the application of the
formalism to various integrals. This implies choosing the regions with
corresponding expansions, covering the integration domain with the
convergence domains and checking the conditions listed above.
Most example integrals have alternatively been evaluated and expanded with
the help of a Mellin--Barnes representation. For these simple one-loop
examples, the asymptotic expansion via Mellin--Barnes integrals can be
easier than the application of the expansion by regions. My experience with
more elaborate multi-loop integrals depending on several small and large
parameters, however, shows that the asymptotic expansion via multiple
Mellin--Barnes integrals is often more difficult than the expansion at the
integrand level within the strategy of regions.

What is the practical use of the formalism presented in this paper,
apart from the generalization of the expansion by regions to cases where
overlap contributions are needed?
In principle, it should be possible also for more complicated integrals to
prove their asymptotic expansion by checking the conditions listed
above. Most of these conditions can indeed be checked in a straightforward
way once the list of regions, expansions and convergence domains is
established. On the other hand, the first obstacle in getting there is
finding all relevant regions such that the convergence domains of their
expansions cover the complete integration domain. And this step might be
quite tedious for complicated multi-loop integrals. It can be easier to
evaluate successfully the contributions from the regions than to prove,
using this formalism, that the evaluation is correct and complete. In
particular, the formalism may need extra regions with scaleless
contributions in order to cover the integration domain with convergence
domains or to ensure the commutation of the expansions.

Even with the formalism at hand, we will certainly not always want to prove
its applicability, if only for practical limitations. But I hope this paper
will convince readers that the expansion by regions is a well-founded
method of asymptotic expansion and that its applicability can at least in
principle be proven case by case.
The standard recipe of the expansion by regions remains valid if overlap
contributions are absent. So the formalism and its conditions provide hints
on the proper choice of regions and regularization schemes which ensure the
homogeneous and unique dependence of the individual contributions on the
expansion parameter.
Finally, the present paper shows that the expansion by regions, in its
generalized form, can be applied to very different kinds of integrals, even
such with finite boundaries or scaleful regularization parameters where the
contributions do not exhibit a homogeneous dependence on the expansion
parameter. Appendix~\ref{app:finiteboundary} illustrates such an example
and points out possible convergence problems arising there.

Having obtained such knowledge on the foundation and generalization of the
expansion by regions, this method is ready to be employed trustfully to
various kinds of asymptotic expansions.

\acknowledgments
\addcontentsline{toc}{section}{Acknowledgments}

This work is supported by the DFG Sonder\-forschungs\-bereich/Transregio~9
``Computer\-gest\"utzte Theoretische Teilchenphysik''.
Diagrams and figures have been drawn with
\texttt{JaxoDraw}~\cite{Binosi:2008ig} which is based on
\texttt{Axodraw}~\cite{Vermaseren:1994je}.

I thank M.~Beneke, J.~Piclum, J.~Rohrwild, P.~Ruiz-Femen\'ia and
V.A.~Smirnov for stimulating and helpful discussions. I also thank
M.~Beneke, J.~Piclum and V.A.~Smirnov for reading the manuscript and
providing useful comments. Last but not least I thank V.A.~Smirnov for
teaching me asymptotic expansions and many tricks for evaluating loop
integrals, and for encouraging me to pursue the research and publication of
this work.

\appendix

\section{Technical details of the evaluations}
\label{app:details}

\subsection{Large-momentum expansion}

This appendix presents details of the evaluation of the off-shell
large-momentum expansion in section~\ref{sec:2point}.

\subsubsection{Large-momentum expansion: soft contributions}
\label{app:2pointscalc}

The soft-region integral
\begin{align}
  \label{eq:2pointscalcint}
  \int\!\rD k \, \frac{(k^2)^{j_1} \, (2k\cdot p)^{j_2}}{
    (k^2 - m^2 + i0)^{n_2}}
\end{align}
in~(\ref{eq:2pointsint}) can be treated as an analytic function of the
summation indices $j_1$ and $j_2$. We first assume $\Rep j_{1,2} < 0$ and
later analytically continue the result to positive values of $j_{1,2}$.
Using alpha parameters
\begin{align}
  \label{eq:alpha}
  \frac{1}{(A+i0)^n} = \frac{e^{-i\pi n/2}}{\Gamma(n)}
    \int_0^\infty\!\rd\alpha \, \alpha^{n-1} \, e^{i\alpha(A+i0)}
  \,,
\end{align}
the integral~(\ref{eq:2pointscalcint}) is written as
\begin{multline}
  \frac{e^{-i\pi(n_2-j_{12})/2}}{
    \Gamma(n_2) \, \Gamma(-j_1) \, \Gamma(-j_2)}
  \int_0^\infty\!\rd\alpha_1 \, \rd\alpha_2 \, \rd\alpha_3 \,
    \alpha_1^{n_2-1} \, \alpha_2^{-j_1-1} \, \alpha_3^{-j_2-1} \,
    e^{i\alpha_1(-m^2+i0)}
\\* {} \times
  \int\!\rD k \, e^{i(\alpha_{12}k^2 + 2\alpha_3 \, p\cdot k + i0)}
  \,,
\end{multline}
with the shorthand notation~(\ref{eq:shorthandsum}). The loop momentum is
integrated via
\begin{align}
  \label{eq:loopintexp}
  \int\! \frac{\rd^d k}{i\pi^{d/2}} \, e^{i(\alpha k^2 + 2p\cdot k + i0)}
  = e^{-i\pi d/4} \, \alpha^{-d/2} \,
    \exp\!\left[i \left(-\frac{p^2}{\alpha} + i0\right)\right]
  ,
\end{align}
yielding the alpha-parameter representation
\begin{multline}
  \mu^{2\eps} \, e^{\eps\gammaE} \,
  \frac{e^{-i\pi(n_2-j_{12}+2-\eps)/2}}{
    \Gamma(n_2) \, \Gamma(-j_1) \, \Gamma(-j_2)}
  \int_0^\infty\!\rd\alpha_1 \, \alpha_1^{n_2-1} \, e^{i\alpha_1(-m^2+i0)}
  \int_0^\infty\!\rd\alpha_2 \, \alpha_2^{-j_1-1} \, \alpha_{12}^{-2+\eps}
\\* {} \times
  \int_0^\infty\!\rd\alpha_3 \, \alpha_3^{-j_2-1} \,
    \exp\!\left[i\alpha_3^2 \left(-\frac{p^2}{\alpha_{12}} +
      i0\right) \right]
\end{multline}
of the soft-region integral. After substituting $\alpha_3 = t^{1/2}$ and
evaluating the $t$-integral by reversing~(\ref{eq:alpha}), the
$\alpha_2$-integral is solved using
\begin{align}
  \label{eq:alphaintBeta}
  \int_0^\infty\!\rd\alpha \, \alpha^{\nu-1} \, (A + B\alpha)^{-\rho}
  = \frac{\Gamma(\nu) \, \Gamma(\rho-\nu)}{\Gamma(\rho)} \,
    B^{-\nu} \, A^{\nu-\rho}
  \,,
\end{align}
followed by the $\alpha_1$-integral via reversing~(\ref{eq:alpha}). The
result reads
\begin{multline}
   \mu^{2\eps} \, e^{\eps\gammaE} \, e^{-i\pi(n_2-j_1-j_2/2)} \,
   (m^2)^{2-n_2-\eps+j_1+j_2/2} \, (-p^2+i0)^{j_2/2} \,
\\* {} \times
   \frac{\Gamma\!\left(-\frac{j_2}{2}\right)
       \Gamma\!\left(2-\eps+j_1+\frac{j_2}{2}\right)
       \Gamma\!\left(n_2-2+\eps-j_1-\frac{j_2}{2}\right)}{
     2 \, \Gamma(n_2) \, \Gamma(-j_2) \,
       \Gamma\!\left(2-\eps+\frac{j_2}{2}\right)}
  \,.
\end{multline}
With the help of the doubling formula for the Gamma
function (see e.g.~\cite{Gradshteyn:2007}) we rewrite
\begin{equation}
  \label{eq:Gammadoubling}
  \frac{\Gamma\!\left(-\frac{j_2}{2}\right)}{\Gamma(-j_2)}
  = \frac{2^{1+j_2} \, \sqrt{\pi}}{\Gamma\!\left(\frac{1-j_2}{2}\right)}
  \,.
\end{equation}
Now we can safely perform the analytic continuation to $j_{1,2} \ge 0$. The
factor $\Gamma\!\left(\frac{1-j_2}{2}\right)$ in the denominator makes the
result vanish for odd~$j_2$, as it should due to the Lorentz structure of
tadpole tensor integrals. For even $j_2$ we can write
\begin{equation}
  \label{eq:Gammaeven}
  \frac{1}{\Gamma\!\left(\frac{1-j_2}{2}\right)}
  = \frac{(-1)^{j_2/2}}{\Gamma\!\left(\frac{1}{2}\right)}
    \left(\tfrac{1}{2}\right)_{j_2/2}
  = \frac{(-1)^{j_2/2}}{\sqrt{\pi}} \,
    \frac{j_2!}{2^{j_2} \left(\frac{j_2}{2}\right)!}
  \,,
\end{equation}
where the relation
\begin{equation}
  \Gamma(\alpha-j)
  = (-1)^j \, \frac{\Gamma(\alpha) \, \Gamma(1-\alpha)}{\Gamma(1-\alpha+j)}
  = (-1)^j \, \frac{\Gamma(\alpha)}{(1-\alpha)_j}
\end{equation}
for integer~$j$ has been used.
Inserting the result into~(\ref{eq:2pointsint}) we obtain, for even~$j_2$,
\begin{multline}
  F^{(s)}_{j_1,j_2} = \mu^{2\eps} \, e^{\eps\gammaE} \,
    e^{-i\pi n_{12}} \, (m^2)^{2-n_2-\eps} \, (-p^2-i0)^{-n_1}
    \left(\frac{m^2}{p^2}\right)^{j_1+j_2/2} \,
\\* {} \times
    \frac{\Gamma\!\left(2-\eps+j_1+\frac{j_2}{2}\right)
         \Gamma\!\left(n_2-2+\eps-j_1-\frac{j_2}{2}\right)}{
       \Gamma(n_1) \, \Gamma(n_2)} \,
    \frac{(-1)^{j_2/2} \, \Gamma(n_1+j_{12})}{
      j_1! \left(\frac{j_2}{2}\right)! \,
        \Gamma\!\left(2-\eps+\frac{j_2}{2}\right)}
\end{multline}
and $F^{(s)}_{j_1,j_2} = 0$ for odd~$j_2$. Rewriting the sum over $j_1,j_2$
as a sum over $j = j_1 + \frac{j_2}{2}$ and $j_2' = \frac{j_2}{2}$
according to~(\ref{eq:2pointsj}), we calculate
\begin{align*}
  \lefteqn{\sum_{j_2'=0}^j \frac{(-1)^{j_2'} \, \Gamma(n_1+j+j_2')}{
      j_2'! \, (j-j_2')! \, \Gamma(2-\eps+j_2')}}
  \qquad
\\* &
  = \frac{1}{j! \, \Gamma(2-n_1-\eps-j)}
    \int_0^1\!\rd x \, x^{n_1+j-1} \, (1-x)^{1-n_1-\eps-j}
    \underbrace{\sum_{j_2'=0}^j \binom{j}{j_2'} (-x)^{j_2'}}_{(1-x)^j}
\\* &
  = \frac{\Gamma(n_1+j) \, \Gamma(2-n_1-\eps)}{
      j! \, \Gamma(2-n_1-\eps-j) \, \Gamma(2-\eps+j)}
\end{align*}
and obtain the result~(\ref{eq:2pointsres}) for $F^{(s)}_j$.

\subsubsection{Large-momentum expansion: overlap contributions}
\label{app:2pointhscalc}

This appendix is dedicated to the extraction of ultraviolet and infrared
$1/\eps$ poles from the scaleless overlap
contributions~(\ref{eq:2pointoverlap}) in the particular case $n_1=1$,
$n_2=2$ where the complete integral~$F$ is finite. We have to evaluate
\begin{align}
  \label{eq:2pointhscalc}
  F^{(h,s)} = \sum_{i,j_1,j_2=0}^\infty  F^{(h,s)}_{i,j_1,j_2}
  = \sum_{i,j_1,j_2=0}^\infty (1+i) \, \frac{j_{12}!}{j_1! \, j_2!} \,
    \frac{(m^2)^i \, (-1)^{j_{12}}}{(p^2)^{1+j_{12}}}
    \int\!\rD k \, \frac{(2k\cdot p)^{j_2}}{(k^2)^{2+i-j_1}}
  \,,
\end{align}
which only consists of vanishing scaleless integrals, but involves
contributions of the form $(1/\eps_\UV - 1/\eps_\IR)$ which we want to
extract. The loop integral can be solved in analogy to the previous
appendix~\ref{app:2pointscalc} by assuming $\Rep j_2 < 0$ before
analytically continuing the result to $j_2 \ge 0$:
\begin{align}
\label{eq:2pointhscalcint}
  \lefteqn{
    \int\!\rD k \, \frac{(2k\cdot p)^{j_2}}{(k^2+i0)^n}
    = \frac{e^{-i\pi(n-j_2)/2}}{\Gamma(n) \, \Gamma(-j_2)}
    \int_0^\infty\!\rd\alpha_1 \, \rd\alpha_2 \,
    \alpha_1^{n-1} \, \alpha_2^{-j_2-1} \,
    \int\!\rD k \, e^{i(\alpha_1 k^2 + 2\alpha_2 \, p\cdot k + i0)}
  } \quad
\nonumber \\*
  &= \mu^{2\eps} \, e^{\eps\gammaE} \,
    \frac{e^{-i\pi(n-j_2+2-\eps)/2}}{\Gamma(n) \, \Gamma(-j_2)}
    \int_0^\infty\!\rd\alpha_1 \, \alpha_1^{n-3+\eps}
    \int_0^\infty\!\rd\alpha_2 \, \alpha_2^{-j_2-1} \,
    \exp\!\left[i\alpha_2^2 \left(-\frac{p^2}{\alpha_1} + i0\right) \right]
\nonumber \\*
  &= \mu^{2\eps} \, e^{\eps\gammaE} \,
    e^{-i\pi(n+2-j_2/2-\eps)/2} \, (-p^2+i0)^{j_2/2} \,
    \frac{\Gamma\!\left(-\frac{j_2}{2}\right)}{2 \, \Gamma(n) \, \Gamma(-j_2)}
    \int_0^\infty\!\rd\alpha_1 \, \alpha_1^{n-3-j_2/2+\eps}
  \,.
\end{align}
The prefactor is rewritten with the help of (\ref{eq:Gammadoubling}) and
(\ref{eq:Gammaeven}), vanishing for odd values of~$j_2$, as it should. The
$\alpha_1$-integral is again scaleless. For extracting the singularities at
$\alpha_1 \to 0$ and $\alpha_1 \to \infty$ we have to separate the
integration at some intermediate scale~$\lambda$. Consider the following
integral for integer~$j$:
\begin{align}
  \int_0^\infty\!\rd\alpha \, \alpha^{j-1+\eps}
  = \int_0^\lambda\!\rd\alpha \, \alpha^{j-1+\eps}
    + \int_\lambda^\infty\!\rd\alpha \, \alpha^{j-1+\eps}
  = \underbrace{\frac{\lambda^{j+\eps}}{j+\eps}}_{\Rep\eps > -j}
    - \underbrace{\frac{\lambda^{j+\eps}}{j+\eps}}_{\Rep\eps < -j}
  = 0
  \,.
\end{align}
The first term is ultraviolet-divergent for $\Rep\eps \le -j$ while the
second term is infrared-divergent for $\Rep\eps \ge -j$. The two terms are
combined using analytic continuation, and the integral vanishes. We are
interested in the separate $1/\eps$ singularities, which only occur for
$j=0$:
\begin{align}
  \label{eq:alphascaleless}
  \int_0^\infty\!\rd\alpha \, \alpha^{j-1+\eps}
  = \delta_{j,0} \left( \frac{1}{\eps_\UV} - \frac{1}{\eps_\IR} \right)
  .
\end{align}
Plugging~(\ref{eq:alphascaleless}) into~(\ref{eq:2pointhscalcint}) we
obtain
\begin{align}
  \label{eq:2pointhscalcintres}
  \int\!\rD k \, \frac{(2k\cdot p)^{j_2}}{(k^2)^n}
  = \delta_{n-2,\, j_2/2}
    \left( \frac{1}{\eps_\UV} - \frac{1}{\eps_\IR} \right)
    (p^2)^{j_2/2} \, \frac{j_2!}{\Gamma(n) \left(\frac{j_2}{2}\right)!}
\end{align}
for even~$j_2$ and zero otherwise. Inserting this
into~(\ref{eq:2pointhscalc}) with $n = 2+i-j_1$, eliminating~$j_1$ with the
Kronecker~delta in~(\ref{eq:2pointhscalcintres}) and replacing $j_2=2j_2'$
yields
\begin{align}
  \label{eq:2pointhscalcsum}
  F^{(h,s)} = \frac{1}{p^2}
    \left( \frac{1}{\eps_\UV} - \frac{1}{\eps_\IR} \right)
    \sum_{i=0}^\infty \left(\frac{m^2}{p^2}\right)^i \, (1+i) \, (-1)^i \,
    \sum_{j_2'=0}^i
    \frac{(-1)^{j_2'} \, (i+j_2')!}{j_2'! \, (i-j_2')! \, (1+j_2')!}
  \,.
\end{align}
The sum over~$j_2'$ can be evaluated in the following way:
\begin{align*}
  \lefteqn{
    \sum_{j_2'=0}^i
    \frac{(-1)^{j_2'} \, (i+j_2')!}{j_2'! \, (i-j_2')! \, (1+j_2')!}
    = \lim_{\delta\to 0} \, \frac{1}{i!} \sum_{j_2'=0}^i \binom{i}{j_2'} \,
    (-1)^{j_2'} \, \frac{\Gamma(1+i+j_2')}{\Gamma(2+j_2'+\delta)}
    } \qquad
\nonumber \\*
  &= \lim_{\delta\to 0} \, \frac{1}{i! \, \Gamma(1-i+\delta)}
    \int_0^1\!\rd x \, x^i \, (1-x)^{\delta-i} \,
    \underbrace{\sum_{j_2'=0}^i \binom{i}{j_2'} \, (-x)^{j_2'}}_{(1-x)^i}
\nonumber \\*
  &= \lim_{\delta\to 0} \,
    \frac{\Gamma(1+\delta)}{\Gamma(1-i+\delta) \, \Gamma(2+i+\delta)}
  = \frac{1}{\Gamma(1-i) \, (1+i)!}
  = \delta_{i,0}
  \,.
\end{align*}
For fixed $i \ge 1$ all terms in~(\ref{eq:2pointhscalcsum}) cancel each
other in the sum over~$j_2'$, resulting in $1/\Gamma(1-i) = 0$. The only
remaining contribution is from $i = 0$, and we obtain
\begin{align}
  F^{(h,s)} = \frac{1}{p^2}
    \left( \frac{1}{\eps_\UV} - \frac{1}{\eps_\IR} \right)
  .
\end{align}
Combining the complete result $F = F^{(h)} + F^{(s)} -
F^{(h,s)}$~(\ref{eq:2pointidentity}), the poles in $F^{(h,s)}$ cancel the
corresponding ones in~(\ref{eq:2point12hsreseps}): $1/\eps_\UV$ cancels the
ultraviolet pole in $F^{(s)}_0$ and $1/\eps_\IR$ cancels the infrared pole in
$F^{(h)}_0$.

\subsection{Threshold expansion}

Details of the threshold expansion in section~\ref{sec:threshold} are dealt
with in this appendix.

\subsubsection{Threshold expansion: check of convergence}
\label{app:thresholdconv}

Let us check that the
expansions~(\ref{eq:thresholdTh})--(\ref{eq:thresholdTp}) of the threshold
expansion in section~\ref{sec:threshold} really converge within their
corresponding domains:
\begin{itemize}
\item In the hard domain~$D_h = \bigl\{ k \in D : \,%
  |\vec k| \gg |\vec p\,| \,\vee\, |k_0| \gg |\vec p\,| \bigr\}$, we either
  have $|\vec k| \gg |\vec p\,|$, then $|\vec p\cdot\vec k| \le |\vec p\,|
  \, |\vec k| \ll \vec k^2$. If this is not the case, then there is $|\vec
  k| \lesssim |\vec p\,| \ll |k_0|$, which implies $|\vec p\cdot\vec k|
  \lesssim |\vec p\,|^2 \ll |q_0 k_0|$. So $|\vec p\cdot\vec k|$ is small
  compared to either $\vec k^2$ or $|q_0 k_0|$ in~(\ref{eq:thresholdTh}),
  and these two terms are present in the denominators even if $T^{(h)}$ is
  applied to $T^{(p)} I_{1,2}$ where $k_0^2$ has been eliminated from there
  before, cf.\ $T^{(h,p)} I_{1,2}$ in~(\ref{eq:thresholdTmult}).

  It remains to be shown that $|\vec p\cdot\vec k|$ is also small compared
  to the combination of terms $|k_0^2 - \vec k^2 \pm q_0 k_0|$, which could
  be spoiled by a cancellation among the individual terms. Keeping track of
  the infinitesimal imaginary part in the propagators expanded
  with~$T^{(h)}$, we look for zeros of the denominators $(k_0^2 - \vec k^2
  \pm q_0 k_0 + i0)$ and $(k_0^2 - \vec k^2 + i0)$ in the complex
  $k_0$-plane where the integration is performed along the real axis. These
  zeros exclusively lie in the upper left and lower right quadrants of the
  complex plane. For $|\vec k| \gg |\vec p\,|$ zeros on different sides of
  the integration contour are well separated from each other. So we can
  bypass all zeros by bending the contour of the \mbox{$k_0$-integration}
  away from the real axis near the zeros. For $|\vec k| \lesssim |\vec
  p\,|$ the zeros may pinch the contour near $k_0=0$. But then $D_h$
  requires $|k_0| \gg |\vec p\,|$, where only two zeros, far apart from
  each other, are found. The same argument also works for the case when
  $T^{(h)}$ is applied to $T^{(p)} I_{1,2}$. Therefore we can always choose
  the integration contour within~$D_h$ such that $|\vec p\cdot\vec k| \ll
  |k_0^2 - \vec k^2 \pm q_0 k_0|$ in~(\ref{eq:thresholdTh}) and $|\vec
  p\cdot\vec k| \ll |{-\vec k^2} \pm q_0 k_0|$ for~$T^{(h,p)} I_{1,2}$
  in~(\ref{eq:thresholdTmult}).

\item In the soft domain~$D_s = \bigl\{ k \in D : \,%
  |\vec k| \lesssim |k_0| \lesssim |\vec p\,| \bigr\}$, we have
  $\vec k^2 \lesssim k_0^2 \lesssim |\vec p\,| \, |k_0| \ll |q_0 k_0|$ and
  $|\vec p\cdot\vec k| \le |\vec p\,| \, |\vec k| \ll |q_0 k_0|$,
  so each of the factors in the numerator of~(\ref{eq:thresholdTs}) is
  small compared to the denominator.

\item In the potential domain~$D_p = \bigl\{ k \in D : \,%
  |k_0| \ll |\vec k| \lesssim |\vec p\,| \bigr\}$, there is $k_0^2 \ll \vec
  k^2$. For the expansion $T^{(p)} I_{1,2}$ in~(\ref{eq:thresholdTp}), we
  additionally need $k_0^2 \ll |{-\vec k^2} \pm q_0 k_0 - 2\vec p\cdot\vec
  k|$. Again the two zeros of $(-\vec k^2 \pm q_0 k_0 - 2\vec p\cdot\vec k
  + i0)$ exclusively lie in the upper left and lower right quadrants of the
  complex $k_0$-plane. The $k_0$-contour is pinched at $k_0 = 0$ in the
  limit $|\vec k| \to 0$, but as $D_p$ requires $|k_0| \ll |\vec k|$, we
  can discard this case when checking the convergence
  of~$T^{(p)}$. Otherwise the two zeros only pinch the $k_0$-contour when
  $\vec k^2 + 2\vec p\cdot\vec k \to 0$, which, for finite~$|\vec k|$,
  requires a certain angular correlation between $\vec k$ and~$\vec p$. So
  the pinching can be avoided if the contour of the angular \mbox{$\vec
  k$-integration} is bent into its complex plane in order to bypass the
  zeros of $\vec k^2 + 2\vec p\cdot\vec k$. The unpinched $k_0$-contour may
  bypass the zeros in the $k_0$-plane by bending away from the real axis.
  If $T^{(p)}$ is applied to $T^{(h)} I_{1,2}$, then the $k_0$-contour is
  only pinched in the limit $|\vec k| \to 0$. So the integration contours
  within~$D_p$ can always be chosen such that $|k_0^2| \ll |{-\vec k^2} \pm
  q_0 k_0 - 2\vec p\cdot\vec k|$ in~(\ref{eq:thresholdTp}) and $|k_0^2| \ll
  |{-\vec k^2} \pm q_0 k_0|$ for~$T^{(h,p)} I_{1,2}$
  in~(\ref{eq:thresholdTmult}).
\end{itemize}

\subsubsection{Threshold expansion: hard contributions}
\label{app:thresholdhcalc}

In the hard-region integrals of $F^{(h)}$~(\ref{eq:thresholdints}), the
$k_0$-integration converges for positive $n_1,n_2,n_3$. We do not need the
analytic regularization here and set $n_1=n_2=n_3=1$ from the start:
\begin{align}
  F^{(h)} = \sum_{j_1,j_2=0}^\infty
    \int\!
    \frac{\rD k \, (2\vec p\cdot\vec k)^{j_{12}}}{
      (k_0^2 - \vec k^2 + q_0 k_0 + i0)^{1+j_1} \,
      (k_0^2 - \vec k^2 - q_0 k_0 + i0)^{1+j_2} \, (k_0^2 - \vec k^2 + i0)}
  \,.
\end{align}
The first two propagators are combined with Feynman parameters, adding the
third propagator afterwards:
\begin{multline}
  F^{(h)} = -\frac{\mu^{2\eps} \, e^{\eps\gammaE}}{i \pi^{d/2}}
    \sum_{j_1,j_2=0}^\infty
    \frac{(2+j_{12})!}{j_1! \, j_2!} \, (-1)^{j_{12}}
    \int_0^1\!\rd x \, \rd y \, x^{j_1} \, (1-x)^{j_2} \, y^{1+j_{12}}
\\* {}\times
    \int_{-\infty}^\infty\!\rd k_0
    \int\!\rd^{d-1}\vec k \,
    \frac{(2\vec p\cdot\vec k)^{j_{12}}}{
      \bigl( \vec k^2 - k_0^2 - (2x-1) y q_0 k_0 - i0 \bigr)^{3+j_{12}}}
  \,.
\end{multline}
The $\vec k$-integral is factorized into a radial $|\vec k|$-integration
and an angular integration. For the latter we use
\begin{align}
  \int\!\rd\Omega_{d-1} =
    \int_{-1}^1\!\rd\cos\theta \, (1-\cos^2\theta)^{(d-4)/2}
    \int\!\rd\Omega_{d-2}
    \,,\qquad
  \int\!\rd\Omega_{d-2} =
    \frac{2 \pi^{(d-2)/2}}{\Gamma\!\left(\frac{d-2}{2}\right)}
    \,,
\end{align}
to distinguish the direction of the external momentum~$\vec p$, writing
$\vec p\cdot\vec k = |\vec p\,| \, |\vec k| \cos\theta$. The angular
integral vanishes for odd~$j_{12}$, and for even~$j_{12}$ the $\vec
k$-integral yields
\begin{multline}
  \label{eq:thresholdhcalckvec}
  (2 |\vec p\,|)^{j_{12}} \, \frac{\pi^{1-\eps}}{\Gamma(1-\eps)}
    \int_0^1\!\rd\cos^2\theta \, (\cos^2\theta)^{(j_{12}-1)/2} \,
      (1-\cos^2\theta)^{-\eps}
\\* \shoveright{ {}\times
    \int_0^\infty\!\rd\vec k^2 \,
      \frac{(\vec k^2)^{(1+j_{12})/2-\eps}}{
        \bigl( \vec k^2 - k_0^2 - (2x-1) y q_0 k_0 - i0 \bigr)^{3+j_{12}}}
  }
\\
  = \pi^{1-\eps} \, (4\vec p\,^2)^{j_{12}/2} \,
    \frac{\Gamma\!\left(\frac{1+j_{12}}{2}\right)
        \Gamma\!\left(\frac{3+j_{12}}{2}+\eps\right)}{
      (2+j_{12})!} \,
    \frac{1}{\bigl( -k_0^2 - (2x-1) y q_0 k_0 - i0 \bigr)^{(3+j_{12})/2+\eps}}
  \,.
\end{multline}
The $k_0$-integral over the last factor in~(\ref{eq:thresholdhcalckvec}) is
performed with a shift $k_0 \to k_0' = k_0 + \left(x-\frac{1}{2}\right) y
q_0$:
\begin{multline}
  \int_0^\infty\!\rd{k_0'}^2 \,
    \frac{({k_0'}^2)^{-1/2}}{
      \Bigl( (-1-i0) \, {k_0'}^2 + \left(x-\frac{1}{2}\right)^2 y^2 q_0^2
      \Bigr)^{(3+j_{12})/2+\eps}}
\\*
  = \frac{\sqrt{\pi} \, \Gamma\!\left(1+\eps+\frac{j_{12}}{2}\right)}{
      \Gamma\!\left(\frac{3+j_{12}}{2}+\eps\right)} \,
    \underbrace{(-1-i0)^{-1/2}}_{= i} \,
    \Bigl( \left(x-\tfrac{1}{2}\right)^2 y^2 q_0^2 \Bigr)^{-1-\eps-j_{12}/2}
  \,.
\end{multline}
By introducing $j = j_{12}/2$, the summation over $j_1,j_2$ with
even~$j_{12}$ is rewritten as
\begin{multline}
  F^{(h)} =
    -\frac{\mu^{2\eps} \, e^{\eps\gammaE}}{
        \sqrt{\pi} \, (q_0^2)^{1+\eps}}
      \int_0^1\! \frac{\rd y}{y^{1+2\eps}}
      \sum_{j=0}^\infty
      \left(\frac{4\vec p\,^2}{q_0^2}\right)^j \,
      \frac{\Gamma\!\left(\frac{1}{2}+j\right) \Gamma(1+\eps+j)}{(2j)!}
\\* {}\times
      \int_0^1\! \frac{\rd x}{\left|x-\frac{1}{2}\right|^{2+2\eps+2j}}
      \underbrace{\sum_{j_1=0}^{2j} \binom{2j}{j_1} \,
          x^{j_1} \, (1-x)^{2j-j_1}}_{
        = [x+(1-x)]^{2j} = 1}
  \,.
\end{multline}
The $x$- and $y$-integrations are easily solved. By substituting
\begin{align}
  (2j)! = \frac{2^{2j}}{\sqrt{\pi}} \,
    \Gamma\!\left(\tfrac{1}{2}+j\right) j!
\end{align}
the result~$F^{(h)}$ in~(\ref{eq:thresholdres}) is obtained.

\subsubsection{Threshold expansion: potential contributions}
\label{app:thresholdpcalc}

The potential-region integrals of $F^{(p)}$~(\ref{eq:thresholdints})
diverge at $k_0 \to \pm\infty$ if $j_{123} > 0$. For evaluating the
potential contribution to all orders in $j_1,j_2,j_3$ we therefore need to
keep general complex propagator powers $n_1$ and $n_2$ as analytic
regulators.
The first two propagators of $F^{(p)}$~(\ref{eq:thresholdints}) are
combined with Feynman parameters:
\begin{multline}
  F^{(p)} = \frac{\mu^{2\eps} \, e^{\eps\gammaE}}{i \pi^{d/2}} \,
    \frac{e^{-i\pi n_{123}}}{\Gamma(n_1) \, \Gamma(n_2)} \,
    \sum_{j_1,j_2,j_3=0}^\infty
    \frac{\Gamma(n_{12}+j_{12}) \, (n_3)_{j_3}}{j_1! \, j_2! \, j_3!}
    \int_0^1\!\rd x \, x^{n_1+j_1-1} \, (1-x)^{n_2+j_2-1}
\\* {} \times
    \int\! \frac{\rd^{d-1}\vec k}{(\vec k^2)^{n_3+j_3}}
    \int_{-\infty}^\infty\!\rd k_0 \,
    \frac{(k_0^2)^{j_{123}}}{
      \bigl( \vec k^2 + 2\vec p\cdot\vec k + (1-2x) q_0 k_0 - i0 \bigr)
      ^{n_{12}+j_{12}}}
  \,.
\end{multline}
Partitioning the $k_0$-integral into the two intervals $(-\infty,0)$ and
$(0,\infty)$, we obtain
\begin{multline}
  \label{eq:thresholdpcalck0}
  \int_0^\infty\!\rd k_0 \,
    \frac{k_0^{2j_{123}}}{
      \Bigl( (\vec k^2 + 2\vec p\cdot\vec k - i0) + (1-2x-i0) q_0 k_0 \Bigr)
      ^{n_{12}+j_{12}}}
\\* \shoveright{
  + \int_0^\infty\!\rd(-k_0) \,
    \frac{(-k_0)^{2j_{123}}}{
      \Bigl( (\vec k^2 + 2\vec p\cdot\vec k - i0) + (2x-1-i0) q_0 (-k_0)
      \Bigr)^{n_{12}+j_{12}}}
  }
\\ \shoveleft{
  = \frac{(2j_{123})! \, \Gamma(n_{12}-1-j_{12}-2j_3)}{
      \Gamma(n_{12}+j_{12})} \,
    \frac{1}{(\vec k^2 + 2\vec p\cdot\vec k - i0)^{n_{12}-1-j_{12}-2j_3}} \,
    \frac{1}{q_0^{1+2j_{123}}}
  }
\\* {} \times
    \left(
      \frac{1}{(1-2x-i0)^{1+2j_{123}}}
      - \frac{1}{(1-2x+i0)^{1+2j_{123}}}
    \right)
  .
\end{multline}
Under the $x$-integral, this term extracts the residue at $x=\tfrac{1}{2}$:
\begin{align}
  &\int_0^1\!\rd x \, x^{n_1+j_1-1} \, (1-x)^{n_2+j_2-1}
    \left(
      \frac{1}{(1-2x-i0)^{1+2j_{123}}}
      - \frac{1}{(1-2x+i0)^{1+2j_{123}}}
    \right)
\nonumber \\* & \qquad
  = -2i\pi \, \Res\left.
    \frac{x^{n_1+j_1-1} \, (1-x)^{n_2+j_2-1}}{(1-2x)^{1+2j_{123}}}
    \right|_{x=\frac{1}{2}}
\nonumber \\ & \qquad
  = \frac{i\pi}{4^{\,j_{123}} \, (2j_{123})!}
    \left(\frac{\partial}{\partial t}\right)^{2j_{123}}
    \left(\tfrac{1}{2}+t\right)^{n_1+j_1-1}
    \left(\tfrac{1}{2}-t\right)^{n_2+j_2-1}
    \Bigr|_{t=0}
 \,,
\end{align}
with $t = x-\frac{1}{2}$.
The $\vec k$-integration
\begin{align}
  \int\! \frac{\rd^{d-1}\vec k}{
    (\vec k^2 + 2\vec p\cdot\vec k - i0)^{n_{12}-1-j_{12}-2j_3} \,
    (\vec k^2)^{n_3+j_3}}
\end{align}
is obviously scaleless if $n_1=n_2=1$ and any of the $j_1,j_2,j_3$ is
larger than zero (cf.~\cite{Beneke:1997zp}). But exactly in this case the
Gamma function in the numerator of~(\ref{eq:thresholdpcalck0}) would be
ill-defined, so we still need analytic regulators and continue the
calculation with general $n_1,n_2,n_3$.
Combining the two denominators with Feynman parameters, the $\vec
k$-integral yields
\begin{multline}
  \label{eq:thresholdpcalckvec}
  \frac{\Gamma(n_{123}-1-j_{123})}{\Gamma(n_{12}-1-j_{12}-2j_3) \,
      \Gamma(n_3+j_3)}
    \int_0^1\!\rd y \, y^{n_{12}-2-j_{12}-2j_3} \, (1-y)^{n_3+j_3-1}
\\* \shoveright{ {}\times
    \int\! \frac{\rd^{d-1}\vec k}{
    \bigl( (\vec k + y\vec p\,)^2 - y^2 \vec p\,^2 - i0
    \bigr)^{n_{123}-1-j_{123}}}
  }
\\ \shoveleft{
  = \frac{\pi^{(d-1)/2} \,
        \Gamma\!\left(n_{123}-\frac{5}{2}+\eps-j_{123}\right)}{
      \Gamma(n_{12}-1-j_{12}-2j_3) \, \Gamma(n_3+j_3)} \,
    (-\vec p\,^2 - i0)^{\frac{5}{2}-n_{123}-\eps+j_{123}}
  }
\\* \shoveright{ {}\times
    \int_0^1\!\rd y \, y^{3-n_{12}-2n_3-2\eps+j_{12}} \, (1-y)^{n_3+j_3-1}
  }
\\ \shoveleft{
  = \frac{\pi^{(d-1)/2} \,
        \Gamma\!\left(n_{123}-\frac{5}{2}+\eps-j_{123}\right)
        \Gamma(4-n_{12}-2n_3-2\eps+j_{12})}{
      \Gamma(n_{12}-1-j_{12}-2j_3) \,
        \Gamma(4-n_{123}-2\eps+j_{123})}
  }
\\* {}\times
    (p^2 - i0)^{\frac{5}{2}-n_{123}-\eps+j_{123}}
  \,.
\end{multline}
This cancels the singularity from~(\ref{eq:thresholdpcalck0}) when
$(n_{12}-1-j_{12}-2j_3)$ is zero or a negative integer. The complete
contribution
\begin{align}
  F^{(p)} &= \frac{\mu^{2\eps} \, e^{\eps\gammaE}}{
        \sqrt{q^2} \, (p^2-i0)^{n_{123}-5/2+\eps}} \, 
    \frac{e^{-i\pi n_{123}} \, \sqrt{\pi}}{\Gamma(n_1) \, \Gamma(n_2)}
\nonumber \\* &{}\times
    \sum_{j_1,j_2,j_3=0}^\infty
    \left(\frac{p^2}{4q^2}\right)^{j_{123}} \,
    \frac{(n_3)_{j_3} \,
        \Gamma\!\left(n_{123}-\frac{5}{2}+\eps-j_{123}\right)
        \Gamma(4-n_{12}-2n_3-2\eps+j_{12})}{
      j_1! \, j_2! \, j_3! \, \Gamma(4-n_{123}-2\eps+j_{123})}
\nonumber \\* &{}\times
    \left(\frac{\partial}{\partial t}\right)^{2j_{123}}
      \left(\tfrac{1}{2}+t\right)^{n_1+j_1-1}
      \left(\tfrac{1}{2}-t\right)^{n_2+j_2-1}
      \Bigr|_{t=0}
\end{align}
is regularized dimensionally. We can switch off the analytic regularization
at this point and set $n_1=n_2=n_3=1$:
\begin{multline}
  \label{eq:thresholdpcalcres0}
  F^{(p)} = -\frac{e^{\eps\gammaE} \, \sqrt{\pi}}{
        \sqrt{q^2 \, (p^2-i0)}}
      \left(\frac{\mu^2}{p^2-i0}\right)^\eps
    \sum_{j_1,j_2,j_3=0}^\infty
    \left(\frac{p^2}{4q^2}\right)^{j_{123}} \,
    \frac{\Gamma\!\left(\frac{1}{2}+\eps-j_{123}\right)
        \Gamma(-2\eps+j_{12})}{
      j_1! \, j_2! \, \Gamma(1-2\eps+j_{123})}
\\* {}\times
    \left(\frac{\partial}{\partial t}\right)^{2j_{123}}
      \left(\tfrac{1}{2}+t\right)^{j_1}
      \left(\tfrac{1}{2}-t\right)^{j_2}
      \Bigr|_{t=0}
  \,.
\end{multline}
Now it is obvious that the potential region gets no contributions from
$j_{123} > 0$: The maximal power of the polynomial in~$t$ in the last line
of~(\ref{eq:thresholdpcalcres0}) is $(j_1+j_2)$, so this polynomial
vanishes upon the $2(j_1+j_2+j_3)$ derivatives with respect to~$t$. The
only non-vanishing contribution originates from $j_1=j_2=j_3=0$ and is
reported in~(\ref{eq:thresholdres}).

The leading-order potential contribution (from $j_1=j_2=j_3=0$) is
well-defined without analytic regularization. So for checking the result we
can set $n_1=n_2=n_3=1$ from the start. We solve the $k_0$-integration by
closing its contour in the imaginary infinity and picking up one of the two
poles:
\begin{align}
  F^{(p)}_0 &= \frac{\mu^{2\eps} \, e^{\eps\gammaE}}{i \pi^{d/2}}
    \int\! \frac{\rd^{d-1}\vec k}{\vec k^2}
    \int_{-\infty}^\infty\!
    \frac{\rd k_0}{
      (q_0 k_0 - \vec k^2 - 2\vec p\cdot\vec k + i0) \,
      (q_0 k_0 + \vec k^2 + 2\vec p\cdot\vec k - i0)}
\nonumber \\
  &= -\frac{1}{q_0} \,
    \frac{\mu^{2\eps} \, e^{\eps\gammaE}}{\pi^{d/2-1}}
    \int\! \frac{\rd^{d-1}\vec k}{
        (\vec k^2 + 2\vec p\cdot\vec k - i0) \, \vec k^2}
\nonumber \\
  &= -\frac{1}{\sqrt{q^2 \, (p^2-i0)}}
    \left(\frac{\mu^2}{p^2-i0}\right)^\eps \,
    \frac{e^{\eps\gammaE} \, \sqrt{\pi} \, \Gamma(\frac{1}{2}+\eps) \,
      \Gamma(-2\eps)}{\Gamma(1-2\eps)}
  \,,
\end{align}
using~(\ref{eq:thresholdpcalckvec}) with $n_1=n_2=n_3=1$ and
$j_1=j_2=j_3=0$. This agrees with the result obtained
from~(\ref{eq:thresholdpcalcres0}).

\subsection{Sudakov form factor}
\label{app:Sudakovcalc}

This appendix sketches the evaluation of the contributions to the vertex
correction in the Sudakov limit from section~\ref{sec:Sudakov}.

\subsubsection{Sudakov form factor: scaleless contributions}
\label{app:Sudakovcalcscaleless}

First we have a look at the contributions from the Glauber and
collinear-plane regions and at the overlap contributions which all turn out
to be scaleless within dimensional and analytic regularization.

\paragraph{Glauber contribution.}
According to the expansion $T^{(g)}$~(\ref{eq:SudakovTg}) the Glauber
contribution reads
\begin{multline}
  F^{(g)} = \frac{\mu^{2\eps} \, e^{\eps\gammaE}}{2i \pi^{d/2}}
    \sum_{j_1,j_2,j_3=0}^\infty
    \frac{(n_1)_{j_1} \, (n_2)_{j_2} \, (n_3)_{j_3}}{j_1! \, j_2! \, j_3!}
    \, (-1)^{j_{123}}
    \int\! \frac{\rd^{d-2}\vec k_\perp}{
        (-\vec k_\perp^2 - m^2 + i0)^{n_3+j_3}}
\\* {}\times
    \int_{-\infty}^\infty\! \frac{\rd k^+ \, (k^+)^{j_{123}}}{
        (-\vec k_\perp^2 + Q k^+ + i0)^{n_1+j_1}}
    \int_{-\infty}^\infty\! \frac{\rd k^- \, (k^-)^{j_{123}}}{
        (-\vec k_\perp^2 + Q k^- + i0)^{n_2+j_2}}
  \,.
\end{multline}
The $k^\pm$-integrals can be rewritten via $t^\pm = Q k^\pm
- \vec k_\perp^2$:
\begin{multline}
  \label{eq:Sudakovcalckpmscaleless}
  \int_{-\infty}^\infty\! \frac{\rd k^\pm \, (k^\pm)^{j_{123}}}{
      (-\vec k_\perp^2 + Q k^\pm + i0)^{n_{1,2}+j_{1,2}}}
  = \frac{1}{Q^{1+j_{123}}} \int_{-\infty}^\infty\!\rd t^\pm \,
    \frac{(t^\pm + \vec k_\perp^2)^{j_{123}}}{
        (t^\pm + i0)^{n_{1,2}+j_{1,2}}}
\\*
  = \frac{1}{Q^{1+j_{123}}} \sum_{i=0}^{j_{123}} \binom{j_{123}}{i} \,
    (\vec k_\perp^2)^{j_{123}-i}
    \int_{-\infty}^\infty\! \frac{\rd t^\pm}{
        (t^\pm + i0)^{n_{1,2}+j_{1,2}-i}}
  \,.
\end{multline}
These $t^\pm$-integrals are scaleless, however. Within analytic
regularization, we may assume that $\Rep n_1 > 1+j_{23}$ and $\Rep n_2 >
1+j_{13}$ (separately for every term under the sum $\sum_{j_1,j_2,j_3}$),
which is required for the convergence of the $t^\pm$-integrals and also
allows to close the integration contour in the upper half of the complex
$t^\pm$-plane. The only pole at $t^\pm = -i0$ and the branch cut slightly
below the negative real axis are then located outside of the closed
contour. So both $t^\pm$-integrals and therefore the $k^\pm$-integrals
vanish. The $\vec k_\perp$-integral is regularized dimensionally. We get
\begin{align}
  F^{(g)} = 0
  \,.
\end{align}

\paragraph{Collinear-plane contribution.}
From $T^{(cp)}$~(\ref{eq:SudakovTcp}) we get
\begin{multline}
  F^{(cp)} = \frac{\mu^{2\eps} \, e^{\eps\gammaE}}{2i \pi^{d/2}}
    \sum_{j_1,\ldots,j_4=0}^\infty
    \frac{(n_1)_{j_1} \, (n_2)_{j_2} \, (n_3)_{j_{34}}}{
      j_1! \cdots j_4!} \,
    (m^2)^{j_4}
    \int\!\rd^{d-2}\vec k_\perp \, (\vec k_\perp^2)^{j_{123}}
\\* {}\times
    \int_{-\infty}^\infty\! \frac{\rd k^+ \, \rd k^-}{
      \bigl(k^+ (k^-+Q) + i0\bigr)^{n_1+j_1} \,
      \bigl((k^++Q) k^- + i0\bigr)^{n_2+j_2} \,
      (k^+ k^- + i0)^{n_3+j_{34}}}
  \,.
\end{multline}
Here the $\vec k_\perp$-integral is scaleless:
\begin{align}
  \label{eq:Sudakovcalckperpscaleless}
  \int\!\rd^{d-2}\vec k_\perp \, (\vec k_\perp^2)^{j_{123}}
  &= \frac{2 \pi^{1-\eps}}{\Gamma(1-\eps)}
    \int_0^\infty\!\rd |\vec k_\perp| \, |\vec k_\perp|^{1+2j_{123}-2\eps}
  = 0
  \,.
\end{align}
In the $|\vec k_\perp|$-integral the singularities at $|\vec k_\perp| \to
0$ and $|\vec k_\perp| \to \infty$ are regularized dimensionally
(by~$\eps$) and cancel each other.
The $k^+$-$k^-$-integral is regularized analytically. It yields the
contribution
\begin{align}
  2i\pi \,
  \frac{e^{-i\pi n_{123}} \, (-1)^{j_{1234}}}{(Q^2)^{n_{123}+j_{1234}-1}} \,
  \frac{\Gamma(n_{123}+j_{1234}-1) \, \Gamma(1-n_{13}-j_{134}) \,
      \Gamma(1-n_{23}-j_{234})}{
    \Gamma(n_1+j_1) \, \Gamma(n_2+j_2) \, \Gamma(2-n_{123}-j_{1234})}
  \,,
\end{align}
which is non-zero and well-defined for $n_{1,2,3} \not\in \Zb$. Altogether
we have
\begin{align}
  F^{(cp)} = 0
  \,.
\end{align}

\paragraph{Hard--collinear overlap contributions.}
For a Lorentz-invariant integration, the 1-col\-lin\-ear expansion
$T^{(1c)}$~(\ref{eq:SudakovT1c}) can be rewritten as
\begin{align}
  \label{eq:SudakovT1cLI}
  T^{(1c)} I_2
  = \sum_{j_1,j_2=0}^\infty
      \frac{(n_2)_{j_{12}}}{j_1! \, j_2!} \,
      \frac{(-k^+ k^-)^{j_1} \, (\vec k_\perp^2)^{j_2}}{
        (Q k^-)^{n_2+j_{12}}}
  = \sum_{j=0}^\infty \frac{(n_2)_j}{j!} \,
      \frac{(-k^2)^j}{(2p_2\cdot k)^{n_2+j}}
  \,,
\end{align}
where terms in the numerator with the same $j = j_1+j_2$ have been combined
via the binomial theorem.
Putting~(\ref{eq:SudakovT1cLI}) together with
$T^{(h)}$~(\ref{eq:SudakovTh}), the overlap contribution reads
\begin{multline}
  \label{eq:SudakovFh1c}
  F^{(h,1c)} = \sum_{j_1,j_2=0}^\infty
    \frac{(n_2)_{j_1} \, (n_3)_{j_2}}{j_1! \, j_2!} \,
    (-1)^{j_1} \, (m^2)^{j_2}
\\* {}\times
    \int\! \frac{\rD k}{(k^2 + 2p_1\cdot k)^{n_1} \,
        (2p_2\cdot k)^{n_2+j_1} \, (k^2)^{n_3+j_2-j_1}}
  \,.
\end{multline}
We evaluate this loop integral for general propagator powers~$n_{1,2,3}$
with the help of alpha parameters~(\ref{eq:alpha}):
\begin{multline}
  \int\! \frac{\rD k}{(k^2 + 2p_1\cdot k)^{n_1} \,
        (2p_2\cdot k)^{n_2} \, (k^2)^{n_3}}
\\*
  = \left( \prod_{i=1}^3 \frac{e^{-i\pi n_i/2}}{\Gamma(n_i)}
      \int_0^\infty\!\rd\alpha_i \, \alpha_i^{n_i-1} \right)
    \int\!\rD k \, \exp\Bigl[ i \Bigl(
      \alpha_{13} k^2 + 2k \cdot (\alpha_1 p_1 + \alpha_2 p_2) \Bigr)
      \Bigr]
  \,.
\end{multline}
Performing the loop integration via~(\ref{eq:loopintexp}), the
$\alpha_2$-integral through reversing~(\ref{eq:alpha}) and then the
$\alpha_1$-integral using~(\ref{eq:alphaintBeta}), a scaleless
$\alpha_3$-integral remains:
\begin{align}
  \mu^{2\eps} \, e^{\eps\gammaE} \,
  \frac{e^{-i\pi (n_{13}+2n_2+2-\eps)/2}}{(Q^2)^{n_2}} \,
  \frac{\Gamma(n_1-n_2) \, \Gamma(2-n_1-\eps)}{
    \Gamma(n_1) \, \Gamma(n_3) \, \Gamma(2-n_2-\eps)}
  \int_0^\infty\!\rd\alpha_3 \, \alpha_3^{n_{13}-3+\eps}
  = 0
  \,.
\end{align}
Plugging this result into~(\ref{eq:SudakovFh1c}) with $n_2 \to n_2+j_1$ and
$n_3 \to n_3+j_2-j_1$, we obtain
\begin{align}
  F^{(h,1c)} = 0
  \,.
\end{align}
The two collinear regions are symmetric to each other via exchanging $p_1
\leftrightarrow p_2$ and $n_1 \leftrightarrow n_2$. Therefore we also have
\begin{align}
  F^{(h,2c)} = 0
  \,.
\end{align}

\paragraph{Collinear--collinear overlap contribution.}
Rewriting both collinear expansions $T^{(1c)}$~(\ref{eq:SudakovT1c}) and
$T^{(2c)}$~(\ref{eq:SudakovT2c}) via~(\ref{eq:SudakovT1cLI}), the
collinear--collinear overlap contribution reads
\begin{align}
  \label{eq:SudakovF1c2c}
  F^{(1c,2c)} &= \sum_{j_1,j_2=0}^\infty
    \frac{(n_1)_{j_1} \, (n_2)_{j_2}}{j_1! \, j_2!} \, (-1)^{j_{12}}
    \int\! \frac{\rD k \, (k^2)^{j_{12}}}{
      (2p_1\cdot k)^{n_1+j_1} \, (2p_2\cdot k)^{n_2+j_2} \,
      (k^2-m^2)^{n_3}}
\nonumber \\
  &= \sum_{j_1,j_2=0}^\infty
    \frac{(n_1)_{j_1} \, (n_2)_{j_2}}{j_1! \, j_2!} \, (-1)^{j_{12}}
    \sum_{i=0}^{j_{12}} \binom{j_{12}}{i} \, (m^2)^{j_{12}-i}
\nonumber \\* &\qquad {}\times
    \int\! \frac{\rD k}{
      (2p_1\cdot k)^{n_1+j_1} \, (2p_2\cdot k)^{n_2+j_2} \,
      (k^2-m^2)^{n_3-i}}
  \,,
\end{align}
where $(k^2)^{j_{12}} = \bigl((k^2-m^2)+m^2\bigr)^{j_{12}}$ in the
numerator is expanded according to the binomial theorem. Again using alpha
parameters~(\ref{eq:alpha}),
\begin{multline}
  \int\! \frac{\rD k}{
      (2p_1\cdot k)^{n_1} \, (2p_2\cdot k)^{n_2} \,
      (k^2-m^2)^{n_3}}
  = \left( \prod_{i=1}^3 \frac{e^{-i\pi n_i/2}}{\Gamma(n_i)}
      \int_0^\infty\!\rd\alpha_i \, \alpha_i^{n_i-1} \right)
    e^{i\alpha_3 (-m^2)}
 \\* {}\times
   \int\!\rD k \, \exp\Bigl[ i \Bigl(
      \alpha_3 k^2 + 2k \cdot (\alpha_1 p_1 + \alpha_2 p_2) \Bigr)
      \Bigr]
  \,.
\end{multline}
Performing the integrations over $k$, $\alpha_1$ and $\alpha_3$, a
scaleless $\alpha_2$-integral remains,
\begin{align}
  \mu^{2\eps} \, e^{\eps\gammaE} \,
  \frac{e^{-i\pi (3n_1+n_2+2n_3)/2}}{(Q^2)^{n_1} \, (m^2)^{n_{13}-2+\eps}} \,
  \frac{\Gamma(n_{13}-2+\eps)}{\Gamma(n_2) \, \Gamma(n_3)}
  \int_0^\infty\!\rd\alpha_2 \, \alpha_2^{n_2-n_1-1}
  = 0
  \,,
\end{align}
which is regularized analytically. Therefore
\begin{align}
  F^{(1c,2c)} = 0
  \,.
\end{align}

\paragraph{Other overlap contributions.}
The remaining overlap contributions in~(\ref{eq:Sudakovidred}) are
combinations of expansions which already individually lead to scaleless
integrals. So it is obvious that these contributions are scaleless as well:
\begin{align}
  \label{eq:Sudakovotherscaleless}
  F^{(h,1c,2c)} &= \sum_{j_1,j_2,j_3=0}^\infty
    \frac{(n_1)_{j_1} \, (n_2)_{j_2} \, (n_3)_{j_3}}{
      j_1! \, j_2! \, j_3!} \,
    (-1)^{j_{12}} \, (m^2)^{j_3}
\nonumber \\* &\qquad {}\times
    \int\! \frac{\rD k}{
      (2p_1\cdot k)^{n_1+j_1} \, (2p_2\cdot k)^{n_2+j_2} \,
      (k^2)^{n_3+j_3-j_{12}}}
    = 0
    \,,
\nonumber \\
  F^{(h,g)} &= \frac{\mu^{2\eps} \, e^{\eps\gammaE}}{2i \pi^{d/2}}
    \sum_{j_1,\ldots,j_4=0}^\infty
    \frac{(n_1)_{j_1} \, (n_2)_{j_2} \, (n_3)_{j_{34}}}{
      j_1! \cdots j_4!} \,
    (-1)^{j_{123}} \, (m^2)^{j_4}
    \int\! \frac{\rd^{d-2}\vec k_\perp}{
        (-\vec k_\perp^2)^{n_3+j_{34}}}
\nonumber \\* &\qquad {}\times
    \int_{-\infty}^\infty\! \frac{\rd k^+ \, (k^+)^{j_{123}}}{
        (-\vec k_\perp^2 + Q k^+ + i0)^{n_1+j_1}}
    \int_{-\infty}^\infty\! \frac{\rd k^- \, (k^-)^{j_{123}}}{
        (-\vec k_\perp^2 + Q k^- + i0)^{n_2+j_2}}
    = 0
    \,,
\nonumber \\
  F^{(1c,g)} &= \frac{\mu^{2\eps} \, e^{\eps\gammaE}}{2i \pi^{d/2}}
    \sum_{j_1,\ldots,j_4=0}^\infty
    \frac{(n_1)_{j_1} \, (n_2)_{j_{23}} \, (n_3)_{j_4}}{
      j_1! \cdots j_4!} \,
    \frac{(-1)^{j_{124}}}{Q^{n_2+j_{23}}}
    \int\! \frac{\rd^{d-2}\vec k_\perp \, (\vec k_\perp^2)^{j_3}}{
        (-\vec k_\perp^2 - m^2)^{n_3+j_4}}
\nonumber \\* &\qquad {}\times
    \int_{-\infty}^\infty\! \frac{\rd k^+ \, (k^+)^{j_{124}}}{
        (-\vec k_\perp^2 + Q k^+ + i0)^{n_1+j_1}}
    \int_{-\infty}^\infty\! \frac{\rd k^-}{(k^- + i0)^{n_2+j_3-j_{14}}}
    = 0
    \,,
\nonumber \\
  F^{(2c,g)} &= 0 \quad
    [\text{via symmetry related to } F^{(1c,g)}]
    \,,
\nonumber \\
  F^{(h,1c,g)} &= \frac{\mu^{2\eps} \, e^{\eps\gammaE}}{2i \pi^{d/2}}
    \sum_{j_1,\ldots,j_5=0}^\infty
    \frac{(n_1)_{j_1} \, (n_2)_{j_{23}} \, (n_3)_{j_{45}}}{
      j_1! \cdots j_5!} \,
    \frac{(-1)^{j_{1234}} \, (m^2)^{j_5}}{Q^{n_2+j_{23}}}
    \int\! \frac{\rd^{d-2}\vec k_\perp}{(-\vec k_\perp^2)^{n_3+j_{45}-j_3}}
\nonumber \\* &\qquad {}\times
    \int_{-\infty}^\infty\! \frac{\rd k^+ \, (k^+)^{j_{124}}}{
        (-\vec k_\perp^2 + Q k^+ + i0)^{n_1+j_1}}
    \int_{-\infty}^\infty\! \frac{\rd k^-}{(k^- + i0)^{n_2+j_3-j_{14}}}
    = 0
    \,,
\nonumber \\
  F^{(h,2c,g)} &= 0 \quad
    [\text{via symmetry related to } F^{(h,1c,g)}]
    \,,
\nonumber \\
  F^{(1c,2c,g)} &= \frac{\mu^{2\eps} \, e^{\eps\gammaE}}{2i \pi^{d/2}}
    \sum_{j_1,\ldots,j_5=0}^\infty
    \frac{(n_1)_{j_{12}} \, (n_2)_{j_{34}} \, (n_3)_{j_5}}{
      j_1! \cdots j_5!} \,
    \frac{(-1)^{j_{135}}}{Q^{n_{12}+j_{1234}}}
    \int\! \frac{\rd^{d-2}\vec k_\perp \, (\vec k_\perp^2)^{j_{24}}}{
        (-\vec k_\perp^2 - m^2)^{n_3+j_5}}
\nonumber \\* &\qquad {}\times
    \int_{-\infty}^\infty\! \frac{\rd k^+}{(k^+ + i0)^{n_1+j_2-j_{35}}}
    \int_{-\infty}^\infty\! \frac{\rd k^-}{(k^- + i0)^{n_2+j_4-j_{15}}}
    = 0
    \,,
\nonumber \\
  F^{(h,1c,2c,g)} &= \frac{\mu^{2\eps} \, e^{\eps\gammaE}}{2i \pi^{d/2}}
    \sum_{j_1,\ldots,j_6=0}^\infty
    \frac{(n_1)_{j_{12}} \, (n_2)_{j_{34}} \, (n_3)_{j_{56}}}{
      j_1! \cdots j_6!} \,
    \frac{(-1)^{j_{12345}} \, (m^2)^{j_6}}{Q^{n_{12}+j_{1234}}}
\nonumber \\* &\qquad {}\times
    \int\! \frac{\rd^{d-2}\vec k_\perp}{
      (-\vec k_\perp^2)^{n_3+j_{56}-j_{24}}}
    \int_{-\infty}^\infty\! \frac{\rd k^+}{(k^+ + i0)^{n_1+j_2-j_{35}}}
    \int_{-\infty}^\infty\! \frac{\rd k^-}{(k^- + i0)^{n_2+j_4-j_{15}}}
    = 0
    \,,
\nonumber \\
  F^{(h,cp)} &= F^{(cp)} = 0 \quad
    [\text{same expansion}]
    \,,
\nonumber \\
  F^{(1c,cp)} &= \frac{\mu^{2\eps} \, e^{\eps\gammaE}}{2i \pi^{d/2}}
    \sum_{j_1,\ldots,j_5=0}^\infty
    \frac{(n_1)_{j_1} \, (n_2)_{j_{23}} \, (n_3)_{j_{45}}}{
      j_1! \cdots j_5!} \,
    \frac{(-1)^{j_2} \, (m^2)^{j_5}}{Q^{n_2+j_{23}}}
    \int\!\rd^{d-2}\vec k_\perp \, (\vec k_\perp^2)^{j_{134}}
\nonumber \\* &\qquad {}\times
    \int_{-\infty}^\infty\! \frac{\rd k^+ \, \rd k^-}{
      \bigl(k^+ (k^-+Q) + i0\bigr)^{n_1+j_1} \,
      (k^- + i0)^{n_2+j_{23}} \,
      (k^+ k^- + i0)^{n_3+j_{45}-j_2}}
    = 0
    \,,
\nonumber \\
  F^{(2c,cp)} &= 0 \quad
    [\text{via symmetry related to } F^{(1c,cp)}]
    \,,
\nonumber \\
  F^{(h,1c,cp)} &= F^{(1c,cp)} = 0 \quad
    [\text{same expansion}]
    \,,
\nonumber \\
  F^{(h,2c,cp)} &= F^{(2c,cp)} = 0 \quad
    [\text{same expansion}]
    \,,
\nonumber \\
  F^{(1c,2c,cp)} &= \frac{\mu^{2\eps} \, e^{\eps\gammaE}}{2i \pi^{d/2}}
    \sum_{j_1,\ldots,j_6=0}^\infty
    \frac{(n_1)_{j_{12}} \, (n_2)_{j_{34}} \, (n_3)_{j_{56}}}{
      j_1! \cdots j_6!} \,
    \frac{(-1)^{j_{13}} \, (m^2)^{j_6}}{Q^{n_{12}+j_{1234}}}
    \int\!\rd^{d-2}\vec k_\perp \, (\vec k_\perp^2)^{j_{245}}
\nonumber \\* &\qquad {}\times
    \int_{-\infty}^\infty\! \frac{\rd k^+ \, \rd k^-}{
      (k^+ + i0)^{n_1+j_2-j_3} \,
      (k^- + i0)^{n_2+j_4-j_1} \,
      (k^+ k^- + i0)^{n_3+j_{56}}}
    = 0
    \,,
\nonumber \\
  F^{(h,1c,2c,cp)} &= F^{(1c,2c,cp)} = 0 \quad
    [\text{same expansion}]
    \,.
\end{align}
These contributions are all well-defined and scaleless through dimensional
and analytic regularization.

\subsubsection{Sudakov form factor: hard contribution}
\label{app:Sudakovhcalc}

The hard-region integrals in $F^{(h)}$~(\ref{eq:Sudakovints}),
\begin{align}
    \int\! \frac{\rD k}{
      (k^2+2p_1\cdot k)^{n_1} \, (k^2+2p_2\cdot k)^{n_2} \,
      (k^2)^{n_3+j}}
    \,,
\end{align}
are easily evaluated using Feynman parameters, yielding
\begin{multline}
  \frac{\Gamma(n_{123}+j)}{\Gamma(n_1) \, \Gamma(n_2) \, \Gamma(n_3+j)}
    \int_0^1\!\rd x_1 \, \rd x_2 \, \rd x_3 \, \delta(1-x_{123}) \,
    x_1^{n_1-1} \, x_2^{n_2-1} \, x_3^{n_3+j-1}
\\* \shoveright{{}\times
    \int\! \frac{\rD k}{
      \bigl( (k + x_1 p_1 + x_2 p_2)^2 - x_1 x_2 Q^2 \bigr)^{n_{123}+j}}
  }
\\ \shoveleft{
  = \mu^{2\eps} \, e^{\eps\gammaE} \, e^{-i\pi n_{123}} \, (-1)^j \,
    (Q^2)^{2-n_{123}-\eps-j} \,
    \frac{\Gamma(n_{123}-2+\eps+j)}{
      \Gamma(n_1) \, \Gamma(n_2) \, \Gamma(n_3+j)}
  }
\\* {}\times
    \underbrace{
      \int_0^1\!\rd x_1 \, \rd x_2 \, \rd x_3 \, \delta(1-x_{123}) \,
      x_1^{1-n_{23}-\eps-j} \, x_2^{1-n_{13}-\eps-j} \, x_3^{n_3+j-1}
    }_{\Gamma(2-n_{23}-\eps-j) \, \Gamma(2-n_{13}-\eps-j) \, \Gamma(n_3+j)
      / \Gamma(4-n_{123}-2\eps-j)}
  \,.
\end{multline}
The complete contribution from the hard region is shown
in~(\ref{eq:Sudakovresgen}). In the case $n_1=n_2=n_3=1$ this is
\begin{align}
  \label{eq:Sudakovhcalcres111}
  F^{(h)} &= -\frac{1}{Q^2} \left(\frac{\mu^2}{Q^2}\right)^\eps \,
    e^{\eps\gammaE}
    \sum_{j=0}^\infty \left(-\frac{m^2}{Q^2}\right)^j \,
    \frac{\Gamma(1+\eps+j) \, \Gamma^2(-\eps-j)}{\Gamma(1-2\eps-j)}
\nonumber \\
  &= -\frac{1}{Q^2} \left(\frac{\mu^2}{Q^2}\right)^\eps \,
    \frac{e^{\eps\gammaE} \, \Gamma(1+\eps) \, \Gamma^2(-\eps)}{
      \Gamma(1-2\eps)}
    \underbrace{\sum_{j=0}^\infty \left(\frac{m^2}{Q^2}\right)^j \,
      \frac{(2\eps)_j}{(1+\eps)_j}
    }_{\hyperF21(2\eps,1;1+\eps; \, m^2/Q^2)}
  \,.
\end{align}
For the hypergeometric function we use the representations
\begin{align}
  \label{eq:F21int}
  \hyperF21(\alpha,\beta;\gamma;z)
  = \sum_{j=0}^\infty z^j \, \frac{(\alpha)_j \, (\beta)_j}{j! \, (\gamma)_j}
  = \frac{\Gamma(\gamma)}{\Gamma(\beta) \, \Gamma(\gamma-\beta)}
    \int_0^1\!\rd t \,
    \frac{t^{\beta-1} \, (1-t)^{\gamma-\beta-1}}{
      (1-tz)^\alpha}
\end{align}
and extract the singularity of the $t$-integral at $t \to 1$:
\begin{align}
  \label{eq:SudakovhcalcF21}
  \hyperF21(2\eps,1;1+\eps;z)
  &= \frac{\Gamma(1+\eps)}{\Gamma(\eps)}
    \left(
      \int_0^1\!\rd t \, (1-t)^{\eps-1} \, \Bigl[
        (1-tz)^{-2\eps} - (1-z)^{-2\eps} \Bigr]
      \right.
\nonumber \\* &\qquad\qquad\qquad \left. {}
      + (1-z)^{-2\eps} \int_0^1\!\rd t \, (1-t)^{\eps-1}
      \right)
\nonumber \\
  &= -2\eps^2 \int_0^1\! \frac{\rd t}{1-t} \,
      \ln\!\left(\frac{1-tz}{1-z}\right)
    + (1-z)^{-2\eps}
    + \Oc(\eps^3)
\nonumber \\
  &= 1 - 2\eps \, \ln(1-z)
    + \eps^2 \, \Bigl( \ln^2(1-z) - 2\,\Li2(z) \Bigr)
    + \Oc(\eps^3)
  \,,
\end{align}
where $\Li2$ is the dilogarithm function. Expanding also the prefactor
in~(\ref{eq:Sudakovhcalcres111}), the result reported
in~(\ref{eq:SudakovFhres111}) is produced.

\subsubsection{Sudakov form factor: collinear contributions}
\label{app:Sudakovccalc}

The 1-collinear contribution $F^{(1c)}$ in~(\ref{eq:Sudakovints}) reads
\begin{multline}
  \label{eq:SudakovccalcF1c}
  F^{(1c)}
  = \sum_{j=0}^\infty
    \frac{(n_2)_j}{j!} \, (-1)^j
    \sum_{i=0}^j \binom{j}{i} \, (m^2)^{j-i}
\\* {}\times
    \int\! \frac{\rD k}{
      (k^2 + 2p_1\cdot k)^{n_1} \, (2p_2\cdot k)^{n_2+j} \,
      (k^2-m^2)^{n_3-i}}
  \,,
\end{multline}
where $(k^2)^j = \bigl((k^2-m^2)+m^2\bigr)^j$ in the numerator has been
expanded according to the binomial theorem. We evaluate this loop integral
for general propagator powers $n_{1,2,3}$ using alpha
parameters~(\ref{eq:alpha}):
\begin{multline}
  \int\! \frac{\rD k}{(k^2 + 2p_1\cdot k)^{n_1} \,
        (2p_2\cdot k)^{n_2} \, (k^2-m^2)^{n_3}}
  = \left( \prod_{i=1}^3 \frac{e^{-i\pi n_i/2}}{\Gamma(n_i)}
      \int_0^\infty\!\rd\alpha_i \, \alpha_i^{n_i-1} \right)
    e^{i\alpha_3 (-m^2)}
\\* {}\times
    \int\!\rD k \, \exp\Bigl[ i \Bigl(
      \alpha_{13} k^2 + 2k \cdot (\alpha_1 p_1 + \alpha_2 p_2) \Bigr)
      \Bigr]
  \,.
\end{multline}
The loop and alpha-parameter integrations are performed with
(\ref{eq:loopintexp}), (\ref{eq:alpha}) and~(\ref{eq:alphaintBeta}). They
yield
\begin{align}
  \mu^{2\eps} \, e^{\eps\gammaE} \, e^{-i\pi n_{123}} \,
  (m^2)^{2-n_{13}-\eps} \, (Q^2)^{-n_2} \,
  \frac{\Gamma(n_1-n_2) \, \Gamma(n_{13}-2+\eps) \, \Gamma(2-n_1-\eps)}{
    \Gamma(n_1) \, \Gamma(n_3) \, \Gamma(2-n_2-\eps)}
  \,.
\end{align}
Plugging this result with $n_2 \to n_2 + j$ and $n_3 \to n_3 - i$
into~(\ref{eq:SudakovccalcF1c}), we obtain
\begin{multline}
  F^{(1c)} = \mu^{2\eps} \, e^{\eps\gammaE} \, e^{-i\pi n_{123}} \,
    (m^2)^{2-n_{13}-\eps} \, (Q^2)^{-n_2} \,
    \frac{\Gamma(2-n_1-\eps)}{\Gamma(n_1) \, \Gamma(n_2)}
    \sum_{j=0}^\infty \left(\frac{m^2}{Q^2}\right)^j \,
\\* {}\times
    \frac{\Gamma(n_2+j) \, \Gamma(n_1-n_2-j)}{
      j! \, \Gamma(2-n_2-\eps-j)}
    \sum_{i=0}^j \binom{j}{i} \, (-1)^i \,
    \frac{\Gamma(n_{13}-2+\eps-i)}{\Gamma(n_3-i)}
  \,.
\end{multline}
The sum over~$i$ is evaluated as follows:
\begin{align}
  \lefteqn{
    \sum_{i=0}^j \binom{j}{i} \, (-1)^i \,
    \frac{\Gamma(n_{13}-2+\eps-i)}{\Gamma(n_3-i)}
  } \qquad
\nonumber \\*
  &= \frac{1}{\Gamma(2-n_1-\eps)}
    \int_0^1\!\rd x \, x^{n_{13}-3+\eps} \, (1-x)^{1-n_1-\eps}
    \underbrace{\sum_{i=0}^j \binom{j}{i} \left(-\frac{1}{x}\right)^i}_{
      (-1)^j \, x^{-j} \, (1-x)^j}
\nonumber \\
  &= (-1)^j \,
    \frac{\Gamma(n_{13}-2+\eps-j) \, \Gamma(2-n_1-\eps+j)}{
      \Gamma(2-n_1-\eps) \, \Gamma(n_3)}
  \,,
\end{align}
producing the result which is reported in~(\ref{eq:Sudakovresgen}). The
2-collinear contribution $F^{(2c)}$ can be obtained from~$F^{(1c)}$ by
simply exchanging $n_1 \leftrightarrow n_2$.

As explained in section~\ref{sec:Sudakov}, we need analytic regularization
for the collinear contributions, and we choose to calculate the case
$n_1=1+\delta$, $n_2=1-\delta$ and $n_3=1$ in a Laurent expansion about
$\delta=0$ up to the finite order~$\delta^0$. Using the results from above,
we obtain
\begin{align}
  \label{eq:Sudakovccalcres111}
  F^{(1c)} &= -\frac{1}{Q^2} \left(\frac{\mu^2}{m^2}\right)^\eps \,
    \left(\frac{Q^2}{m^2}\right)^\delta \,
    \frac{e^{\eps\gammaE}}{\Gamma(1+\delta) \, \Gamma(1-\delta)}
    \sum_{j=0}^\infty \left(-\frac{m^2}{Q^2}\right)^j
\nonumber \\* &\qquad {}\times
    \frac{\Gamma(1-\delta+j) \, \Gamma(2\delta-j) \,
        \Gamma(\eps+\delta-j) \, \Gamma(1-\eps-\delta+j)}{
      j! \, \Gamma(1-\eps+\delta-j)}
\nonumber \\
  &= -\frac{1}{Q^2} \left(\frac{\mu^2}{m^2}\right)^\eps \,
    \left(\frac{Q^2}{m^2}\right)^\delta \,
    \frac{e^{\eps\gammaE} \, \Gamma(\eps+\delta) \, \Gamma(1-\eps-\delta)}{
      \Gamma(1-\eps+\delta)} \,
    \frac{\Gamma(2\delta)}{\Gamma(1+\delta)}
\nonumber \\* &\qquad {}\times
    \underbrace{
      \sum_{j=0}^\infty \left(\frac{m^2}{Q^2}\right)^j \,
      \frac{(1-\delta)_j \, (\eps-\delta)_j}{j! \, (1-2\delta)_j}
    }_{\hyperF21(\eps-\delta,1-\delta; 1-2\delta; \, m^2/Q^2)}
  \,.
\end{align}
Using the integral representation~(\ref{eq:F21int}) the hypergeometric
function is expanded about $\delta=0$ like in~(\ref{eq:SudakovhcalcF21}):
\begin{multline}
  \hyperF21(\eps-\delta, 1-\delta; 1-2\delta; z)
  = (1-z)^{-\eps} \, \Biggl(
    1 + \delta \, \ln(1-z)
\\* {}
      + \delta \int_0^1\! \frac{\rd t}{t} \left[
        1 - \left(1 + \frac{tz}{1-z}\right)^{-\eps} \right]
    \Biggr)
    + \Oc(\delta^2)
  \,.
\end{multline}
The 1-collinear contribution, expanded in~$\delta$, but valid for
any~$\eps$, reads
\begin{align}
  F^{(1c)} &= -\frac{1}{2 Q^2} \left(\frac{\mu^2}{m^2}\right)^\eps \,
    e^{\eps\gammaE} \, \Gamma(\eps)
    \left(1-\frac{m^2}{Q^2}\right)^{-\eps} \,
    \Biggl(
      \frac{1}{\delta}
      + \ln\!\left(\frac{Q^2}{m^2}\right)
      + \ln\!\left(1-\frac{m^2}{Q^2}\right)
\nonumber \\* &\qquad\qquad{}
      + \int_0^1\! \frac{\rd t}{t} \left[
        1 - \left(1 + t \, \frac{m^2}{Q^2-m^2}\right)^{-\eps} \right]
      - \gammaE + \psi(\eps) - 2\,\psi(1-\eps)
    \Biggr)
\nonumber \\* &\qquad{}
    + \Oc(\delta)
  \,,
\end{align}
where $\psi(z) = \Gamma'(z)/\Gamma(z)$ is the digamma function. We expand
additionally about $\eps=0$, keeping the same prefactor $(\mu^2/Q^2)^\eps$
as in the hard contribution:
\begin{align}
  \label{eq:Sudakovccalcres111exp}
  F^{(1c)} &= -\frac{1}{2 Q^2} \left(\frac{\mu^2}{Q^2}\right)^\eps \,
    \Biggl(
      \frac{1}{\delta} \left[
        \frac{1}{\eps} + \ln\!\left(\frac{Q^2}{m^2}\right)
        - \ln\!\left(1-\frac{m^2}{Q^2}\right) \right]
      - \frac{1}{\eps^2}
      + \frac{2}{\eps} \, \ln\!\left(1-\frac{m^2}{Q^2}\right)
\nonumber \\* &\qquad{}
      + \frac{1}{2} \, \ln^2\!\left(\frac{Q^2}{m^2}\right)
      + \ln\!\left(\frac{Q^2}{m^2}\right)
        \ln\!\left(1-\frac{m^2}{Q^2}\right)
      - \ln^2\!\left(1-\frac{m^2}{Q^2}\right)
      + \Li2\!\left(\frac{m^2}{Q^2}\right)
\nonumber \\* &\qquad{}
      + \frac{5\pi^2}{12}
    \Biggr)
    + \Oc(\delta) + \Oc(\eps)
  \,.
\end{align}
Using the symmetry between the two collinear regions upon $p_1
\leftrightarrow p_2$ and $n_1 \leftrightarrow n_2$, the 2-collinear
contribution~$F^{(2c)}$ can be obtained
from~(\ref{eq:Sudakovccalcres111exp}) by replacing $\delta \to -\delta$.

\subsection{Forward scattering}
\label{app:forward}

\subsubsection{Forward scattering: evaluation with analytic regulators}
\label{app:forwardanalytic}

In this appendix the evaluation of the
contributions~(\ref{eq:forwardidred}) to the forward-scattering integral
(\ref{eq:forward}) or~(\ref{eq:forwardlc}) in section~\ref{sec:forward} is
sketched, using the propagator powers~$n_{1,2,3,4}$ as analytic regulators.

\paragraph{Hard contribution.}
The integral over $T^{(h)}_0 I$~(\ref{eq:forwardTh}) is written in a
Lorentz-invariant way:
\begin{align}
  F^{(h)}_0 &= \frac{1}{2} \int\! \frac{\rD k}{(k^2)^{n_{12}}}
      \left( \frac{1}{(k^2 - 2p_1\cdot k)^{n_3}}
        + \frac{1}{(k^2 + 2p_1\cdot k)^{n_3}} \right)
\nonumber \\* &\qquad {}\times
      \left( \frac{1}{(k^2 + 2p_2\cdot k)^{n_4}}
        + \frac{1}{(k^2 - 2p_2\cdot k)^{n_4}} \right)
\nonumber \\*
    &= F^{(h)}_{+;0} + F^{(h)}_{-;0}
  \,,
\nonumber \\
  F^{(h)}_{\pm;0} &= \int\!
      \frac{\rD k}{(k^2)^{n_{12}} \, (k^2 - 2p_1\cdot k)^{n_3} \,
        (k^2 \pm 2p_2\cdot k)^{n_4}}
  \,,
\end{align}
where terms including the factor $(k^2 + 2p_1\cdot k)^{-n_3}$ have been
transformed via $k \to -k$ under the integral. The two
terms~$F^{(h)}_{\pm;0}$ directly correspond to the diagrams in
figure~\ref{fig:forward} (p.~\pageref{fig:forward}). They are evaluated in
a straightforward way using Feynman parameters and yield
\begin{multline}
  F^{(h)}_{\pm;0} = \mu^{2\eps} \, e^{\eps\gammaE} \,
    e^{-i\pi n_{1234}} \, (\mp Q^2-i0)^{2-n_{1234}-\eps}
\\* {}\times
    \frac{\Gamma(n_{1234}-2+\eps) \, \Gamma(2-n_{123}-\eps) \,
        \Gamma(2-n_{124}-\eps)}{
      \Gamma(n_3) \, \Gamma(n_4) \, \Gamma(4-n_{1234}-2\eps)}
  \,.
\end{multline}
The complete result reads
\begin{multline}
  \label{eq:forwardcalchres}
  F^{(h)}_0 = \mu^{2\eps} \, e^{\eps\gammaE} \,
    e^{-i\pi n_{1234}} \, (Q^2)^{2-n_{1234}-\eps} \,
    (1 + e^{i\pi(n_{1234}+\eps)})
\\* {}\times
    \frac{\Gamma(n_{1234}-2+\eps) \, \Gamma(2-n_{123}-\eps) \,
        \Gamma(2-n_{124}-\eps)}{
      \Gamma(n_3) \, \Gamma(n_4) \, \Gamma(4-n_{1234}-2\eps)}
  \,.
\end{multline}

\paragraph{Collinear contributions.}
We also write the integral over $T^{(1c)}_0 I$~(\ref{eq:forwardT1c}) as a
Lorentz-invariant integral using the $d$-dimensional vector $r_\perp = r +
(p_1-p_2) \, r^2/Q^2$ satisfying \mbox{$p_{1,2}\cdot r_\perp = 0$} and
$r_\perp^2 = -\vec r_\perp^{\,2}$. With this we express
\begin{align}
  \vec r_\perp \cdot \vec k_\perp
  = -r\cdot k + \frac{r^+ k^- + r^- k^+}{2}
  = -r_\perp\cdot k
  \,,
\end{align}
using~(\ref{eq:lightconescalar}), and obtain the integral
\begin{align}
  \label{eq:forwardcalccF01cint}
  F^{(1c)}_0 &= \frac{1}{2} \int\!
      \frac{\rD k}{(k^2)^{n_1} \,
        \bigl( (k - r_\perp)^2 - 2(r_\perp^2/Q^2) \, p_2\cdot k \bigr)^{n_2}}
  \nonumber \\* &\qquad {}\times
      \left(
        \frac{1}{( k^2 - 2p_1\cdot k)^{n_3}}
        + \frac{1}{\bigl( k^2 - 2r_\perp\cdot k + 2p_1\cdot k
            - 2(r_\perp^2/Q^2) \, p_2\cdot k \bigr)^{n_3}}
      \right)
  \nonumber \\* &\qquad {}\times
      \left(
        \frac{1}{(2p_2\cdot k)^{n_4}} + \frac{1}{(-2p_2\cdot k)^{n_4}}
      \right)
  .
\end{align}
Let us first evaluate only the contribution from the first term in each
round bracket of~(\ref{eq:forwardcalccF01cint}), which is separately
well-defined through analytic regularization. Using alpha
parameters~(\ref{eq:alpha}),
\begin{multline}
  \frac{1}{2} \int\!
      \frac{\rD k}{(k^2)^{n_1} \,
        \bigl( (k - r_\perp)^2 - 2(r_\perp^2/Q^2) \, p_2\cdot k
          \bigr)^{n_2} \,
        ( k^2 - 2p_1\cdot k)^{n_3} \, (2p_2\cdot k)^{n_4}}
\\ \shoveleft{
  = \frac{1}{2}
    \left( \prod_{i=1}^4 \frac{e^{-i\pi n_i/2}}{\Gamma(n_i)}
      \int_0^\infty\!\rd\alpha_i \, \alpha_i^{n_i-1} \right)
    e^{i\alpha_2(r_\perp^2+i0)}
  }
\\* {}\times
    \int\!\rD k \, \exp\!\left[ i \Biggl(
      \alpha_{123} k^2
      - 2k \cdot \left[
        \alpha_2 r_\perp + \alpha_3 p_1
        + \left(\frac{r_\perp^2}{Q^2} \, \alpha_2 - \alpha_4\right) p_2
          \right]
      + i0
      \Biggr) \right]
  .
\end{multline}
After performing the loop integral with~(\ref{eq:loopintexp}) and the
$\alpha_4$-integral via~(\ref{eq:alpha}) we obtain
\begin{multline}
  \label{eq:forwardcalccF01cffalpha}
  \frac{1}{2} \,
  \frac{\mu^{2\eps} \, e^{\eps\gammaE} \, e^{-i\pi(n_{123}+2-\eps)/2}}{
    \Gamma(n_1) \, \Gamma(n_2) \, \Gamma(n_3) \, (Q^2+i0)^{n_4}}
  \int_0^\infty\!\rd\alpha_1 \, \rd\alpha_2 \, \rd\alpha_3 \,
  \alpha_1^{n_1-1} \, \alpha_2^{n_2-1} \, \alpha_3^{n_3-n_4-1} \,
  \alpha_{123}^{n_4-2+\eps}
\\* {}\times
  \exp\!\left(i \, \frac{\alpha_1 \alpha_2}{\alpha_{123}} \,
    (r_\perp^2 + i0)\right)
  .
\end{multline}
We can solve this integral by multiplying it with unity in the form of
\begin{align}
  1 = \int_0^\infty\!\rd\eta \,
      \delta\biggl(\eta-\sum_{j\in S}\alpha_j\biggr)
  , \qquad \text{with }
  \emptyset \ne S \subset \{1,2,3\}
  \,,
\end{align}
where the sum of alpha parameters in the delta function runs over an
arbitrary non-empty subset of $\alpha_{1,2,3}$. Then, changing the order of
integration, all alpha parameters are rescaled as $\alpha_i \to
\eta\alpha_i$ under the $\eta$-integral and the delta function is rewritten
as $(1/\eta) \, \delta(1-\sum_{j\in S}\alpha_j)$. Performing the
$\eta$-integral first using~(\ref{eq:alpha}) yields
\begin{multline}
  \frac{1}{2} \,
  \mu^{2\eps} \, e^{\eps\gammaE} \, e^{-i\pi n_{1234}} \,
  (-r_\perp^2-i0)^{2-n_{123}-\eps} \, (-Q^2-i0)^{-n_4} \,
  \frac{\Gamma(n_{123}-2+\eps)}{
    \Gamma(n_1) \, \Gamma(n_2) \, \Gamma(n_3)}
\\* {}\times
  \int_0^\infty\!\rd\alpha_1 \, \rd\alpha_2 \, \rd\alpha_3 \,
  \delta\biggl(1-\sum_{j\in S}\alpha_j\biggr) \,
  \alpha_1^{1-n_{23}-\eps} \, \alpha_2^{1-n_{13}-\eps} \,
  \alpha_3^{n_3-n_4-1} \, \alpha_{123}^{n_{1234}-4+2\eps}
  \,.
\end{multline}
Now the three-fold integral is of the form
\begin{align}
  \label{eq:alphadeltamult}
  \int_0^\infty\rd\alpha_1 \cdots \rd\alpha_n \,
  \delta\biggl(1-\sum_{j\in S}\alpha_j\biggr) \,
  \frac{\alpha_1^{\nu_1-1} \cdots \alpha_n^{\nu_n-1}}{
    (\alpha_1 + \ldots + \alpha_n)^{\nu_1+\ldots+\nu_n}}
  = \frac{\Gamma(\nu_1) \cdots \Gamma(\nu_n)}{\Gamma(\nu_1+\ldots+\nu_n)}
  \,,
\end{align}
valid for any $\emptyset \ne S \subset \{1,\ldots,n\}$, and we obtain
\begin{multline}
  \label{eq:forwardcalccF01cffres}
  \frac{1}{2} \,
  \mu^{2\eps} \, e^{\eps\gammaE} \, e^{-i\pi n_{1234}} \,
  (\vec r_\perp^{\,2})^{2-n_{123}-\eps} \, (-Q^2-i0)^{-n_4} \,
  \Gamma(n_3-n_4)
\\* {}\times
  \frac{\Gamma(n_{123}-2+\eps) \, \Gamma(2-n_{13}-\eps) \,
      \Gamma(2-n_{23}-\eps)}{
    \Gamma(n_1) \, \Gamma(n_2) \, \Gamma(n_3) \,
      \Gamma(4-n_{1234}-2\eps)}
\end{multline}
for the contribution from the first term in each factor
of~(\ref{eq:forwardcalccF01cint}). Analytic regularization is needed for
this contribution, otherwise $\Gamma(n_3-n_4)$ would be ill-defined.

The contribution from the first $\times$ second term
in~(\ref{eq:forwardcalccF01cint}), i.e.\ with $(-2p_2\cdot k)^{-n_4}$
instead of $(2p_2\cdot k)^{-n_4}$, can be calculated
from~(\ref{eq:forwardcalccF01cffres}) by replacing $p_2 \to -p_2$ and
therefore $Q^2 \to -Q^2$. (The combination $p_2/Q^2$ remains invariant
under this transformation.)
The remaining two contributions, involving the second term in the first
round bracket of~(\ref{eq:forwardcalccF01cint}), are obtained from the
previous two contributions via the replacement $k \to r_\perp +
(r_\perp^2/Q^2) \, p_2 - k$ and $n_1 \leftrightarrow n_2$ under the loop
integral, which leaves the expression~(\ref{eq:forwardcalccF01cffres})
invariant.
The complete result reads
\begin{multline}
  \label{eq:forwardcalccF01cres}
  F^{(1c)}_0 = \mu^{2\eps} \, e^{\eps\gammaE} \, e^{-i\pi n_{1234}} \,
  (\vec r_\perp^{\,2})^{2-n_{123}-\eps} \, (Q^2)^{-n_4} \,
  (1 + e^{i\pi n_4}) \, \Gamma(n_3-n_4)
\\* {}\times
  \frac{\Gamma(n_{123}-2+\eps) \, \Gamma(2-n_{13}-\eps) \,
      \Gamma(2-n_{23}-\eps)}{
    \Gamma(n_1) \, \Gamma(n_2) \, \Gamma(n_3) \,
      \Gamma(4-n_{1234}-2\eps)}
  \,.
\end{multline}

The 2-collinear contribution,
\begin{align}
  \label{eq:forwardcalccF02cint}
  F^{(2c)}_0 &= \frac{1}{2} \int\!
      \frac{\rD k}{(k^2)^{n_1} \,
        \bigl( (k - r_\perp)^2 + 2(r_\perp^2/Q^2) \, p_1\cdot k \bigr)^{n_2}}
      \left(
        \frac{1}{(-2p_1\cdot k)^{n_3}} + \frac{1}{(2p_1\cdot k)^{n_3}}
      \right)
  \nonumber \\* &{}\times
      \left(
        \frac{1}{( k^2 + 2p_2\cdot k)^{n_4}}
        + \frac{1}{\bigl( k^2 - 2r_\perp\cdot k
            + 2(r_\perp^2/Q^2) \, p_1\cdot k - 2p_2\cdot k \bigr)^{n_4}}
      \right)
  ,
\end{align}
results from~(\ref{eq:forwardcalccF01cint}) by exchanging $p_1
\leftrightarrow -p_2$ and $n_3 \leftrightarrow n_4$. It is thus obtained
from~(\ref{eq:forwardcalccF01cres}) via $n_3 \leftrightarrow n_4$ and reads
\begin{multline}
  \label{eq:forwardcalccF02cres}
  F^{(2c)}_0 = \mu^{2\eps} \, e^{\eps\gammaE} \, e^{-i\pi n_{1234}} \,
  (\vec r_\perp^{\,2})^{2-n_{124}-\eps} \, (Q^2)^{-n_3} \,
  (1 + e^{i\pi n_3}) \, \Gamma(n_4-n_3)
\\* {}\times
  \frac{\Gamma(n_{124}-2+\eps) \, \Gamma(2-n_{14}-\eps) \,
      \Gamma(2-n_{24}-\eps)}{
    \Gamma(n_1) \, \Gamma(n_2) \, \Gamma(n_4) \,
      \Gamma(4-n_{1234}-2\eps)}
  \,.
\end{multline}

\paragraph{Glauber contribution.}
The integral over $T^{(g)}_0 I$~(\ref{eq:forwardTg}) reads
\begin{align}
  \label{eq:forwardcalcgint}
  F^{(g)}_0 &= \frac{1}{2} \,
    \frac{\mu^{2\eps} \, e^{\eps\gammaE}}{2i\pi^{d/2}}
    \int\! \frac{\rd^{d-2}\vec k_\perp}{
      (-\vec k_\perp^2)^{n_1} \,
      \bigl( -(\vec k_\perp - \vec r_\perp)^2 \bigr)^{n_2}}
\nonumber \\* &\qquad {}\times
    \int_{-\infty}^\infty\!\rd k^+
    \left(
      \frac{1}{( -Q k^+ - \vec k_\perp^2 + i0 )^{n_3}}
      + \frac{1}{( Q k^+ - \vec k_\perp^2
          + 2\vec r_\perp\cdot\vec k_\perp + i0 )^{n_3}}
    \right)
\nonumber \\* &\qquad {}\times
    \int_{-\infty}^\infty\!\rd k^-
    \left(
      \frac{1}{( Q k^- - \vec k_\perp^2 + i0 )^{n_4}}
      + \frac{1}{( -Q k^- - \vec k_\perp^2
          + 2\vec r_\perp\cdot\vec k_\perp + i0 )^{n_4}}
    \right)
  .
\end{align}
Analytic regularization makes the $k^\pm$-integrations over the individual
terms well-defined and allows to assume $\Rep n_{3,4} > 1$. Then we can
close the integration contours at $i\infty$ or $-i\infty$, choosing
separately for each term the side where no pole and no branch cut lies
within the closed contour. So all $k^+$- and $k^-$-integrals vanish as
scaleless integrals, cf.~(\ref{eq:Sudakovcalckpmscaleless}). The $\vec
k_\perp$-integral is regularized dimensionally. Therefore,
\begin{align}
  F^{(g)}_0 = 0
  \,.
\end{align}

\paragraph{Collinear-plane contribution.}
Integrating over $T^{(cp)}_0 I$~(\ref{eq:forwardTcp}) yields
\begin{align}
  \label{eq:forwardcalccpint}
  F^{(cp)}_0 &= \frac{1}{2} \,
    \frac{\mu^{2\eps} \, e^{\eps\gammaE}}{2i\pi^{d/2}}
    \int\!\rd^{d-2}\vec k_\perp
    \int_{-\infty}^\infty\!
    \frac{\rd k^+ \, \rd k^-}{(k^+ k^- + i0)^{n_{12}}}
\nonumber \\* &\qquad {}\times
    \left(
      \frac{1}{\bigl( k^+ (k^--Q) + i0 \bigr)^{n_3}}
      + \frac{1}{\bigl( k^+ (k^-+Q) + i0 \bigr)^{n_3}}
    \right)
\nonumber \\* &\qquad {}\times
    \left(
      \frac{1}{\bigl( (k^++Q) k^- + i0 \bigr)^{n_4}}
      + \frac{1}{\bigl( (k^+-Q) k^- + i0 \bigr)^{n_4}}
    \right)
  .
\end{align}
As the integrand is independent of~$\vec k_\perp$, the $\vec
k_\perp$-integration is scaleless through dimensional regularization,
cf.~(\ref{eq:Sudakovcalckperpscaleless}). The $k^\pm$-integrals are
regularized analytically. So
\begin{align}
  F^{(cp)}_0 = 0
  \,.
\end{align}

\paragraph{Hard--collinear overlap contributions.}
The hard--1-collinear overlap contribution reads
\begin{multline}
  F^{(h,1c)}_0 = \frac{1}{2} \int\!
      \frac{\rD k}{(k^2)^{n_{12}}}
      \left(
        \frac{1}{( k^2 - 2p_1\cdot k)^{n_3}}
        + \frac{1}{( k^2 + 2p_1\cdot k)^{n_3}}
      \right)
\\* {}\times
      \left(
        \frac{1}{(2p_2\cdot k)^{n_4}} + \frac{1}{(-2p_2\cdot k)^{n_4}}
      \right)
  .
\end{multline}
It corresponds to $F^{(1c)}_0$~(\ref{eq:forwardcalccF01cint}) with $r_\perp
= 0$. According to~(\ref{eq:forwardcalccF01cffalpha}), we get
\begin{align}
  F^{(h,1c)}_0 &=
    \frac{\mu^{2\eps} \, e^{\eps\gammaE} \, e^{-i\pi(n_{123}+2-\eps)/2}}{
      \Gamma(n_1) \, \Gamma(n_2) \, \Gamma(n_3)} \,
    \Bigl( (Q^2+i0)^{-n_4} + (-Q^2+i0)^{-n_4} \Bigr)
\nonumber \\* &\qquad {}\times
    \int_0^\infty\!\rd\alpha_1 \, \rd\alpha_2 \, \rd\alpha_3 \,
    \alpha_1^{n_1-1} \, \alpha_2^{n_2-1} \, \alpha_3^{n_3-n_4-1} \,
    \alpha_{123}^{n_4-2+\eps}
\nonumber \\
  &= \frac{\mu^{2\eps} \, e^{\eps\gammaE} \, e^{-i\pi(n_{123}+2-\eps)/2}}{
      (Q^2)^{n_4}} \,
    (1 + e^{-i\pi n_4}) \,
    \frac{\Gamma(2-n_{124}-\eps)}{\Gamma(n_3) \, \Gamma(2-n_4-\eps)}
    \int_0^\infty\!\rd\alpha_3 \, \alpha_3^{n_{123}-3+\eps}
\nonumber \\*
  &= 0
  \,.
\end{align}
This integral is scaleless by dimensional regularization. Analytic
regularization is only required if the $\alpha_3$-integration is performed
first.
Similarly, via the symmetry $p_1 \leftrightarrow -p_2$ and $n_3
\leftrightarrow n_4$,
\begin{align}
  F^{(h,2c)}_0 = 0
  \,.
\end{align}

\paragraph{Collinear--collinear overlap contribution.}
The 1-collinear--2-collinear overlap contribution reads
\begin{multline}
  \label{eq:forwarccalcF01c2cint}
  F^{(1c,2c)}_0 = \frac{1}{2} \int\!
    \frac{\rD k}{(k^2)^{n_1} \,
      \bigl( (k - r_\perp)^2 \bigr)^{n_2}}
    \left(
      \frac{1}{(-2p_1\cdot k)^{n_3}} + \frac{1}{(2p_1\cdot k)^{n_3}}
    \right)
\\* {}\times
    \left(
      \frac{1}{(2p_2\cdot k)^{n_4}} + \frac{1}{(-2p_2\cdot k)^{n_4}}
    \right)
  .
\end{multline}
The first term of each round bracket yields the contribution
\begin{multline}
  \frac{1}{2}
    \left( \prod_{i=1}^4 \frac{e^{-i\pi n_i/2}}{\Gamma(n_i)}
      \int_0^\infty\!\rd\alpha_i \, \alpha_i^{n_i-1} \right)
    e^{i\alpha_2(r_\perp^2+i0)}
\\* \shoveright{ {}\times
    \int\!\rD k \, \exp\Bigl( i \bigl[
      \alpha_{12} k^2
      - 2k \cdot (\alpha_2 r_\perp + \alpha_3 p_1 - \alpha_4 p_2)
      + i0
      \bigr] \Bigr)
  }
\\ \shoveleft{
  = \frac{1}{2}
    \frac{\mu^{2\eps} \, e^{\eps\gammaE} \, e^{-i\pi(n_{123}+2-\eps)/2}}{
      \Gamma(n_1) \, \Gamma(n_2) \, \Gamma(n_3) \, (Q^2+i0)^{n_4}}
    \int_0^\infty\!\rd\alpha_3 \, \alpha_3^{n_3-n_4-1}
  }
\\* {}\times
    \int_0^\infty\!\rd\alpha_1 \, \rd\alpha_2 \,
    \alpha_1^{n_1-1} \, \alpha_2^{n_2-1} \, \alpha_{12}^{n_4-2+\eps}
    \exp\!\left(i \, \frac{\alpha_1 \alpha_2}{\alpha_{12}} \,
      (r_\perp^2 + i0)\right)
  .
\end{multline}
The $\alpha_{1,2}$-integrations are regularized dimensionally, while the
$\alpha_3$-integral is scaleless by analytic regularization. The remaining
contributions from~(\ref{eq:forwarccalcF01c2cint}) are obtained via $p_1
\to -p_1$ and/or $p_2 \to -p_2$ and vanish as well. Therefore,
\begin{align}
  F^{(1c,2c)}_0 = 0
  \,.
\end{align}

\paragraph{Other overlap contributions.}
All other overlap contributions in~(\ref{eq:forwardidred}) are expansions
of scaleless integrals evaluated above, so they are scaleless as well:
\begin{align}
  \label{eq:forwardcalcotherint}
  F^{(h,1c,2c)}_0 &= \frac{1}{2} \int\!
      \frac{\rD k}{(k^2)^{n_{12}}}
      \left(
        \frac{1}{(-2p_1\cdot k)^{n_3}} + \frac{1}{(2p_1\cdot k)^{n_3}}
      \right)
  \nonumber \\* &\qquad {}\times
      \left(
        \frac{1}{(2p_2\cdot k)^{n_4}} + \frac{1}{(-2p_2\cdot k)^{n_4}}
      \right)
    = 0
    \,,
\nonumber \\
  F^{(h,g)}_0 &= \frac{1}{2} \,
      \frac{\mu^{2\eps} \, e^{\eps\gammaE}}{2i\pi^{d/2}}
      \int\! \frac{\rd^{d-2}\vec k_\perp}{
        (-\vec k_\perp^2)^{n_{12}}}
  \nonumber \\* &\qquad {}\times
      \int_{-\infty}^\infty\!\rd k^+
      \left(
        \frac{1}{( -Q k^+ - \vec k_\perp^2 + i0 )^{n_3}}
        + \frac{1}{( Q k^+ - \vec k_\perp^2 + i0 )^{n_3}}
      \right)
  \nonumber \\* &\qquad {}\times
      \int_{-\infty}^\infty\!\rd k^-
      \left(
        \frac{1}{( Q k^- - \vec k_\perp^2 + i0 )^{n_4}}
        + \frac{1}{( -Q k^- - \vec k_\perp^2 + i0 )^{n_4}}
      \right)
    = 0
    \,,
\nonumber \\
  F^{(1c,g)}_0 &= \frac{1}{2} \,
      \frac{\mu^{2\eps} \, e^{\eps\gammaE}}{2i\pi^{d/2}} \,
      \frac{1}{Q^{n_4}}
      \int\! \frac{\rd^{d-2}\vec k_\perp}{
        (-\vec k_\perp^2)^{n_1} \,
        \bigl( -(\vec k_\perp - \vec r_\perp)^2 \bigr)^{n_2}}
  \nonumber \\* &\qquad {}\times
      \int_{-\infty}^\infty\!\rd k^+
      \left(
        \frac{1}{( -Q k^+ - \vec k_\perp^2 + i0 )^{n_3}}
        + \frac{1}{( Q k^+ - \vec k_\perp^2
            + 2\vec r_\perp\cdot\vec k_\perp + i0 )^{n_3}}
      \right)
  \nonumber \\* &\qquad {}\times
      \int_{-\infty}^\infty\!\rd k^-
      \left(
        \frac{1}{( k^- + i0 )^{n_4}} + \frac{1}{( -k^- + i0 )^{n_4}}
      \right)
    = 0
    \,,
\nonumber \\
  F^{(2c,g)}_0 &= 0 \quad
    \text{[via symmetry related to $F^{(1c,g)}_0$]}
    \,,
\nonumber \\
  F^{(h,1c,g)}_0 &= \frac{1}{2} \,
      \frac{\mu^{2\eps} \, e^{\eps\gammaE}}{2i\pi^{d/2}} \,
      \frac{1}{Q^{n_4}}
      \int\! \frac{\rd^{d-2}\vec k_\perp}{
        (-\vec k_\perp^2)^{n_{12}}}
  \nonumber \\* &\qquad {}\times
      \int_{-\infty}^\infty\!\rd k^+
      \left(
        \frac{1}{( -Q k^+ - \vec k_\perp^2 + i0 )^{n_3}}
        + \frac{1}{( Q k^+ - \vec k_\perp^2 + i0 )^{n_3}}
      \right)
  \nonumber \\* &\qquad {}\times
      \int_{-\infty}^\infty\!\rd k^-
      \left(
        \frac{1}{( k^- + i0 )^{n_4}} + \frac{1}{( -k^- + i0 )^{n_4}}
      \right)
    = 0
    \,,
\nonumber \\
  F^{(h,2c,g)}_0 &= 0 \quad
    \text{[via symmetry related to $F^{(h,1c,g)}_0$]}
    \,,
\nonumber \\
  F^{(1c,2c,g)}_0 &= \frac{1}{2} \,
      \frac{\mu^{2\eps} \, e^{\eps\gammaE}}{2i\pi^{d/2}} \,
      \frac{1}{Q^{n_{34}}}
      \int\! \frac{\rd^{d-2}\vec k_\perp}{
        (-\vec k_\perp^2)^{n_1} \,
        \bigl( -(\vec k_\perp - \vec r_\perp)^2 \bigr)^{n_2}}
  \nonumber \\* &\qquad {}\times
      \int_{-\infty}^\infty\!\rd k^+
      \left(
        \frac{1}{( -k^+ + i0 )^{n_3}}
        + \frac{1}{( k^+ + i0 )^{n_3}}
      \right)
  \nonumber \\* &\qquad {}\times
      \int_{-\infty}^\infty\!\rd k^-
      \left(
        \frac{1}{( k^- + i0 )^{n_4}} + \frac{1}{( -k^- + i0 )^{n_4}}
      \right)
    = 0
    \,,
\nonumber \\
  F^{(h,1c,2c,g)}_0 &= \frac{1}{2} \,
      \frac{\mu^{2\eps} \, e^{\eps\gammaE}}{2i\pi^{d/2}} \,
      \frac{1}{Q^{n_{34}}}
      \int\! \frac{\rd^{d-2}\vec k_\perp}{
        (-\vec k_\perp^2)^{n_{12}}}
      \int_{-\infty}^\infty\!\rd k^+
      \left(
        \frac{1}{( -k^+ + i0 )^{n_3}}
        + \frac{1}{( k^+ + i0 )^{n_3}}
      \right)
  \nonumber \\* &\qquad {}\times
      \int_{-\infty}^\infty\!\rd k^-
      \left(
        \frac{1}{( k^- + i0 )^{n_4}} + \frac{1}{( -k^- + i0 )^{n_4}}
      \right)
    = 0
    \,,
\nonumber \\
  F^{(h,cp)}_0 &= F^{(cp)}_0 = 0 \quad
    \text{[same expansion]}
    \,,
\nonumber \\
  F^{(1c,cp)}_0 &= \frac{1}{2} \,
      \frac{\mu^{2\eps} \, e^{\eps\gammaE}}{2i\pi^{d/2}} \,
      \frac{1}{Q^{n_4}}
      \int\!\rd^{d-2}\vec k_\perp
      \int_{-\infty}^\infty\!
      \frac{\rd k^+ \, \rd k^-}{(k^+ k^- + i0)^{n_{12}}}
  \nonumber \\* &\qquad {}\times
      \left(
        \frac{1}{\bigl( k^+ (k^--Q) + i0 \bigr)^{n_3}}
        + \frac{1}{\bigl( k^+ (k^-+Q) + i0 \bigr)^{n_3}}
      \right)
  \nonumber \\* &\qquad {}\times
      \left(
        \frac{1}{( k^- + i0 )^{n_4}} + \frac{1}{( -k^- + i0 )^{n_4}}
      \right)
    = 0
    \,,
\nonumber \\
  F^{(2c,cp)}_0 &= 0 \quad
    \text{[via symmetry related to $F^{(1c,cp)}_0$]}
    \,,
\nonumber \\
  F^{(h,1c,cp)}_0 &= F^{(1c,cp)}_0 = 0 \quad
    \text{[same expansion]}
    \,,
\nonumber \\
  F^{(h,2c,cp)}_0 &= F^{(2c,cp)}_0 = 0 \quad
    \text{[same expansion]}
    \,,
\nonumber \\
  F^{(1c,2c,cp)}_0 &= \frac{1}{2} \,
      \frac{\mu^{2\eps} \, e^{\eps\gammaE}}{2i\pi^{d/2}} \,
      \frac{1}{Q^{n_{34}}}
      \int\!\rd^{d-2}\vec k_\perp
      \int_{-\infty}^\infty\!
      \frac{\rd k^+ \, \rd k^-}{(k^+ k^- + i0)^{n_{12}}}
  \nonumber \\* &\qquad {}\times
      \left(
        \frac{1}{( -k^+ + i0 )^{n_3}} + \frac{1}{( k^+ + i0 )^{n_3}}
      \right)
      \left(
        \frac{1}{( k^- + i0 )^{n_4}} + \frac{1}{( -k^- + i0 )^{n_4}}
      \right)
    = 0
    \,,
\nonumber \\
  F^{(h,1c,2c,cp)}_0 &= F^{(1c,2c,cp)}_0 = 0 \quad
    \text{[same expansion]}
    \,.
\end{align}
All these integrals are well-defined via dimensional and analytic
regularization.

\subsubsection{Forward scattering: evaluation without analytic regulators}
\label{app:forward1111}

Here the evaluation of the contributions~(\ref{eq:forwardidred}) to the
forward-scattering integral in section~\ref{sec:forward} is repeated with
analytic regularization switched off. This means that the propagator powers
are set to the fixed values $n_1=n_2=n_3=n_4=1$ from the start.

\paragraph{Hard contribution.}
The evaluation of the hard contribution in
appendix~\ref{app:forwardanalytic} is also valid without analytic
regularization. From~(\ref{eq:forwardcalchres}) we get
\begin{align}
  \label{eq:forward1111calcF0hres}
  F^{(h)}_0 = \frac{1}{(Q^2)^2}
    \left(\frac{\mu^2}{Q^2}\right)^\eps \,
    (1 + e^{i\pi\eps}) \,
    \frac{e^{\eps\gammaE} \, \Gamma(2+\eps) \, \Gamma^2(-1-\eps)}{
      \Gamma(-2\eps)}
  \,.
\end{align}

\paragraph{Collinear contributions.}
The 1-collinear contribution without analytic regulators reads
\begin{align}
  \label{eq:forward1111calccint}
  F^{(1c)}_0 &= \frac{1}{2Q} \,
      \frac{\mu^{2\eps} \, e^{\eps\gammaE}}{2i\pi^{d/2}}
      \int\!\rd^{d-2}\vec k_\perp \!
      \int_{-\infty}^\infty\!
      \frac{\rd k^+ \, \rd k^-}{
        (k^+ k^- - \vec k_\perp^2 + i0) \,
        \bigl( (k^+ - r^+_0) k^-
          - (\vec k_\perp - \vec r_\perp)^2 + i0 \bigr)}
\nonumber \\* &\quad {}\times
      \left(
        \frac{1}{k^+ (k^--Q) - \vec k_\perp^2 + i0}
        + \frac{1}{k^+ (k^-+Q) - r^+_0 k^- - \vec k_\perp^2
          + 2\vec r_\perp\cdot\vec k_\perp + i0}
      \right)
\nonumber \\* &\quad {}\times
      \left(
        \frac{1}{k^- + i0} + \frac{1}{-k^- + i0}
      \right)
  .  
\end{align}
The factor in the last line of~(\ref{eq:forward1111calccint}) vanishes
everywhere under the $k^-$-integral except for the pole at $k^-=0$ which
lies below the real axis in the first term and above the real axis in the
second term. The only remaining contribution arises from the closed
integration contour around this pole, i.e.\ from the residue at $k^-=0$,
which allows us to express
\begin{align}
  \label{eq:deltacontribution}
  \frac{1}{k^- + i0} + \frac{1}{-k^- + i0}
  = \frac{1}{k^- + i0} - \frac{1}{k^- - i0}
  = -2i\pi \, \delta(k^-)
  \,.
\end{align}
The 1-collinear contribution becomes
\begin{multline}
  F^{(1c)}_0 = -\frac{1}{2Q} \,
      \frac{\mu^{2\eps} \, e^{\eps\gammaE}}{\pi^{1-\eps}}
      \int\! \frac{\rd^{d-2}\vec k_\perp}{
        \vec k_\perp^2 \, (\vec k_\perp - \vec r_\perp)^2}
\\* {}\times
      \int_{-\infty}^\infty\!\rd k^+
      \left(
        \frac{1}{-Q k^+ - \vec k_\perp^2 + i0}
        + \frac{1}{Q k^+ - \vec k_\perp^2
          + 2\vec r_\perp\cdot\vec k_\perp + i0}
      \right)
  .  
\end{multline}
The $k^+$-integral converges at $|k^+|\to\infty$ because the leading
behaviour for large~$|k^+|$ is cancelled between the two terms, such that
the integrand falls off like $1/(k^+)^2$ towards infinity. We close the
integration contour at $+i\infty$ or at $-i\infty$, taking into account the
residue of the one pole within the contour. In general, we get
\begin{align}
  \label{eq:kpmintconvergeres}
  \int_{-\infty}^\infty\!\rd k^+
      \left(
        \frac{1}{-Q k^+ + A + i0} + \frac{1}{Q k^+ + B + i0}
      \right)
  = -\frac{2i\pi}{Q}
  \,,
\end{align}
for real $A,B$ and $Q>0$, which is in particular independent of $A$
and~$B$.
The $\vec k_\perp$-integral is solved using a Feynman parameter,
\begin{multline}
  \label{eq:kperprperpint}
  \int\! \frac{\rd^{d-2}\vec k_\perp}{
      \vec k_\perp^2 \, (\vec k_\perp - \vec r_\perp)^2}
  = \int_0^1\!\rd x \,
    \int\! \frac{\rd^{d-2}\vec k_\perp}{
      \bigl( (\vec k_\perp - x \, \vec r_\perp)^2
        + x(1-x) \, \vec r_\perp^{\,2} \bigr)^2}
\\
  = \pi^{1-\eps} \, \Gamma(1+\eps)
    \int_0^1\!\rd x \, \bigl( x(1-x) \, \vec r_\perp^{\,2} \bigr)^{-1-\eps}
  =  \frac{\pi^{1-\eps}}{(\vec r_\perp^{\,2})^{1+\eps}} \,
    \frac{\Gamma(1+\eps) \, \Gamma^2(-\eps)}{\Gamma(-2\eps)}
  \,.
\end{multline}
The complete 1-collinear contribution reads
\begin{align}
  \label{eq:forward1111calcF01cres}
  F^{(1c)}_0 = \frac{i\pi}{\vec r_\perp^{\,2} \, Q^2}
      \left(\frac{\mu^2}{\vec r_\perp^{\,2}}\right)^\eps \,
      \frac{e^{\eps\gammaE} \, \Gamma(1+\eps) \, \Gamma^2(-\eps)}{
        \Gamma(-2\eps)}
  \,.  
\end{align}

The 2-collinear contribution,
\begin{align}
  \label{eq:forward1111calccF02cint}
  F^{(2c)}_0 &= \frac{1}{2Q} \,
      \frac{\mu^{2\eps} \, e^{\eps\gammaE}}{2i\pi^{d/2}}
      \int\!\rd^{d-2}\vec k_\perp \!
      \int_{-\infty}^\infty\!
      \frac{\rd k^+ \, \rd k^-}{
        (k^+ k^- - \vec k_\perp^2 + i0) \,
        \bigl( k^+ (k^- - r^-_0)
          - (\vec k_\perp - \vec r_\perp)^2 + i0 \bigr)}
\nonumber \\* &\quad {}\times
      \left(
        \frac{1}{(k^++Q) k^- - \vec k_\perp^2 + i0}
        + \frac{1}{(k^+-Q) k^- - k^+ r^-_0 - \vec k_\perp^2
          + 2\vec r_\perp\cdot\vec k_\perp + i0}
      \right)
\nonumber \\* &\quad {}\times
      \left(
        \frac{1}{-k^+ + i0} + \frac{1}{k^+ + i0}
      \right)
  ,
\end{align}
is obtained from $F^{(1c)}_0$~(\ref{eq:forward1111calccint}) by replacing
$k^+ \leftrightarrow -k^-$ and $r^+_0 \to -r^-_0 = r^+_0$. So the same
result is reproduced,
\begin{align}
  \label{eq:forward1111calcF02cres}
  F^{(2c)}_0 = F^{(1c)}_0
    = \frac{i\pi}{\vec r_\perp^{\,2} \, Q^2}
      \left(\frac{\mu^2}{\vec r_\perp^{\,2}}\right)^\eps \,
      \frac{e^{\eps\gammaE} \, \Gamma(1+\eps) \, \Gamma^2(-\eps)}{
        \Gamma(-2\eps)}
  \,.  
\end{align}

\paragraph{Glauber contribution.} The Glauber-region
integral~(\ref{eq:forwardcalcgint}) for $n_1=n_2=n_3=n_4=1$ reads
\begin{align}
  \label{eq:forward1111calcgint}
  F^{(g)}_0 &= \frac{1}{2} \,
    \frac{\mu^{2\eps} \, e^{\eps\gammaE}}{2i\pi^{d/2}}
    \int\! \frac{\rd^{d-2}\vec k_\perp}{
      \vec k_\perp^2 \, (\vec k_\perp - \vec r_\perp)^2}
\nonumber \\* &\qquad {}\times
    \int_{-\infty}^\infty\!\rd k^+
    \left(
      \frac{1}{-Q k^+ - \vec k_\perp^2 + i0}
      + \frac{1}{Q k^+ - \vec k_\perp^2
          + 2\vec r_\perp\cdot\vec k_\perp + i0}
    \right)
\nonumber \\* &\qquad {}\times
    \int_{-\infty}^\infty\!\rd k^-
    \left(
      \frac{1}{Q k^- - \vec k_\perp^2 + i0}
      + \frac{1}{-Q k^- - \vec k_\perp^2
          + 2\vec r_\perp\cdot\vec k_\perp + i0}
    \right)
  .
\end{align}
The $k^\pm$-integrals converge because the leading behaviour for $|k^\pm|
\to \infty$ is cancelled between the two terms of each round bracket. All
integrals needed for~(\ref{eq:forward1111calcgint}) are known from
(\ref{eq:kpmintconvergeres}) and~(\ref{eq:kperprperpint}). The result is
\begin{align}
  \label{eq:forward1111calcF0gres}
  F^{(g)}_0 = \frac{i\pi}{\vec r_\perp^{\,2} \, Q^2}
      \left(\frac{\mu^2}{\vec r_\perp^{\,2}}\right)^\eps \,
      \frac{e^{\eps\gammaE} \, \Gamma(1+\eps) \, \Gamma^2(-\eps)}{
        \Gamma(-2\eps)}
  \,.  
\end{align}

\paragraph{Collinear--collinear overlap contribution.}
\begin{align}
  \label{eq:forward1111calcccint}
  F^{(1c,2c)}_0 &= \frac{1}{2Q^2} \,
      \frac{\mu^{2\eps} \, e^{\eps\gammaE}}{2i\pi^{d/2}}
      \int\!\rd^{d-2}\vec k_\perp \!
      \int_{-\infty}^\infty\!
      \frac{\rd k^+ \, \rd k^-}{
        (k^+ k^- - \vec k_\perp^2 + i0) \,
        \bigl( k^+ k^- - (\vec k_\perp - \vec r_\perp)^2 + i0 \bigr)}
\nonumber \\* &\qquad {}\times
      \underbrace{\left(
        \frac{1}{-k^+ + i0} + \frac{1}{k^+ + i0}
      \right)}_{-2i\pi \, \delta(k^+)}
      \,
      \underbrace{\left(
        \frac{1}{k^- + i0} + \frac{1}{-k^- + i0}
      \right)}_{-2i\pi \, \delta(k^-)}
  \,,
\end{align}
where the two round brackets are replaced according
to~(\ref{eq:deltacontribution}). Using~(\ref{eq:kperprperpint}), the
\mbox{1-collinear}--\mbox{2-collinear} overlap contribution yields
\begin{align}
  \label{eq:forward1111calcF01c2cres}
  F^{(1c,2c)}_0 = \frac{i\pi}{\vec r_\perp^{\,2} \, Q^2}
      \left(\frac{\mu^2}{\vec r_\perp^{\,2}}\right)^\eps \,
      \frac{e^{\eps\gammaE} \, \Gamma(1+\eps) \, \Gamma^2(-\eps)}{
        \Gamma(-2\eps)}
  \,.  
\end{align}

\paragraph{Collinear--Glauber overlap contributions.}
The 1-collinear--Glauber overlap contribution
\begin{align}
  F^{(1c,g)}_0 &= \frac{1}{2Q} \,
      \frac{\mu^{2\eps} \, e^{\eps\gammaE}}{2i\pi^{d/2}}
      \int\! \frac{\rd^{d-2}\vec k_\perp}{
        \vec k_\perp^2 \, (\vec k_\perp - \vec r_\perp)^2}
      \int_{-\infty}^\infty\!\rd k^-
      \left(
        \frac{1}{k^- + i0} + \frac{1}{-k^- + i0}
      \right)
  \nonumber \\* &\qquad {}\times
      \int_{-\infty}^\infty\!\rd k^+
      \left(
        \frac{1}{-Q k^+ - \vec k_\perp^2 + i0}
        + \frac{1}{Q k^+ - \vec k_\perp^2
            + 2\vec r_\perp\cdot\vec k_\perp + i0}
      \right)
  ,  
\end{align}
its symmetric counterpart~$F^{(2c,g)}_0$ (obtained via $k^+ \leftrightarrow
-k^-$), and the threefold overlap contribution
\begin{align}
  F^{(1c,2c,g)}_0 &= \frac{1}{2Q^2} \,
      \frac{\mu^{2\eps} \, e^{\eps\gammaE}}{2i\pi^{d/2}}
      \int\! \frac{\rd^{d-2}\vec k_\perp}{
        \vec k_\perp^2 \, (\vec k_\perp - \vec r_\perp)^2}
      \int_{-\infty}^\infty\!\rd k^+
      \left(
        \frac{1}{-k^+ + i0} + \frac{1}{k^+ + i0}
      \right)
  \nonumber \\* &\qquad {}\times
      \int_{-\infty}^\infty\!\rd k^-
      \left(
        \frac{1}{k^- + i0} + \frac{1}{-k^- + i0}
      \right)
\end{align}
are evaluated using the relations (\ref{eq:deltacontribution}),
(\ref{eq:kpmintconvergeres}) and~(\ref{eq:kperprperpint}). They yield
identical results,
\begin{align}
  \label{eq:forward1111calcoverlapres}
  F^{(1c,g)}_0 = F^{(2c,g)}_0 = F^{(1c,2c,g)}_0
    =\frac{i\pi}{\vec r_\perp^{\,2} \, Q^2}
      \left(\frac{\mu^2}{\vec r_\perp^{\,2}}\right)^\eps \,
      \frac{e^{\eps\gammaE} \, \Gamma(1+\eps) \, \Gamma^2(-\eps)}{
        \Gamma(-2\eps)}
  \,.  
\end{align}

\paragraph{Contributions involving the collinear-plane region.}
The contributions $F^{(cp)}_0$ (\ref{eq:forwardcalccpint}) and
$F^{(h,cp)}_0$, $F^{(1c,cp)}_0$, $F^{(2c,cp)}_0$, $F^{(h,1c,cp)}_0$,
$F^{(h,2c,cp)}_0$, $F^{(1c,2c,cp)}_0$,
$F^{(h,1c,2c,cp)}_0$~(\ref{eq:forwardcalcotherint}) all involve the
scaleless integral $\int\!\rd^{d-2}\vec k_\perp$ (with no further
integrand) which vanishes through dimensional regularization. This is still
true without analytic regularization, when the propagator powers are fixed
to $n_1=n_2=n_3=n_4=1$. But the \mbox{$k^\pm$-integrals} in these
contributions are problematic without analytic regulators. The integrands
fall off sufficiently fast at $|k^\pm| \to \infty$, but the collinear-plane
expansion makes poles above and below the real \mbox{$k^\pm$-axis} collapse
at $k^\pm = 0$. So the integration contour is pinched between poles which
are infinitesimally close to each other. These $k^\pm$-integrals are
regularized analytically in the evaluation of
appendix~\ref{app:forwardanalytic}, but here they are singular.

We may regularize the pinching of poles in the $k^\pm$-integrals without
resorting to analytic regularization by keeping a finite (instead of
infinitesimal) imaginary part~$i\delta$ (with $\delta > 0$) in
all propagators, e.g.
\begin{align}
  \label{eq:forward1111calcF0cp}
  F^{(cp)}_0 &= \frac{1}{2} \,
    \frac{\mu^{2\eps} \, e^{\eps\gammaE}}{2i\pi^{d/2}}
    \int\!\rd^{d-2}\vec k_\perp
    \int_{-\infty}^\infty\!
    \frac{\rd k^+ \, \rd k^-}{(k^+ k^- + i\delta)^2}
\nonumber \\* &\qquad {}\times
    \left(
      \frac{1}{k^+ (k^--Q) + i\delta}
      + \frac{1}{k^+ (k^-+Q) + i\delta}
    \right)
\nonumber \\* &\qquad {}\times
    \left(
      \frac{1}{(k^++Q) k^- + i\delta}
      + \frac{1}{(k^+-Q) k^- + i\delta}
    \right)
  .
\end{align}
When $\delta$ is fixed at some finite positive value, the poles above
and below the integration contours are separated from each other and the
collinear-plane integral is well-defined. The result from the
$k^\pm$-integrations is singular in the limit $\delta \to 0$, but it
is multiplied by the scaleless $\vec k_\perp$-integral, so the complete
contribution vanishes. This is true for all contributions involving the
collinear-plane region:
\begin{align}
  F^{(cp)}_0 &= F^{(h,cp)}_0 = F^{(1c,cp)}_0 = F^{(2c,cp)}_0
  = F^{(h,1c,cp)}_0 = F^{(h,2c,cp)}_0 = F^{(1c,2c,cp)}_0
\nonumber \\*
  &= F^{(h,1c,2c,cp)}_0
  = 0
  \,.
\end{align}
For consistency, the same regularization by a finite imaginary
part~$i\delta$ should be applied to the contributions from the other
regions as well, but there the limit $\delta \to 0$ (with $\delta > 0$) is
regular when it is performed before the limit $d \to 4$ from dimensional
regularization.

In addition, as we have already noted after the
identity~(\ref{eq:forwardidred}), all contributions involving the
collinear-plane expansion~$T^{(cp)}$ cancel each other in pairs at the
integrand level, according to~(\ref{eq:Sudakovcpcancel}), because the hard
expansion~$T^{(h)}$ does not change the integrand when it is applied in
addition to~$T^{(cp)}$:
\begin{multline}
  \bigl( F^{(cp)}_0 - F^{(h,cp)}_0 \bigr)
  - \bigl( F^{(1c,cp)}_0 - F^{(h,1c,cp)}_0 \bigr)
  - \bigl( F^{(2c,cp)}_0 - F^{(h,2c,cp)}_0 \bigr)
\\*
  + \bigl( F^{(1c,2c,cp)}_0 - F^{(h,1c,2c,cp)}_0 \bigr)
  = 0
  \,.
\end{multline}
So even if individual collinear-plane contributions were singular, their
sum would vanish at the integrand level.
Also, the collinear-plane contribution~$F^{(cp)}_0$ is parametrically of
order $(\vec r_\perp^{\,2})^{1-\eps} \, (Q^2)^{-3}$
(\ref{eq:forwardscaling1111}) and thus suppressed by two powers of $\vec
r_\perp^{\,2}/Q^2$ with respect to $F^{(1c)}_0$, $F^{(2c)}_0$ and
$F^{(g)}_0$. So it does not contribute to the leading-order result~$F_0$.

\paragraph{Overlap contributions involving the hard region.}
The overlap contributions involving the hard expansion and any of the
collinear or Glauber regions read
\begin{align}
  \label{eq:forward1111calchoverlap}
  F^{(h,1c)}_0 &= \frac{1}{2Q} \,
      \frac{\mu^{2\eps} \, e^{\eps\gammaE}}{2i\pi^{d/2}}
      \int\!\rd^{d-2}\vec k_\perp \!
      \int_{-\infty}^\infty\!
      \frac{\rd k^+ \, \rd k^-}{
        (k^+ k^- - \vec k_\perp^2 + i0)^2}
      \left(
        \frac{1}{k^- + i0} + \frac{1}{-k^- + i0}
      \right)
\nonumber \\* &\qquad {}\times
      \left(
        \frac{1}{k^+ (k^--Q) - \vec k_\perp^2 + i0}
        + \frac{1}{k^+ (k^-+Q) - \vec k_\perp^2 + i0}
      \right)
    ,
\nonumber \\
  F^{(h,g)}_0 &= \frac{1}{2} \,
    \frac{\mu^{2\eps} \, e^{\eps\gammaE}}{2i\pi^{d/2}}
    \int\! \frac{\rd^{d-2}\vec k_\perp}{(\vec k_\perp^2)^2}
    \int_{-\infty}^\infty\!\rd k^+
    \left(
      \frac{1}{-Q k^+ - \vec k_\perp^2 + i0}
      + \frac{1}{Q k^+ - \vec k_\perp^2 + i0}
    \right)
\nonumber \\* &\qquad {}\times
    \int_{-\infty}^\infty\!\rd k^-
    \left(
      \frac{1}{Q k^- - \vec k_\perp^2 + i0}
      + \frac{1}{-Q k^- - \vec k_\perp^2 + i0}
    \right)
    ,
\nonumber \\
  F^{(h,1c,2c)}_0 &= \frac{1}{2Q^2} \,
      \frac{\mu^{2\eps} \, e^{\eps\gammaE}}{2i\pi^{d/2}}
      \int\!\rd^{d-2}\vec k_\perp \!
      \int_{-\infty}^\infty\!
      \frac{\rd k^+ \, \rd k^-}{(k^+ k^- - \vec k_\perp^2 + i0)^2}
\nonumber \\* &\qquad {}\times
      \left(
        \frac{1}{-k^+ + i0} + \frac{1}{k^+ + i0}
      \right)
      \left(
        \frac{1}{k^- + i0} + \frac{1}{-k^- + i0}
      \right)
    ,
\nonumber \\
  F^{(h,1c,g)}_0 &= \frac{1}{2Q} \,
      \frac{\mu^{2\eps} \, e^{\eps\gammaE}}{2i\pi^{d/2}}
      \int\! \frac{\rd^{d-2}\vec k_\perp}{(\vec k_\perp^2)^2}
      \int_{-\infty}^\infty\!\rd k^+
      \left(
        \frac{1}{-Q k^+ - \vec k_\perp^2 + i0}
        + \frac{1}{Q k^+ - \vec k_\perp^2 + i0}
      \right)
  \nonumber \\* &\qquad {}\times
      \int_{-\infty}^\infty\!\rd k^-
      \left(
        \frac{1}{k^- + i0} + \frac{1}{-k^- + i0}
      \right)
    ,
\nonumber \\
  F^{(h,1c,2c,g)}_0 &= \frac{1}{2Q^2} \,
      \frac{\mu^{2\eps} \, e^{\eps\gammaE}}{2i\pi^{d/2}}
      \int\! \frac{\rd^{d-2}\vec k_\perp}{(\vec k_\perp^2)^2}
      \int_{-\infty}^\infty\!\rd k^+
      \left(
        \frac{1}{-k^+ + i0} + \frac{1}{k^+ + i0}
      \right)
\nonumber \\* &\qquad {}\times
      \int_{-\infty}^\infty\!\rd k^-
      \left(
        \frac{1}{k^- + i0} + \frac{1}{-k^- + i0}
      \right)
    ,
\end{align}
omitting the symmetric integrals $F^{(h,2c)}_0$ and $F^{(h,2c,g)}_0$
obtained from the integrals above via $(1c) \to (2c)$ with $k^+
\leftrightarrow -k^-$.
The $k^\pm$-integrals are evaluated using (\ref{eq:deltacontribution})
and~(\ref{eq:kpmintconvergeres}). All contributions
in~(\ref{eq:forward1111calchoverlap}) identically yield
\begin{align}
  \label{eq:forward1111calchoverlapres}  
  F^{(h,1c)}_0 &= F^{(h,2c)}_0 = F^{(h,g)}_0 = F^{(h,1c,2c)}_0
    = F^{(h,1c,g)}_0 = F^{(h,2c,g)}_0 = F^{(h,1c,2c,g)}_0
\nonumber \\*
  &= \frac{i\pi}{Q^2} \,
    \frac{\mu^{2\eps} \, e^{\eps\gammaE}}{\pi^{1-\eps}}
    \int\! \frac{\rd^{d-2}\vec k_\perp}{(\vec k_\perp^2)^2}
    = 0
  \,,
\end{align}
which is a scaleless integral through dimensional regularization.

\section{Expansion by regions with a finite boundary}
\label{app:finiteboundary}

This appendix illustrates the behaviour of the expansion by regions and the
formalism of sections \ref{sec:formalism} and~\ref{sec:formalismnc} when a
finite integration boundary is involved. Consider the one-dimensional
integral
\begin{align}
  \label{eq:finborig}
  F = \int_0^B\!\rd k \, I
  \quad \text{with} \quad
  I = \frac{1}{(k+m)^\alpha \, (k+M)^\beta}
  \quad \text{and} \quad
  0 < m \ll M \ll B
  \,.
\end{align}

\subsection{Direct evaluation with a finite boundary}
\label{app:finiteboundarydirect}

The integration domain $D = [0,B]$ in~(\ref{eq:finborig}) involves the
finite, non-zero boundary~$B$, which introduces an additional scale into
the problem. Let us define two regions,
\begin{itemize}
\item the \textbf{soft region \boldmath$(s)$}, characterized by $k \sim m$,
  with the expansion
  \begin{align}
    \label{eq:finbTs}
    T^{(s)} \equiv \sum_{j_1=0}^\infty T^{(s)}_{j_1} \,,\qquad
    T^{(s)}_{j_1} I =
      \frac{(\beta)_{j_1} \, (-1)^{j_1}}{j_1! \, M^{\beta+j_1}} \,
      \frac{k^{j_1}}{(k+m)^\alpha}
    \,,
  \end{align}
  converging absolutely for $k \in D_s = [0,\lambda]$ with
  $m \ll \lambda \ll M$,
\item and the \textbf{hard region \boldmath$(h)$}, characterized by $k \sim
  M$, with the expansion
  \begin{align}
    \label{eq:finbTh}
    T^{(h)} \equiv \sum_{j_2=0}^\infty T^{(h)}_{j_2} \,,\qquad
    T^{(h)}_{j_2} I = \frac{(\alpha)_{j_2} \, (-m)^{j_2}}{j_2!} \,
      \frac{1}{k^{\alpha+j_2} \, (k+M)^\beta}
    \,,
  \end{align}
  converging absolutely for $k \in D_h = (\lambda,B]$.
\end{itemize}
We have $D_s \cap D_h = \emptyset$, $D_s \cup D_h = D$ and
\begin{align}
  \label{eq:finbTsh}
  T^{(s)} T^{(h)} &= T^{(h)} T^{(s)} \equiv T^{(s,h)}
  \equiv \sum_{j_1,j_2=0}^\infty T^{(s,h)}_{j_1,j_2}
  \,,
\nonumber \\
  T^{(s,h)}_{j_1,j_2} I
  &= \frac{(\beta)_{j_1} \, (\alpha)_{j_2} \, (-1)^{j_1} \, (-m)^{j_2}}{
      j_1! \, j_2! \, M^{\beta+j_1}} \,
    \frac{1}{k^{\alpha+j_2-j_1}}
  \,.
\end{align}
Also the integrals
\begin{align}
  \label{eq:finbints}
  F^{(s)}_{j_1} &= \frac{(\beta)_{j_1} \, (-1)^{j_1}}{j_1! \, M^{\beta+j_1}}
    \int_0^B\!\frac{\rd k \, k^{j_1}}{(k+m)^\alpha}
    \,,
\nonumber \\*
  F^{(h)}_{j_2} &= \frac{(\alpha)_{j_2} \, (-m)^{j_2}}{j_2!}
    \int_0^B\!\frac{\rd k}{k^{\alpha+j_2} \, (k+M)^\beta}
    \,,
\nonumber \\*
  F^{(s,h)}_{j_1,j_2} &=
    \frac{(\beta)_{j_1} \, (\alpha)_{j_2} \, (-1)^{j_1} \, (-m)^{j_2}}{
      j_1! \, j_2! \, M^{\beta+j_1}}
    \int_0^B\!\frac{\rd k}{k^{\alpha+j_2-j_1}}
\end{align}
are well-defined if we use the denominator power~$\alpha$ as analytic
regulator, choosing its value for each of the integrals $F^{(h)}_{j_2}$ and
$F^{(s,h)}_{j_1,j_2}$ such that they converge at $k \to 0$.  So the
conditions \ref{item:EbRcondD}, \ref{item:EbRcondcomm}
and~\ref{item:EbRcondreg} of the formalism in section~\ref{sec:formalism}
hold. But we have a problem with the convergence of the summations as
required in condition~\ref{item:EbRcondsums}: The series
$\sum_{j_1=0}^\infty F^{(s)}_{j_1}$ and $\sum_{j_1=0}^\infty
F^{(s,h)}_{j_1,j_2}$ (for fixed~$j_2$) diverge, because these two integrals
involve contributions from $k \sim B$ producing powers $(B/M)^{j_1}$ (with
$B/M \gg 1$).\footnote{%
  Note, however, that the series $\sum_{j_2=0}^\infty F^{(h)}_{j_2}$ and
  $\sum_{j_2=0}^\infty F^{(s,h)}_{j_1,j_2}$ (for fixed~$j_1$) are
  convergent, although contributions $(m/k)^{j_2}$ with $k \to 0$ are
  involved. But the analytic regularization prevents contributions from the
  zero-boundary $k=0$ and makes the integrals yield results scaling only
  with $k \sim M$ and $k \sim B$.}

With $F^{(s)}$ and $F^{(s,h)}$ being divergent and therefore
condition~\ref{item:EbRcondsums} violated, we cannot naively claim the
master identity~(\ref{eq:formalismidentity}), here $F = F^{(s)} + F^{(h)} -
F^{(s,h)}$, to be true. However, the difference $F^{(s)} - F^{(s,h)}$ is
finite when the terms are combined before summing over~$j_1$. This can be
seen by writing
\begin{align}
  \label{eq:finbextend}
  \int_0^B\!\frac{\rd k \, k^{j_1}}{(k+m)^\alpha}
  &= \int_0^\infty\!\frac{\rd k \, k^{j_1}}{(k+m)^\alpha}
    - \int_B^\infty\!\frac{\rd k \, k^{j_1}}{(k+m)^\alpha}
\nonumber \\*
  &= \frac{j_1! \, \Gamma(\alpha-1-j_1)}{\Gamma(\alpha) \, m^{\alpha-1-j_1}}
    - \sum_{j_2=0}^\infty \frac{(\alpha)_{j_2} \, (-m)^{j_2}}{j_2!}
      \int_B^\infty\!\frac{\rd k}{k^{\alpha+j_2-j_1}}
  \,,
\end{align}
using the power~$\alpha$ as analytic regulator also at $k \to \infty$ and
expanding the second term safely for $m \ll B \le k$. The two contributions
combined yield
\begin{multline}
  \label{eq:finbcomb}
  F^{(s)}_{j_1} - \sum_{j_2=0}^\infty F^{(s,h)}_{j_1,j_2}
  = \frac{(\beta)_{j_1} \, \Gamma(\alpha-1-j_1)}{
      \Gamma(\alpha) \, m^{\alpha-1} \, M^\beta}
    \left(-\frac{m}{M}\right)^{j_1}
\\* {}
    - \frac{(\beta)_{j_1} \, (-1)^{j_1}}{j_1! \, M^{\beta+j_1}} \,
      \sum_{j_2=0}^\infty
      \frac{(\alpha)_{j_2} \, (-m)^{j_2}}{j_2!}
      \left( \int_0^B\!\frac{\rd k}{k^{\alpha+j_2-j_1}}
        + \int_B^\infty\!\frac{\rd k}{k^{\alpha+j_2-j_1}} \right)
  .
\end{multline}
The sum of the two integrals in the second line of~(\ref{eq:finbcomb}) is
zero (using analytic regularization), because
\begin{align}
  \label{eq:finbscaleless}
  \int_0^B\!\frac{\rd k}{k^{\alpha+j_2-j_1}} &=
    \frac{B^{1-\alpha+j_1-j_2}}{1-\alpha+j_1-j_2}
    \quad \text{for } \Rep\alpha < 1+j_1-j_2 \,,
\nonumber \\*
  \int_B^\infty\!\frac{\rd k}{k^{\alpha+j_2-j_1}} &=
    -\frac{B^{1-\alpha+j_1-j_2}}{1-\alpha+j_1-j_2}
    \quad \text{for } \Rep\alpha > 1+j_1-j_2 \,,
\end{align}
or simply because the two integrals add up to a scaleless integral from 0
to~$\infty$.

The upper limit of the integrals~$F^{(h)}_{j_2}$ in~(\ref{eq:finbints}) can
be extended to~$\infty$ like in~(\ref{eq:finbextend}), and the second term
is evaluated in analogy to~(\ref{eq:finbscaleless}). In total the result is
\begin{align}
  \label{eq:finbres}
  F &= \sum_{j_1=0}^\infty \left(
        F^{(s)}_{j_1} - \sum_{j_2=0}^\infty F^{(s,h)}_{j_1,j_2} \right)
      + \sum_{j_2=0}^\infty F^{(h)}_{j_2}
\nonumber \\
  &= \frac{1}{m^{\alpha-1} \, M^\beta} \, \frac{1}{\alpha-1}
      \sum_{j_1=0}^\infty \left(\frac{m}{M}\right)^{j_1} \,
      \frac{(\beta)_{j_1}}{(2-\alpha)_{j_1}}
\nonumber \\* &\qquad{}
    + \frac{1}{M^{\alpha+\beta-1}} \,
      \frac{\Gamma(1-\alpha)}{\Gamma(\beta)}
      \sum_{j_2=0}^\infty \left(\frac{m}{M}\right)^{j_2} \,
      \frac{\Gamma(\alpha+\beta-1+j_2)}{j_2!}
\nonumber \\* &\qquad{}
    - \frac{1}{B^{\alpha+\beta-1}} \sum_{j_1,j_2=0}^\infty
      \left(-\frac{M}{B}\right)^{j_1} \left(-\frac{m}{B}\right)^{j_2} \,
      \frac{(\beta)_{j_1} \, (\alpha)_{j_2}}{
        j_1! \, j_2! \, (\alpha+\beta-1+j_{12})}
  \,.
\end{align}
The series expansions in~(\ref{eq:finbres}) converge and reproduce the exact
result of the integral~$F$ in~(\ref{eq:finborig}). Note that due to the
finite integration boundary at $k=B$ the individual integrals
$F^{(s)}_{j_1}$ and $F^{(h)}_{j_2}$ in~(\ref{eq:finbints}) do not yield
homogeneous functions of $m$, $M$ and~$B$, and therefore the overlap
contributions $F^{(s,h)}_{j_1,j_2}$ are not scaleless. Additional
expansions of $F^{(s)}_{j_1}$ and $F^{(h)}_{j_2}$ as
in~(\ref{eq:finbextend}) are needed to arrive at the
form~(\ref{eq:finbres}) where all summation terms are homogeneous functions
of $m$, $M$ and~$B$.

Individual terms in the result~(\ref{eq:finbres}) are singular when
$\alpha$ is a positive integer or when $(\alpha+\beta-1)$ is zero or a
negative integer. One can check, however, that all these singularities
cancel between the terms in~(\ref{eq:finbres}) such that the total result
is finite as it should be.

The difficulties with the convergence of the expansions $F^{(s)}$ and
$F^{(s,h)}$ may be avoided if one is only interested in a leading-order
expansion. According to~(\ref{eq:formalismidentity0}), the leading-order
approximation to~$F$ is given by
\begin{align}
  F_0 = F^{(s)}_0 + F^{(h)}_0 - F^{(s,h)}_{0,0}
      = \frac{1}{M^\beta} \int_0^B\!\frac{\rd k}{(k+m)^\alpha}
        + \int_0^B\!\frac{\rd k}{k^\alpha \, (k+M)^\beta}
        - \frac{1}{M^\beta} \int_0^B\!\frac{\rd k}{k^\alpha}
  \,,
\end{align}
with the zeroth-order terms from~(\ref{eq:finbints}). It does not require
the convergence of the full series. The first two contributions to the
leading-order approximation~$F_0$ are not homogeneous functions of $m$, $M$
and~$B$; they may be further approximated using~(\ref{eq:finbextend}) and
keeping only the leading term of the additional expansion. This leads to a
different leading-order approximation with homogeneous contributions:
\begin{align}
  \tilde F_0
    &= \frac{1}{M^\beta} \int_0^\infty\!\frac{\rd k}{(k+m)^\alpha}
      + \int_0^\infty\!\frac{\rd k}{k^\alpha \, (k+M)^\beta}
      - \int_B^\infty\!\frac{\rd k}{k^{\alpha+\beta}}
      - \frac{1}{M^\beta} \int_0^\infty\!\frac{\rd k}{k^\alpha}
\nonumber \\*
    &= \frac{1}{m^{\alpha-1} \, M^\beta} \, \frac{1}{\alpha-1}
      + \frac{1}{M^{\alpha+\beta-1}} \,
      \frac{\Gamma(1-\alpha) \, \Gamma(\alpha+\beta-1)}{\Gamma(\beta)}
      - \frac{1}{B^{\alpha+\beta-1}} \,
        \frac{1}{\alpha+\beta-1}
  \,,
\end{align}
which corresponds to the leading-order term (with $j_1=j_2=0$)
in~(\ref{eq:finbres}).

\subsection{Evaluation by extending to an infinite boundary}
\label{app:finiteboundarytheta}

There is an alternative way to treat the integral~(\ref{eq:finborig}) with
finite boundaries in the expansion by regions, which, however, requires the
formalism for non-commuting expansions introduced in
section~\ref{sec:formalismnc}.

Let us write the original integral as
\begin{align}
  \label{eq:finborig2}
  F = \int_0^\infty\!\rd k \, \hat I
  \quad \text{with} \quad
  \hat I = \frac{\theta(B-k)}{(k+m)^\alpha \, (k+M)^\beta}
  \,,
\end{align}
with infinite upper boundary of the integration domain $\hat D=[0,\infty)$,
expressing the restriction $k \le B$ through the Heaviside step
function~$\theta$. With a third factor and the new scale~$B$ in the
integrand~$\hat I$, we add a third region. Let us keep $D_s = [0,\lambda]$,
but redefine $D_h = (\lambda,\Lambda]$ with $m \ll \lambda \ll M \ll
\Lambda \ll B$. So, in the soft and hard regions, we have $k \le \Lambda
\ll B$, which allows us to use
\begin{align}
  \theta(B-k) = 1 \,,\quad k < B \,,
\end{align}
there.\footnote{%
  Formally this is the expansion $\theta(B-k) = \theta(B) - k \, \delta(B) +
  \frac{k^2}{2} \, \delta'(B) + \ldots \equiv 1$ with $\theta(B)=1$ and
  $\delta^{(j)}(B)=0$ for $B>0$ to all orders~$j$.}
So the soft and hard expansions of the new integrand, $T^{(s)}$ and
$T^{(h)}$, as well as their combined double expansion $T^{(s,h)}$ exactly
correspond to the known expressions (\ref{eq:finbTs}), (\ref{eq:finbTh})
and~(\ref{eq:finbTsh}). Additionally we now have
\begin{itemize}
\item the \textbf{ultrahard region \boldmath$(uh)$}, characterized by $k
  \sim B$, with the expansion
  \begin{align}
    \label{eq:finbTuh}
    T^{(uh)} \equiv \sum_{j_1,j_2=0}^\infty T^{(uh)}_{j_1,j_2} \,,\qquad
    T^{(uh)}_{j_1,j_2} \hat I =
      \frac{(\beta)_{j_1} \, (\alpha)_{j_2} \, (-M)^{j_1} \, (-m)^{j_2}}{
        j_1! \, j_2!} \,
      \frac{\theta(B-k)}{k^{\alpha+\beta+j_{12}}}
    \,,
  \end{align}
  converging absolutely for $k \in D_{uh} = (\Lambda,\infty)$.
\end{itemize}
We have chosen the new convergence domains such that $D_s \cap D_h = D_s
\cap D_{uh} = D_h \cap D_{uh} = \emptyset$ and $D_s \cup D_h \cup D_{uh} =
\hat D$.
The ultrahard expansion commutes with the hard expansion,
\begin{align}
  \label{eq:finbThuh}
  T^{(h)} T^{(uh)} &= T^{(uh)} T^{(h)} \equiv T^{(h,uh)}
  \equiv \sum_{j_1,j_2=0}^\infty T^{(h,uh)}_{j_1,j_2}
  \,,
\nonumber \\
  T^{(h,uh)}_{j_1,j_2} \hat I &=
    \frac{(\beta)_{j_1} \, (\alpha)_{j_2} \, (-M)^{j_1} \, (-m)^{j_2}}{
      j_1! \, j_2!} \,
    \frac{1}{k^{\alpha+\beta+j_{12}}}
  \,,
\end{align}
but not with the soft expansion:
\begin{align}
  \label{eq:finbTsuh}
  T^{(uh)} T^{(s)} &\equiv T^{(uh \leftarrow s)}
    = T^{(s,h)}
    \,,
\nonumber \\
  T^{(s)} T^{(uh)} &\equiv T^{(s \leftarrow uh)}
    = T^{(h,uh)}
  \,.
\end{align}
Therefore we need the formalism for non-commuting expansions as developed
in section~\ref{sec:formalismnc}. The subset of regions with non-commuting
expansions is $\Rnc = \{s,uh\}$, and condition~\ref{item:EbRcondcommnc}
(p.~\pageref{item:EbRcondcommnc}) holds with $\Rc = \{h\}$.
Also condition~\ref{item:EbRcondnc} (p.~\pageref{item:EbRcondnc}) is
fulfilled because the hard expansion does not further change the
integrands which are already doubly expanded with $T^{(s)}$ and $T^{(uh)}$:
\begin{equation}
  T^{(uh \leftarrow s,h)} = T^{(uh \leftarrow s)}
  \,,\qquad
  T^{(s \leftarrow uh,h)} = T^{(s \leftarrow uh)}
  \,.
\end{equation}

The integrals needed within this formalism read
\begin{align}
  \label{eq:finb2ints}
  F^{(s)}_{j_1} &= \frac{(\beta)_{j_1} \, (-1)^{j_1}}{j_1! \, M^{\beta+j_1}}
      \int_0^\infty\!\frac{\rd k \, k^{j_1}}{(k+m)^\alpha}
    = \frac{1}{m^{\alpha-1} \, M^\beta} \left(\frac{m}{M}\right)^{j_1} \,
      \frac{1}{\alpha-1} \, \frac{(\beta)_{j_1}}{(2-\alpha)_{j_1}}
    \,,
\nonumber \\
  F^{(h)}_{j_2} &= \frac{(\alpha)_{j_2} \, (-m)^{j_2}}{j_2!}
      \int_0^\infty\!\frac{\rd k}{k^{\alpha+j_2} \, (k+M)^\beta}
\nonumber \\* &
    = \frac{1}{M^{\alpha+\beta-1}}
      \left(\frac{m}{M}\right)^{j_2} \,
      \frac{\Gamma(1-\alpha)}{\Gamma(\beta)} \,
      \frac{\Gamma(\alpha+\beta-1+j_2)}{j_2!}
    \,,
\nonumber \\
  F^{(uh)}_{j_1,j_2} &=
      \frac{(\beta)_{j_1} \, (\alpha)_{j_2} \, (-M)^{j_1} \, (-m)^{j_2}}{
        j_1! \, j_2!} \,
      \int_0^\infty\!\rd k \, \frac{\theta(B-k)}{k^{\alpha+\beta+j_{12}}}
\nonumber \\* &
    = -\frac{1}{B^{\alpha+\beta-1}}
      \left(-\frac{M}{B}\right)^{j_1} \left(-\frac{m}{B}\right)^{j_2} \,
      \frac{(\beta)_{j_1} \, (\alpha)_{j_2}}{
        j_1! \, j_2! \, (\alpha+\beta-1+j_{12})}
    \,,
\nonumber \\
  F^{(s,h)}_{j_1,j_2} &=
      \frac{(\beta)_{j_1} \, (\alpha)_{j_2} \, (-1)^{j_1} \, (-m)^{j_2}}{
        j_1! \, j_2! \, M^{\beta+j_1}}
      \int_0^\infty\!\frac{\rd k}{k^{\alpha+j_2-j_1}}
    = 0
    \,,
\nonumber \\
  F^{(h,uh)}_{j_2,j_1} &=
      \frac{(\beta)_{j_1} \, (\alpha)_{j_2} \, (-M)^{j_1} \, (-m)^{j_2}}{
        j_1! \, j_2!} \,
      \int_0^\infty\!\frac{\rd k}{k^{\alpha+\beta+j_{12}}}
    = 0
    \,,
\end{align}
i.e.\ all combinations of expansions which commute with each other. All
integrals in~(\ref{eq:finb2ints}) are well-defined via the analytic
regulator~$\alpha$ (condition~\ref{item:EbRcondreg}), and the summations
over $j_1,j_2$ converge absolutely for $m \ll M \ll B$
(condition~\ref{item:EbRcondsums}).

According to~(\ref{eq:formalismncidentity}), the integral can be expressed
as
\begin{align}
  \label{eq:finb2identity}
  F &= F^{(s)} + F^{(h)} + F^{(uh)}
    - F^{(s,h)} - F^{(h,uh)}
\nonumber \\* &\qquad{}
    - F^{(uh \leftarrow s)}_{[uh]} + F^{(uh \leftarrow s,h)}_{[uh]}
    - F^{(s \leftarrow uh)}_{[s]} + F^{(s \leftarrow uh,h)}_{[s]}
  \,.
\end{align}
As condition~\ref{item:EbRcondnc} holds, the extra terms in the second line
of~(\ref{eq:finb2identity}) cancel with each other. So according
to~(\ref{eq:formalismncidred}),
\begin{align}
  \label{eq:finb2idred}
  F &= F^{(s)} + F^{(h)} + F^{(uh)}
    - F^{(s,h)} - F^{(h,uh)}
  \,.
\end{align}
The first three terms with single expansions are series of homogeneous
functions of $m$, $M$ and~$B$, so the overlap contributions in the last two
terms yield scaleless integrals and vanish, see~(\ref{eq:finb2ints}). The
final result shortens to
\begin{align}
  \label{eq:finb2idred2}
  F &= F^{(s)} + F^{(h)} + F^{(uh)}
  \,,
\end{align}
reproducing the result~(\ref{eq:finbres}) from the previous calculation,
when the expressions in~(\ref{eq:finb2ints}) are added together.

Let us have a short look at the extra terms in~(\ref{eq:finb2identity})
which we have dropped:
\begin{align}
  F^{(uh \leftarrow s)}_{j_2,j_1 \, [uh]}
    &= F^{(uh \leftarrow s,h)}_{j_2,j_1 \, [uh]}
    = \frac{(\beta)_{j_1} \, (\alpha)_{j_2} \, (-1)^{j_1} \, (-m)^{j_2}}{
        j_1! \, j_2! \, M^{\beta+j_1}}
      \int_\Lambda^\infty\!\frac{\rd k}{k^{\alpha+j_2-j_1}}
\nonumber \\*
    &= -\frac{1}{\Lambda^{\alpha-1} \, M^\beta}
      \left(-\frac{\Lambda}{M}\right)^{j_1}
      \left(-\frac{m}{\Lambda}\right)^{j_2} \,
      \frac{(\beta)_{j_1} \, (\alpha)_{j_2}}{
        j_1! \, j_2! \, (1-\alpha+j_1-j_2)}
    \,,
\nonumber \\
  F^{(s \leftarrow uh)}_{j_1,j_2 \, [s]}
    &= F^{(s \leftarrow uh,h)}_{j_1,j_2 \, [s]}
    = \frac{(\beta)_{j_1} \, (\alpha)_{j_2} \, (-M)^{j_1} \, (-m)^{j_2}}{
        j_1! \, j_2!} \,
      \int_0^\lambda\!\frac{\rd k}{k^{\alpha+\beta+j_{12}}}
\nonumber \\*
    &= -\frac{1}{\lambda^{\alpha+\beta-1}}
      \left(-\frac{M}{\lambda}\right)^{j_1}
      \left(-\frac{m}{\lambda}\right)^{j_2} \,
      \frac{(\beta)_{j_1} \, (\alpha)_{j_2}}{
        j_1! \, j_2! \, (\alpha+\beta-1+j_{12})}
    \,.
\end{align}
The summations of these terms over~$j_1$ diverge, because $\Lambda/M \gg 1$
and $M/\lambda \gg 1$. So by choosing to expand the integral
$F$~(\ref{eq:finborig}) via extending its upper boundary to infinity and
introducing a third region, we still cannot completely avoid the problem of
diverging series expansions encountered in
section~\ref{app:finiteboundarydirect}. But at least we have isolated the
divergence problems in terms which disappear from the final
result~(\ref{eq:finb2idred2}) in a well-understood way.

Note that also other series expansions which appear in intermediate steps
of the derivation, in particular $\sum_{j_1} F^{(s,h)}_{j_1,j_2\,[h]}$ and
$\sum_{j_1} F^{(h,uh)}_{j_2,j_1\,[h]}$, diverge because the domain~$D_h$
has boundaries touching both the soft and the ultrahard domain. So we can
only sum the expansions over~$j_1$ up to~$\infty$ once the terms have been
combined into the integrals~(\ref{eq:finb2ints}) performed over the
complete integration domain. For those the convergence of the series
expansions is required by condition~\ref{item:EbRcondsums} in
section~\ref{sec:formalism}.

\bibliographystyle{JHEP-bj} 
\bibliography{ExpansionByRegions}

\end{document}